\let\CCKernelLabel\label
\documentclass[aps,prl,twocolumn,10pt,amsmath,amssymb,superscriptaddress,nofootinbib,floatfix,longbibliography]{revtex4-2}
\usepackage[T1]{fontenc}
\usepackage[english]{babel}

\usepackage{array}[=2016-10-06]
\usepackage{graphicx,bm,booktabs,mathtools,microtype,amsthm,array}
\usepackage{stmaryrd}
\usepackage[dvipsnames]{xcolor}
\usepackage{tikz}
\usetikzlibrary{decorations.markings}
\usepackage{listings}

\makeatletter
\@ifl@t@r\fmtversion{2023-06-01}{%
  \let\label\CCKernelLabel
  \let\ltx@label\CCKernelLabel
}{}
\makeatother
\usepackage[colorlinks=true,allcolors=MidnightBlue,pdfusetitle]{hyperref}

\addto\captionsenglish{\renewcommand{\figurename}{Fig.}}
\makeatletter
\renewcommand*{\fnum@figure}{\figurename\thefigure}
\makeatother

\allowdisplaybreaks

\newcommand{\lsim}{{\;\raise0.3ex\hbox{$<$\kern-0.75em\raise-1.1ex\hbox{$\sim$}}\;}}
\newcommand{\gsim}{{\;\raise0.3ex\hbox{$>$\kern-0.75em\raise-1.1ex\hbox{$\sim$}}\;}}
\newcommand{\bea}{\begin{equation}}
\newcommand{\eea}{\end{equation}}
\newcommand{\beq}{\begin{equation}}
\newcommand{\eeq}{\end{equation}}
\newcommand{\ba}{\begin{array}}
\newcommand{\ea}{\end{array}}
\newcommand{\beqa}{\begin{equation}}
\newcommand{\eeqa}{\end{equation}}
\newcommand{\beqs}{\begin{subequations}}
\newcommand{\eeqs}{\end{subequations}}

\def\baa{\begin{array}}
\def\eaa{\end{array}}

\mathchardef\minus="002D
\def\nn{\nonumber}

\renewcommand{\rm}{\mathrm}
\def\({\left(}
\def\){\right)}
\def\[{\left[}
\def\]{\right]}

\def\to{\rightarrow}

\newcommand{\di}{\mathrm{d}}

\def\vs{\vspace*{1mm}}

\def\hp{\hspace*{0.3mm}}
\def\hs{\hspace*{0.3mm}}

\def\hsm{\hspace*{-0.3mm}}

\def\fNL{f_{\rm{NL}}}

\def\fNLCC{f_{\rm{NL}}^{\rm{CC}}}
\def\ii  {\rm{i}}

\def\tnu{\tilde{\nu}}
\def\tlambda{\tilde{\lambda}}

\newcommand{\m}{\widetilde M_R}
\newcommand{\Nt}{\widetilde N}
\newcommand{\Pz}{P_\zeta}
\newcommand{\Cc}{\mathbb C_+}
\newcommand{\Ch}{\mathcal C_{\rm{hard}}}
\newcommand{\Cs}{\mathcal C_{\rm{soft}}}
\newcommand{\FP}{\operatorname{FP}}
\newcommand{\AC}{\operatorname{AC}}
\newcommand{\res}{\operatorname{Res}}
\newcommand{\vct}[1]{\boldsymbol{#1}}
\newcommand{\Bphi}{\langle\delta\phi_{\vct k_1^{}}^{}\!\delta\phi_{\vct k_2^{}}^{}\hspace*{-0.3mm}\delta\phi_{\vct k_3^{}}^{}\rangle'}
\newcommand{\Gn}[6][]{G_{#2}\!#1\left[\begin{matrix}#3\,;\,#4\\#5\,;\,#6\end{matrix}\right]}

\AddToHook{cmd/thebibliography/after}{%
  \rightskip=0pt plus 1em\relax
  \interlinepenalty=10000\relax
}
\allowdisplaybreaks[2]
\begin{document}

\title{Complete Cosmological Collider Signatures from Right-handed Neutrino Loops 
\texorpdfstring{\\[1mm]}{ }
with Chemical-Potential Enhancement}
\author{{\sc Linghao Song}}\email{lh.song@sjtu.edu.cn}
\affiliation{Tsung-Dao Lee Institute $\&$ School of Physics and Astronomy, 
\\
Shanghai Jiao Tong University, Shanghai, China}
\author{{\sc Jingtao You}}\email{119760616yjt@sjtu.edu.cn}
\affiliation{Tsung-Dao Lee Institute $\&$ School of Physics and Astronomy, 
\\
Shanghai Jiao Tong University, Shanghai, China}
\author{{\sc Ming Cheng}}\email{mingcheng2017@gmail.com}
\affiliation{Tsung-Dao Lee Institute $\&$ School of Physics and Astronomy, 
\\
Shanghai Jiao Tong University, Shanghai, China}
\author{{\sc Hong-Jian He}}\email{hjhe@sjtu.edu.cn}
\affiliation{Tsung-Dao Lee Institute $\&$ School of Physics and Astronomy, 
\\ 
Shanghai Jiao Tong University, Shanghai, China}
\affiliation{Department of Physics, Tsinghua University, Beijing, China; 
\\
Center for High Energy Physics, Peking University, Beijing, China}
\author{{\sc Xingang Chen}}\email{xingang.chen@cfa.harvard.edu}
\affiliation{Institute for Theory and Computation, 
Harvard-Smithsonian Center for Astrophysics, 
Cambridge, MA, USA}
\begin{abstract}
We study the complete cosmological collider (CC) signatures from right-handed neutrino loops with chemical-potential enhancement.\ 
These CC signals are induced by the inflaton-right-handed-neutrino interaction and can be extracted in the collinear squeezed configuration, which allows the right-handed-neutrino triangle loop to be factorized into the hard and soft parts.\ 
We precisely compute the soft part of the fermion loop for the first time 
by using the complete mode functions and the full loop integral, 
and treat the hard part as a tree-level subdiagram.\ 
We subtract the large-momentum asymptotic terms of the soft loop integrand and retain their finite contributions to the oscillation. We then evaluate the full spinor-helicity and Schwinger-Keldysh sums to obtain a finite, closed analytic formula for the complex clock coefficient.\ 
We further derive a compact formula for the leading CC signals 
in the large-chemical-potential expansion.\ 
We analyze the structure of the soft-part contribution by examining the sum of residues 
of the Mellin poles of the fermion propagators.\
For this, we demonstrate that the first local pole on each soft line contributes half of the asymptotic coefficient on one Schwinger-Keldysh path and the remaining local poles on the second soft line supply the other half.\ 
\end{abstract}
\maketitle
\raggedbottom

\vspace*{3mm}
\noindent
{\bf 1.~Introduction}
\vspace*{1.5mm}

Primordial non-Gaussianity provides a powerful probe of the heavy particle content and interactions during inflation\,\cite{Chen:2010xka,Wang:2013zva,Meerburg:2019qqi,Achucarro:2022qrl},
where such heavy particles include the right-handed neutrinos which are predicted by the
structure of the Standard Model (SM) and realize the neutrino seesaw\,\cite{Minkowski,GRS,*Yanagida:1979as,*Glashow:1979nm,*Mohapatra:1979ia,*Schechter:1980gr}
around the inflaton scale.\ 
These heavy particles can leave distinctive imprints in primordial correlation functions 
through nonanalytic momentum dependence, 
whose oscillatory and angular structures encode information about their masses and spins.\ 
This idea underlies the Cosmological Collider (CC) program\,\cite{Chen:2009we,QSFI,Baumann:2011nk,Noumi:2012vr,AHM}, 
which has motivated the applications in particle physics\,\cite{Chen:2016uwp,Chen:2016hrz,Kumar:2017ecc,Chen,Kumar:2018jxz,Lu:2019tjj,Hook,Hook:2019vcn,Wang:2020ioa,Cui:2021iie,Chen:2023txq,Craig:2024qgy,Bodas:2024hih,You:2026CC,Sundrum:2026ghf}
and whose predictions are increasingly being tested against observations of the cosmic microwave background
and large-scale structure\,\cite{Planck:2013wtn,Sohn:2024xzd,Cabass:2024wob,Goldstein:2024bky,Philcox:2025wts,Kumar:2026ogn,Kumar:2026dih,Philcox:2026tjj}.\ 
Establishing a robust connection between these CC signatures and the underlying particle physics requires not only accurate predictions for their characteristic momentum dependence, but also reliable calculations of their signal amplitudes.

\vs 

Loop signals are particularly relevant for fermions and charged particles, 
which enter the cosmological correlators of bosonic perturbations through 
pair production\,\cite{Chen:2016uwp,Chen:2016hrz,Chen,Hook,Hook:2019vcn,Lu:2019tjj,Cui:2021iie}.\ 
The right-handed neutrinos can contribute to the cosmological correlators
and thus connect the neutrino seesaw mechanism to 
the inflationary measurements\,\cite{You:2024hit,Han:2024qbw,You:2026CC}.\ 
A rolling-inflaton chemical potential can enhance heavy-particle production\,\cite{Adshead,AdsheadPhen,Wang:2019gbi,Sou:2021juh}, 
motivating scalar, vector and fermionic CC scenarios\,\cite{Wang:2020ioa,Wang:2020uic,Bodas:2020yho,Tong:2022cdz,Qin:2022fbv,Bodas:2024hih,Bodas:2025vpb,Aoki:2026olh}.\ 
In this setup, the mass and chemical potential jointly determine the oscillation frequency, 
whereas the amplitude carries information about 
the right-handed-neutrino interactions and determines their production rate.\ 

\vs 

Recent advances in the cosmological bootstrap and study of massive tree-level amplitudes
have helped clarify the analytical structure of cosmological correlators\,\cite{Arkani-Hamed:2018kmz,Baumann:2019oyu,Baumann:2020dch,Baumann:2022jpr,Pimentel:2022fsc,Jazayeri:2022kjy,Xianyu230107047,Xianyu:2023ytd,Liu:2024str,Xianyu:2025lbk,Qin:2025xct,Cheung:2025dmc,Grafe:2026avi}.\ 
Mellin--Barnes methods\,\cite{Sleight:2019mgd,Sleight:2019hfp,Qin:2022fbv,Qin:2024gtr,McCune:2026vqb,Chen:2026khu}, spectral representations and dispersion relations\,\cite{Xianyu:2022jwk,Zhang:2025nzd,Liu:2024xyi,Werth:2024mjg,Liu:2026jzn,Grafe:2026qsm},
Laplace space representation\,
\cite{Belrhali:2026ktb,Belrhali:2026ygh,Belrhali:2026uxn},
and factorization and cutting rules\,\cite{Goodhew:2020hob,Melville:2021lst,Tong:2021wai,Qin:2023bjk,Qin:2023nhv,Ema,Colipi-Marchant:2025oin,Das:2025qsh,Das:2026vfv} provide independent tools for obtaining full results or extracting nonanalytic contributions.\ Recent progress on computing loop diagrams includes analyses of their analytic structure, differential equations, factorization and numerical evaluation\,\cite{Salcedo:2022aal,Lee:2023jby,AguiSalcedo:2023nds,Bhowmick:2025mxh,Pimentel:2026kqc,Baumann:2026atn,Chen:2026dqp,Qin:2026yuo,Borinsky:2026roj,McCune:2026qfy}.\ 
The spinor structure, Majorana contractions, and contour ordering make the direct fermion-loop calculations more involved than their scalar counterparts\,\cite{You:2026CC,Aoki:2026olh,Aoki:2026vbc,Aoki:2026smw,Han:2026FermionBubble}, 
hence a systematic and reliable treatment of both the time and loop-momentum integrations
is highly demanding.

\vs

In this work, we compute the fermion triangle loop contributions 
of the right-handed Majorana neutrinos by including the complete soft modes.\  
Let $k_i\!=\!|\vct k_i|$ denote the magnitudes of the three external comoving momenta. In the squeezed configuration $k_1\simeq k_2\gg k_3$, the signal originates from two soft internal lines (bubble loop)
and a hard subdiagram.\ The squeezed limit justifies expanding the soft lines at their hard endpoints, but $q|\tau_3|$ need not be small at the soft vertex, where $q$ is the magnitude of the loop momentum and $\tau_3$ is the conformal time of that vertex, 
hence we retain their full time dependence.\ 
In the squeezed limit, the spinor structure can be factorized by 
imposing a localized directional cut and hard-endpoint projection.\ 
We fix the finite part of the soft loop integral by extracting the oscillatory coefficient with the original momentum boundary in place. This gives the complex coefficient in terms of the Euler beta function and the usual Meijer $G$ functions.
From this coefficient we obtain both the amplitude and the phase of the leading oscillatory signal.\
This is in contrast with our previous study\,\cite{You:2026CC} 
(see also \cite{Aoki:2026olh}) which applied the late-time approximation 
to both the hard and soft endpoints (leading to a highly suppressed contribution).
For the present study, we further demonstrate how the size of the fermion triangle loop integral 
can be elegantly explained by summing up the contributions of individual Mellin poles,  
which also helps to understand why keeping only the nonlocal leading pole 
misses the major contributions 
in the previous studies\,\cite{Chen,You:2026CC,Aoki:2026olh}.  
Appendices\,\ref{sm:model}-\ref{sm:checks} present our detailed derivations.\

\vspace*{3mm}
\noindent
{\bf 2.~Inflaton Interaction with Right-handed Neutrinos and Loop-induced Bispectrum}
\vspace*{1.5mm}

In our previous works \cite{You:2024hit,Han:2024qbw,You:2026CC}, 
we have formulated the effective interaction between the inflaton ($\phi$) 
and right-handed neutrinos ($N$) with $\phi$ obeying the shift symmetry:  
\begin{equation}
\mathcal L=\ii\Nt^\dagger\bar\sigma^\mu\partial_\mu\Nt
-\frac{\,aM_R\,}{\,2\,}(\Nt\Nt\!+\!\mathrm{h.c.})
-\frac{\,\partial_\mu\phi\,}{\,\Lambda\,}\Nt^\dagger\bar\sigma^\mu\Nt \,,
\label{eq:lag}
\end{equation}
where the scale factor $a$ is given by $a(\tau)\!=\!-1/(H\tau)$. Here $H$ is the inflationary Hubble rate and $\tau<0$ is conformal time.\ 
In the above, $\Nt\!\!=\hsm\! a^{\frac{3}{2}}N$ is the canonical field 
for the right-handed Majorana neutrino of mass $M_R$, $\mathrm{h.c.}$ denotes the Hermitian conjugate, and 
this dimension-5 interaction between $\phi$ and $N$ is suppressed by the cutoff scale $\Lambda$.\ 
We use the notations $\sigma^\mu\!=\!(1,\vct\sigma)$ and $\bar\sigma^\mu\!=\!(1,-\vct\sigma)$, with $\vct\sigma$ the Pauli matrices,
and split the inflaton field into its background field plus the quantum fluctuation, 
$\phi\!=\!\phi_0^{}\!+\!\delta\phi$.\  
The two-component spinor conventions follow Ref.\,\cite{Dreiner:2008tw}.

\vs 

As discussed in Ref.\,\cite{You:2026CC}, the inflaton background $\phi_0^{}$ 
generates a chemical potential $\lambda\!=\!{\,\dot\phi_0}/{\Lambda\,}$, where the dot denotes a derivative with respect to cosmic time, 
through the term $-a\lambda\Nt^\dagger\bar{\sigma}_0\Nt$, 
whereas the fluctuation $\delta\phi$ interacts with the right-handed neutrino 
and thus produces a contribution to the primordial bispectrum of the inflaton $\Bphi$. Here $\delta\phi_{\vct k}$ is a late-time Fourier mode and the prime removes the momentum-conserving delta function.\ 
For the present analysis, we define the following dimensionless physical parameters, 
\begin{equation}
  \label{eq:physical}
 \m=\frac{\,M_R\,}{\,H\,},\quad\tlambda=\frac{\,\lambda\,}{\,H\,},\quad\tnu=\sqrt{\m^2\!+\!\tlambda^2}. 
\end{equation}

There are in total eight diagrams for the one-loop three-point correlator of inflaton\,\cite{You:2026CC}, 
as we show in Fig.\ref{fig:eight}, which are labeled by $j\!=\!1,\cdots\!,8$.\ 
Thus, the three-point inflaton correlator is derived as follows:
{\small
\begin{equation}
\begin{aligned}[b]
\Bphi =\(\!\!\frac{\,-\ii\,}{\,\Lambda\,}\!\)^{\!\!3}\!\!\!
\sum_{a,b,c=\pm1}\!\!\!\! abc\int_{-\infty}^{0}\!\!\!
\di\tau_1^{}\hs\di\tau_2^{}\hs\di\tau_3^{} &
\\[-1pt]
{}\times\prod_{i=1}^{3}\!\mathcal F_{\mu_i a_i}(\vct k_i,\tau_i)\!\!
\int\!\!\!\frac{\,\di^3q\,}{\,(2\pi)^3\,}\sum_{j=1}^{8}\!\mathcal Y_{abc}^{\mu_1\mu_2\mu_3}(j)\,,
&
\end{aligned}
\label{eq:parent}
\end{equation}
} 
\hspace*{-2mm}%
where $(a_1,a_2,a_3)\!=\!(a,b,c)$ specify the Schwinger--Keldysh (SK) branches at the three vertex times $\tau_i$: $+1$ labels the forward branch and $-1$ the backward branch.
The symbol $\mathcal F_{\mu_i a_i}^{}$ is the differentiated external inflaton field 
at the vertex with index $i$ (denoted by v$_i$):   
\begin{equation}
\mathcal F_{\mu a}(\vct k,\tau)=(\partial_{\tau},-\ii\vct k)_\mu
\frac{\,H^2(1-\ii ak\tau)e^{\ii ak\tau}\,}{\,2k^3\,} \,.
\label{eq:externalSource}
\end{equation}
In Eq.\eqref{eq:parent}, $\mathcal Y_{abc}^{\mu_1\mu_2\mu_3}(j)$ 
is the loop-integrand tensor containing three fermion propagators and 
the loop-diagram index $j$.\ Repeated Lorentz indices are summed.\ 
These loop-integrand tensors are presented in Appendix\,\ref{sm:contractions}.\ 
We can further collect the eight Majorana contractions into a single trace.\  
For this we combine the Weyl field $\Nt$ and its charge conjugate into the 
Nambu-Gorkov spinor $\Psi\!=\!\binom{\Nt}{\ii\sigma_2(\Nt^\dagger)^T}$.\ 
Its contour propagator is $S\!=\!\langle T_C\Psi\Psi^\dagger\rangle$, with $T_C$ ordering fields along the SK contour, 
and the vertex contains a matrix, 
$\Gamma^\mu\!=\!\mathrm{diag}(\bar\sigma^\mu,-\sigma^\mu)$.\ 
Expanding the trace over the Nambu blocks recovers the eight Majorana contractions:
{\small
\begin{equation}
\sum_{j=1}^{8}\!\mathcal Y_{abc}^{\mu_1\mu_2\mu_3}(j)
\!=\!-\operatorname{Tr}[\Gamma^{\mu_1}S_{ab}\llbracket 12\rrbracket\Gamma^{\mu_2}S_{bc}\llbracket 23\rrbracket\Gamma^{\mu_3}S_{ca}\!\llbracket 31\rrbracket]\,,
\label{eq:eightNambu}
\end{equation}
}
{\par\parindent=2.5mm
\hspace*{-2.5mm}%
where $S_{ab}\llbracket 12\rrbracket\!=\!S_{ab}(\vct p_{12},\tau_1,\tau_2)$ 
is a $2\!\times\!2$ matrix of $2\!\times\!2$ spinor blocks. The momentum $\vct p_{12}\!=\!\vct q+\vct k_1$ flows along the hard line; Eq.\,\eqref{sm:contractionmomenta} gives the routing on all three lines.\ 
Appendix\,\ref{sm:contractions} presents the propagator construction 
and verifies the block expansion and its normalization.\ 
\par}

In this work, instead of evaluating the full result of the 3-point correlator with the fermion loop 
shown in Eq.\eqref{eq:parent}, we aim to compute the CC signal from this 3-point correlator, 
which means the complete leading oscillatory component of the bispectrum in the squeezed configuration 
($k_3 \!\ll\! k_1 \!\simeq\! k_2$):
{\small
\begin{equation}
\Bphi_{\rm{osc,LO}}={}\frac{\,H^6\,}{\,k_1^6\,}
\left[\Cc\rho^{\ii\hp 2\tnu}\!+\!\Cc^*\rho^{-\ii\hp 2\tnu}\right] \,, \label{eq:target}
\end{equation}
}
{\par\parindent=2.5mm
\hspace*{-2.5mm}%
where $\rho$ is defined as the momentum-ratio $\rho\!=\!k_3/k_1$ and $\Cc$ denotes the coefficient.\  
The magnitude and phase of $\Cc$ determine the oscillatory signal. 
\par}

\vspace*{3mm}
\noindent
{\bf 3.~Factorization of 3-Point Inflaton Correlator}
\vspace*{1.5mm}

To extract the leading CC signal in the squeezed limit $k_3 \!\ll\! k_1 \!\simeq\hsm k_2$, 
we factorize its coefficient into a hard part associated with $k_1$ and a soft part associated with $k_3$, 
as shown in Fig.\,\ref{fig:time}.\ 
This differs from the factorization scheme of our previous work\,\cite{You:2026CC}, 
where the two soft fermion propagators were approximated by their leading nonlocal terms 
obtained from late-time expansions at both endpoints, which factorizes the contribution into three parts: 
a left subdiagram, a right subdiagram, and a bubble integral.

\vs 

\begin{figure}[!t]
\centering
\includegraphics[width=\columnwidth]{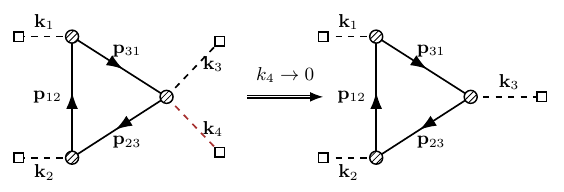}
\vspace*{-7mm}
\caption{%
Auxiliary four-point construction and its folded limit as adopted by Ref.\,\cite{You:2026CC}.\ 
The red dashed line denotes the auxiliary external leg carrying $\vct k_4$; the other dashed lines 
are external inflaton legs, and arrows indicate the fermion-line orientation.\ 
Taking $k_4\!\to\!0$ gives $k_s\!\to\! k_3$ and $r_2\!\to\!1$ and the subsequent squeezed limit 
is $r_1\!=\!k_3/(k_1+k_2)\!\simeq\!\rho/2\!\to\!0$, as in Eq.\eqref{eq:r1-r2}. 
}
\label{fig:folded}
\label{fig:1}
\end{figure}

\begin{figure*}[!t]
\centering
\includegraphics[width=0.8\textwidth]{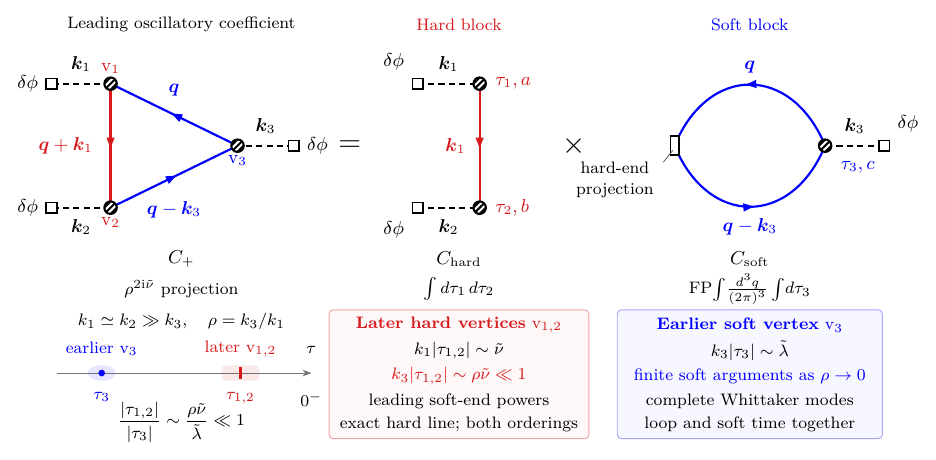}
\vspace*{-5mm}
\caption{%
Hard-soft factorization of the leading squeezed clock coefficient, 
$\Cc\!=\!\Ch\hs\Cs$, where $\Ch$ and $\Cs$ are the hard and soft factors in Eq.\eqref{eq:cut}.\ 
The left diagram represents the coefficient of $\rho^{\ii\hp2\tnu}$ 
after summing over the Majorana contractions and SK branches.\ 
The red hard block contains the leading hard-end powers of the two soft lines 
and the hard propagator at leading order of the soft momentum, 
retaining its full time-dependence and both orderings of $(\tau_1,\tau_2)$.\ 
The blue soft block retains the complete modes at $\tau_3$ 
and combines the soft-time and loop-momentum integrations; $\FP$ denotes the finite part of the soft loop integral, including the oscillatory contributions from its subtracted tail. The open rectangle on the soft block's left denotes the hard-end projection. Dashed lines are external inflaton legs, and hatched circles denote interaction vertices with SK labels $a,b,c$. The illustrated time scales apply in the enhanced large-$\tlambda$ production region, with $q,|\vct q-\vct k_3|=O(k_3)$. The squeezed limit $\rho\to0$ is taken at fixed $(\m,\tlambda)$ before the large-$\tlambda$ expansion.}
\label{fig:time}
\label{fig:2}
\end{figure*}

To assess the approximation of Ref.\,\cite{You:2026CC}, 
we use the auxiliary four-point construction: 
an extra external leg with momentum $\vct k_4$ is attached to the soft vertex, 
as shown in Fig.\,\ref{fig:folded} with the auxiliary coupling and external-line normalization 
given by \cite{You:2026CC}.\ Thus, taking the folded limit yields:
\begin{equation}
\Bphi =\lim_{k_4\to0}\left(-\frac{2k_4}{H}\right)\mathcal B_4^{\rm{aux}}\,,
\end{equation}
where $\mathcal B_4^{\rm{aux}}$ denotes the primed auxiliary four-point correlator. Define $k_s=|\vct k_1+\vct k_2|$, $k_{12}=k_1+k_2$, $k_{34}=k_3+k_4$, and $r_1=k_s/k_{12}$, $r_2=k_s/k_{34}$.\ 
The limit $k_4\!\to\! 0$ gives $k_s\!\to\! k_3$ and $r_2\!\to\!1$.\ 
Then, the squeezed limit of the resulting three-point correlator becomes as follows:
\begin{equation}
\label{eq:r1-r2}
  r_1=\frac{k_3}{k_{12}}\simeq\frac{\rho}{2}\to 0 \hs ,
  \qquad r_2=\frac{k_3}{k_3}=1\hs .
\end{equation}

Thus, the three-point signal is extracted at the folded boundary $r_2^{}\!=\!1\hs$, 
rather than in the joint collapsed limit $r_1,r_2\!\to\! 0$ of the four-point 
correlator, which will lead to divergent sum under the Mellin-Barnes transformation.\ 
To see this, we first consider a nonlocal signal branch of the auxiliary four-point correlator.\ 
Evaluating the relevant Mellin-Barnes residues\,\cite{Qin:2022fbv,Xianyu220501692,Qin:2023bjk} 
gives the following expansion of the nonlocal signal, 
after its common kinematic powers are factored out:
\begin{equation}
\label{eq:Sum-n1234}
\sum_{n_1,n_2,n_3,n_4=0}^{\infty} \hspace*{-6mm}
f(n_1,n_2,n_3,n_4)\,r_1^{n_1+n_3}r_2^{n_2+n_4}\,,
\end{equation}
where $f(n_1,n_2,n_3,n_4)$ denotes the coefficients derived from the residues of the Mellin poles.\  
The nonnegative integers $(n_1,n_3)$ label the hard-end terms and $(n_2,n_4)$ label the soft-end terms.\ 
In the joint collapsed limit ($r_1,r_2\!\to\! 0$), the leading term selects the first residue 
at every endpoint.\ 
In the three-point squeezed limit \eqref{eq:r1-r2}, 
taking $r_1\!\to\! 0$ will select $n_1\!=\!n_3\!=\!0$ for getting nonzero result, 
but the folded-limit result $r_2\!=\!1$ leads to absence of kinematic suppression for the higher powers of 
$n_2\!+\!n_4\!>\!0\hs$.\ Hence, the whole nonlocal coefficient \eqref{eq:Sum-n1234} 
reduces to a sum of the soft-end towers, 
\begin{equation}
\sum_{n_2,n_4=0}^{\infty} \hspace*{-2mm}f(0,n_2,0,n_4)\,,
\end{equation}
which does not equal the single coefficient $f(0,0,0,0)$ retained 
by the conventional leading-pole approximation\,\cite{Chen,Hook,Aoki:2026olh,You:2026CC}.\ 
 
This conventional leading-pole approximation also omitted the local clock contribution.\ 
With the four-point description, the characteristic nonlocal and local oscillatory factors 
are $(r_1^{}r_2^{})^{\pm\ii\hp2\tnu}$ and $(r_1^{}/r_2^{})^{\pm\ii\hp2\tnu}$, respectively.\ 
At $r_2^{}\!=\!1\hs$, both of them reduce to $r_1^{\pm\ii\hp2\tnu}$ and cannot be distinguished 
by their leading squeezed momentum dependence alone, although their propagator and pole decompositions 
remain distinct.\ Hence, both the omitted nonlocal higher-order terms ($n_2\!+\!n_4>0$) and 
the local contribution must be included in the complete three-point clock coefficient.\ 
For the above reasons, in this work we will retain these contributions by keeping the 
{\it complete mode functions} at the soft endpoint, rather than using the late-time expansion 
at both endpoints as adopted by Ref.\,\cite{You:2026CC}.

\vs 

The same conclusion follows from the mode-function arguments in the soft loop-momentum region,  
$p,q\!=\!O(k_3)$, where $q\!=\!|\vct q|$ and $p\!=\!|\vct q\!-\!\vct k_3|$.\ 
In the enhanced production regime, the soft interaction occurs at the time 
$|\tau_3|\!\sim\!\tlambda/k_3$, which is earlier than the hard interactions 
at the time $|\tau_{1,2}|\!\sim\!\tnu/k_1\,$.\ 
For the fixed physical parameters, their time scales satisfy 
$|\tau_{1,2}|/|\tau_3|\!\sim\!\rho\hs\tnu/\tlambda\!\ll\!1\,$.\ 
Thus, the soft-mode arguments at the hard vertices obey 
$q|\tau_{1,2}|\!\hsm\sim\hsm\!\rho\hs\tnu\!\to\!0$, whereas 
$|q\tau_3^{}|\hsm\!=\hsm\!(q/k_3)k_3|\tau_3|$ remains of $O(\tlambda)$ which is large.\ 
The same reasoning applies to $p\!=\!O(k_3^{})\hs$.\ 
This means that the late-time limit $|q\tau_3^{}|\!\ll\! 1$ does not hold.\ 
Hence, the squeezed limit controls the late-time expansion at the hard endpoints, 
but does not justify  the late-time approximation at the soft vertex.

\vs  

Based upon the above observations, We will expand the two soft fermion propagators only 
at their hard endpoints and retain their complete Whittaker modes 
at the soft endpoint $\tau_3^{}\hs$.\ 
Hence, we perform the soft-time and momentum integrations on the {\it complete mode functions.} 

\vs 

In the present analysis, we could still use localized directional 
cutting\,\cite{Tong:2021wai,Qin:2023bjk,Ema} to factorize the soft and hard parts
for extracting the leading clock coefficient.\ 
For this, we choose a fixed dimensionless separation parameter 
$0<\varepsilon_{\rm{sep}}\!\ll\!1$ and restrict the loop momentum smoothly up to 
$q\!\lesssim\!\varepsilon_{\rm{sep}} k_1$, 
with $k_3\!\ll\!\varepsilon_{\rm{sep}} k_1\!\ll\! k_1$, as specified 
in Eq.\eqref{sm:separator}.\ 
Within this region, we replace the two soft propagators by 
$S_{-c,c}\llbracket 23\rrbracket$ and $S_{c,-c}\llbracket 31\rrbracket$,  
while retaining the hard propagator, vertices, and SK weights.\ 
We denote the leading oscillatory coefficient of the resulting 
restricted integral by $\mathbb C_+^{\rm{cut}}(\varepsilon_{\rm{sep}})$.\ 

\vs 

\vs 

The fermion spinor structure then separates the hard and soft vertices.\ 
Selecting a leading hard-end power fixes the ratio of the two components $u$ and $v$ 
of a fermion mode of fixed helicity [Eq.\eqref{sm:modes}], giving a rank-one endpoint factor.\ 
Eq.\eqref{sm:rankone} uses this projection to split the triangle trace into a hard overlap 
(involving vertices-1 and 2) and a soft overlap (involving vertex-3).\ 
After combining the helicity and contour sums, the localized cut and spinor projection 
give the factorization illustrated in Fig.\,\ref{fig:time},
\begin{equation}
\boxed{\mathbb C_+^{\rm{cut}}(\varepsilon_{\rm{sep}})=\Cc=\Ch\,\Cs \,.}
\label{eq:cut}
\end{equation}
The hard factor $\Ch$ couples the leading hard-end powers of the two soft lines 
to the hard external inflaton modes.\ It contains the two hard external legs 
and the hard propagator evaluated at leading order in the soft momentum, with the time integrations over $(\tau_1,\tau_2)$.\ 
On the other hand, the soft factor $\Cs$ contains the external inflaton leg carrying $\vct k_3$ 
and the two complete Whittaker modes at $\tau_3$.\ 
It retains the full soft-time dependence through coupled soft-time and loop-momentum integrations.\ 
The detailed derivation for this factorization is provided in Appendix\,\ref{App:B}.

\vs 

%

\vspace*{3mm}
\noindent
{\bf 4.~Analytical Derivation of Closed-Form Correlator}
\vspace*{1.5mm}

We compute the hard and soft factors in Eq.\eqref{eq:cut} 
to obtain the complete leading squeezed clock coefficient.\ 
For the convenience to derive the closed-form formulas, 
we introduce the following combinations 
of the physical parameters in Eq.\eqref{eq:physical}:
\begingroup
\predisplaypenalty=0\interdisplaylinepenalty=10000
\begin{equation}
\begin{alignedat}[c]{2}
& a=-\ii(\tnu\!-\!\tlambda),\qquad 
&& b=-\ii(\tnu\!+\!\tlambda)\,,\\[2pt]
& A=2\!+\!\ii\hp 2\tnu,            
&& \omega=\frac{\,\sinh(2\pi\tlambda)\sinh(\pi\tlambda)\,}{\,\sinh(\pi\tnu)\,} \hs .
\end{alignedat}
\label{eq:finitepars}
\end{equation}
\endgroup
With these, we compute the hard and soft contributions to the clock coefficient
as follows.

\vspace*{2mm}
\noindent 
{\Large $\bullet$}\,{Hard Contribution:}
\\[1mm]
After the hard-endpoint projection, we express the hard factor in the schematic form:
{\small 
\begin{align}
\hspace*{-8mm}
\Ch={} & \mathcal N_h^{}\!\!\!\!\sum_{a,b=\pm1}\!\!\!\!a\hs b\!
\int_{-\infty}^{0}\!\!\!\!\di\tau_1\di\tau_2\,
\mathcal F_{\mu a}^{}(\vct k_1,\tau_1)
\mathcal F_{\nu b}^{}(-\vct k_1,\tau_2)
\notag
\\[1pt]
\hspace*{-8mm}
& \hspace*{8mm}
\times\!\mathcal P_h^{}(\tau_1,\tau_2)
\!\left[\Gamma^\mu S_{ab}^{}(\vct k_1;\tau_1,\tau_2)\Gamma^\nu\right],
\label{eq:hardIntegral}
\end{align}
}
\hspace*{-3mm}
where $a,b\!=\!\pm 1$ denote the SK branches.\ 
The two $\mathcal{F}$ functions are defined in Eq.\eqref{eq:externalSource}
and contributed by the differentiated external inflaton states, 
with $\vct k_2\!=\!-\vct k_1$ at the leading squeezed order, 
and $S_{ab}^{}$ is the single internal fermion propagator.\ 
The projection $\mathcal P_h^{}$ includes the leading hard-end power terms 
$(-k_1^{}\tau_1^{})^{\ii\tnu}(-k_1\tau_2)^{\ii\tnu}$ 
of the two soft lines, their spinor contractions, and the helicity sum.\ 
The time-independent factor $\mathcal N_h^{}$ collects the coupling and normalization factors.\
We give the explicit formulas for $\mathcal P_h$ and $\mathcal N_h$ in 
Eqs.\eqref{sm:hardIntegralExplicit}-\eqref{sm:hardIntegralNormalization} of Appendix\,\ref{sm:hard}.\  
Both time orderings of $(\tau_1,\tau_2)$ are retained in the SK sum
through the fermion propagator ${S}_{ab}^{}\hs$.

\vs

We derive the hard contribution in detail in in Appendix\,\ref{App:C}.\ 
We use the Nambu representation to combine the contributions of the eight individual diagrams 
of Fig.\,\ref{fig:eight} into a single trace, yielding the following compact expression 
for the hard factor:
{\small
\begin{equation}
\begin{aligned}[b]
\hspace*{-11mm}
\Ch= &-\frac{\,\ii\pi\tnu(A\!-\!1)\tanh(\pi\tnu)\,}{\,8\Lambda^2(A\!-\!1\!-\!\m^2)|\mathrm B(a,b)|^2\,} 
\\[2pt]
\hspace*{-11mm}
& \times\!\hsm\biggl[\!
\frac{\,A^2(A\!-\!1)\sinh(\pi\tlambda)\,}{\,(A\!-\!1\!-\!\m^2)\sinh(\pi\tnu)\,}
\!-\!\frac{\,\tlambda\m^2\,}{\,\tnu\,}\overline{\mathrm B(a,b)}\biggr],
\end{aligned}
\label{eq:hardfinite}
\end{equation}
}
\hspace*{-2.5mm}
where $\mathrm B(a,b)\!=\!\allowbreak\Gamma(a)\Gamma(b)/\Gamma(a\!+\!b)$ is the Beta function, $\Gamma$ is the Euler gamma function, and $\overline{\mathrm B(a,b)}$ is the complex conjugate of $\mathrm B(a,b)$.

\vspace{2mm}
\noindent
{\Large $\bullet$}\,{Soft Contribution:}
\\[1mm]
For the soft factor $\Cs$, we first perform the soft-time integral, 
which contains a product of two Whittaker functions from the two soft fermion lines.\  
On each SK branch $c\!=\!\pm1$, we parametrize the Wick-rotated soft time as 
$\tau_3\!=\!\ii c\,z_3/k_3$ with $z_3^{}\!>\!0\hs$.\ 
Then, the external phase $e^{\ii c k_3\tau_3}$ becomes $e^{-z_3}$ on both branches.\ 
For $p,q,k_3^{}>0$, $\ell\!=\!0,1$, and $\kappa_j\!=\!\pm\tfrac{\,1\,}{\,2\,}\pm\ii\tlambda$, 
the soft-time moment $\mathcal I_{\ell}^{}$ takes the following form, 
with $W_{\kappa,\ii\tnu}^{}$ the Whittaker function and $(\kappa_1^{},\kappa_2^{})$ 
its indices on the $(p,q)$ lines respectively:
\par\noindent\begin{minipage}{\columnwidth}
  {\small
    \begin{subequations}\label{eq:Wresult}
      \begin{flalign}
&\mathcal I_\ell(p,\!q;k_3)\!=\!k_3^{-\ell}\!\int_0^\infty\!\di z_3\,z_3^{\ell-1}e^{-z_3}W_{\kappa_1,\ii\tnu}\!\left(\tfrac{2p}{k_3}z_3\right) \! W_{\kappa_2,\ii\tnu}\!\left(\tfrac{2q}{k_3}z_3\right)&&
\notag
\\[2pt]
&\mathord{=}\;\sum_{m,n=\pm 1}\!\!\!\!
\frac{~R_{\kappa_1,m}R_{\kappa_2,n}\Gamma(d_{mn})~}{(k_3\!+\!p\!+\!q)^{d_{mn}}}(2p)^{\frac{1}{2}+\ii m\tnu}(2q)^{\frac{1}{2}+\ii n\tnu}&&\notag           \\
        &\phantom{=}\,\mathord{\times}\, F_2\!\!\left(\!\!\begin{gathered}
                              d_{mn};\tfrac{\,1\,}{\,2\,}\!+\!\ii m\tnu\!-\!\kappa_1,\tfrac{\,1\,}{\,2\,}\!+\!\ii n\tnu\!-\!\kappa_2;\\[-1pt]
                              1\!+\!\ii\hp 2m\tnu,1\!+\!\ii\hp 2n\tnu;\frac{\,2p\,}{\,k_3\!+\!p\!+\!q\,},\frac{\,2q\,}{\,k_3\!+\!p\!+\!q\,}
                            \end{gathered}\!\right)\!,&&\label{eq:Wseed}\\[1pt]
& R_{\kappa,m}
\!=\!\frac{\,\Gamma(-\ii\hp 2m\tnu)\,}{\,\Gamma(\frac{\,1\,}{\,2\,}\!-\!\ii m\tnu\!-\!\kappa)\,},
\hspace*{5mm}
d_{mn}\!=\ell\!+\!1\!+\!\ii(m\!+\!n)\tnu \hs.
& & \label{eq:Wdefs}
\end{flalign}
\end{subequations}
}
\vspace*{-1mm}  
\end{minipage}
In the above, each Whittaker function is the sum of two Kummer terms [\eqref{sm:Wconnection}].\ The signs $m$ and $n$ select these terms on the $p$ and $q$ lines, respectively, giving the four contributions in Eq.\,\eqref{eq:Wseed}.
Here $F_2^{}$ is Appell's second function.\  
The factor $k_3^{-\ell}$ follows from the change to dimensionless time, 
whereas the moments $\ell\!=\!0,1$ arise from the differentiated inflaton mode.\ 
The weight factors $e^{-z_3}$ and $z_3e^{-z_3}$ originate from 
Eq.\eqref{eq:externalSource}.\ The complete four-term expression is analytically continued from 
$k_3^{}\!>\!p\!+\!q$ to the physical triangle 
$|p\!-\!q|\!\leqslant\! k_3^{}\!\leqslant\! p\!+\!q\hs$,
as discussed in Appendix\,\ref{sm:Wcontinuation}.\

\vs 

To obtain the soft factor $\Cs$, we combine the time moments in Eq.\eqref{eq:Wseed} 
with their spinor weights and integrate over the loop momentum.\ 
This relation takes the following form:
\begin{equation}
  \begin{aligned}[b]
    \Cs\!=\!{}&\FP\!\int\frac{\di^3\vct q}{(2\pi)^3}\sum_{j,\ell}
    \mathcal K_{j\ell}(\vct q,\vct k_3)\,
    \mathcal I_\ell^{(j)}(p,q;k_3)\,,
  \end{aligned}
  \label{eq:softIbridge}
\end{equation}
where $p\!=\!|\vct q\!-\!\vct k_3|$, $q\!=\!|\vct q|$, $\ell\!=\!0,1$, 
and $j\!=\!A,B,C,D$ denotes the four spinor-channels.\ 
The superscript $(j)$ specifies the corresponding pair of Whittaker indices.\ 
The kernels $\mathcal K_{j\ell}$ collect the spinor and contour weights, momentum factors, 
and overall normalization.\ The above formula is derived in Eqs.\eqref{sm:softIbridgeExplicit}-\eqref{sm:softKexplicit}
of Appendix\,\ref{App:D}.\ 
After summing over the four spinor-channels, we subtract the nonintegrable terms 
at large loop momentum and integrate the convergent remainder.\ 
The finite-part prescription ($\FP$) adds back the finite contributions of the subtracted terms 
to the oscillatory coefficient as given by Eq.\,\eqref{sm:finitepart}.

\vs 

After the angular and radial integrations, only ordinary Meijer $G$ functions remain.\ 
We reduce the integral to two square moments in Appendix\,\ref{sm:soft}.\ 
Then, we compute them to obtains $\Cs$, $\mathcal T_c$, 
and $\mathcal H$ in Eqs.\eqref{sm:softFinal}, \eqref{sm:Tfinal}, and \eqref{sm:Hfinal}
as follows: 
{\small\allowdisplaybreaks[4]
    \begin{subequations}\label{eq:finiteall}
      \begin{flalign}
        &\begin{aligned}[b]
           & \Cs=-\frac{\,2^{A-4}\overline{\mathrm B(a,b)}\,}{\,\pi\Lambda\tnu A(A\!-\!1)(A\!+\!2)\sinh(2\pi\tnu)\,}                                               \\[2pt]
           & \hspace{6pt}\quad\,\,\times\biggl[(A\!-\!2)(4\!+\!A\!-\!A^2)\!-\!16A\tlambda^2+\frac{\,\ii\hp 8(\mathcal T_a\!+\!\mathcal T_b)\,}{\,\pi^3(A\!+\!1)\,}\biggr]\,,
        \end{aligned}&&\label{eq:softfinite}\\[3pt]
        &\mathcal T_c\!=\!\sin\pi c\,[(\omega\!-\!\cos\pi c)\mathcal H(A,c)
            +(\omega\!+\!\cos\pi c)\overline{\mathcal H(\bar A,\bar c)}]\,,&&\label{eq:Tfinite}\\[3pt]
        &\mathcal H(A,c)\!=\!\sum_{r=0}^{1}\!\left[\frac{A(1\!-\!A)}{4}\right]^{\!1-r}\!\!\!\!\!\!
\Gn[\!]{-1}{-\frac A2,\frac{1\!-\!A}{2},-r,1\!-\!c}{\varnothing}{0,1,-\frac12}{1\!-\!A\!-\!c}\!\!.
&&
\label{eq:Hfinite}\raisetag{18pt}
\end{flalign}\end{subequations}
  }
{\par\parindent=2.5mm
    \hspace*{-2.5mm}%
The derivations are given in detail in Appendix\,\ref{sm:native}.\ 
In Eqs.\eqref{eq:Tfinite} and \eqref{eq:Hfinite}, $c$ equals $a$ or $b$ as defined in Eq.\eqref{eq:finitepars}.\ 
The ordinary Meijer $G$ function $G_z$ is written in the grouped-parameter notation of Eq.\,\eqref{sm:Gdefinition}. The subscript $z$ is its argument and $\varnothing$ denotes an empty parameter group.\ 
We choose the branch $\log(-1)\!=\!+\ii\pi$, and the conjugate combination in Eq.\,\eqref{eq:Tfinite} 
supplies the opposite boundary value.\  Combining this result with the hard factor in Eq.\eqref{eq:hardfinite},
we derive  $\Cc \hsm\!=\hsm\Ch\hs\Cs$ as a finite combination of one independent Beta value 
and eight Meijer $G$ values, retaining the full finite-mass and chemical-potential dependence 
of the leading squeezed coefficient.
\par}

\vspace*{3mm}
\noindent
{\bf 5.~Analysis of Large-Chemical-Potential Limit}
\vspace*{1.5mm}

We convert the inflaton fluctuation to the curvature perturbation, 
$\zeta\!=\!-H\delta\phi/\dot\phi_0^{}\hs$.\  
We normalize its bispectrum by the dimensionless curvature power spectrum $\Pz(k)\!\equiv\!k^3\langle\zeta_{\vct k}\zeta_{-\vct k}\rangle'/(2\pi^2)\hs$.\  
At leading slow-roll order, we use its nearly scale-independent tree-level value 
$\Pz\!=\!H^4/(4\pi^2\dot\phi_0^2)\hs$.\ 
The dimensionless amplitude $\fNLCC$ of the leading squeezed oscillation 
is defined by the following template for the curvature bispectrum:

\begin{equation}
\hspace*{-4mm}%
\text{\small$\displaystyle
    \langle\zeta_{\vct k_1}\zeta_{\vct k_2}\zeta_{\vct k_3}\rangle'_{\rm{osc}}
    =\frac{\,144\pi^4\Pz^2\,}{\,5k_1^6\,}\fNLCC
    \cos\!\left(\!2\tnu\ln\frac{\,k_1\,}{\,k_3\,}\!+\!\varphi_0\!\right)\!,
  $}
\label{eq:fnltemplate}
\end{equation}
where $\varphi_0\!=\!\arg(-\Cc^*)$ is the oscillation phase for the ratio $k_1/k_3$ 
used in the cosine as given by Eq.\,\eqref{sm:fnlmatch}.\ 
Thus, we express the factorized bispectrum amplitude as follows: 
\begin{equation}
  \boxed{\fNLCC=\frac{\,5H^3\,}{\,9\pi\sqrt{\Pz}\,}|\Ch\,\Cs|\,.}
  \label{eq:fnl}
\end{equation}

We denote the complete oscillatory amplitude of the curvature bispectrum 
by $\fNLCC$ and the nonlocal late-time result of 
Ref.\,\cite{You:2026CC} by $f_{\rm{NL}}^{\rm{NLoc}}$.\  
For the fixed right-handed-neutrino mass $\m$ and large chemical potential $\tlambda\hs$, 
we derive the complete bispectrum amplitude as follows:
\begin{equation}
\label{eq:asymnew}
\fNLCC  \simeq\frac{\,5\pi\,}{\,18\,}\Pz\m^2\tlambda^3
e^{-11\pi\m^2/(4\tlambda)}\,.    
\end{equation}
\begin{figure}[t]
\centering
\includegraphics[width=\columnwidth]{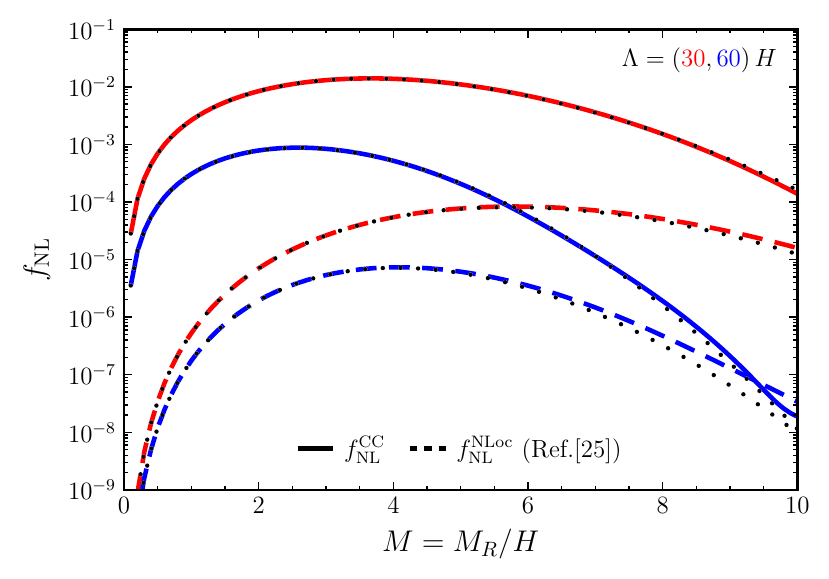}
\vspace*{-8mm}
\caption{Leading oscillatory amplitude $f_{\rm{NL}}^{}$ versus the right-handed neutrino mass $M_R/H$ for $\Lambda/H\!=\!30$ (red) and $60$ (blue).\ 
The solid curves show the complete amplitude $\fNLCC$ of Eq.\,\eqref{eq:fnl}, 
and the dashed curves show the nonlocal late-time amplitude $f_{\rm{NL}}^{\rm{NLoc}}$ of Ref.\cite{You:2026CC}.\ 
The black dots on the solid curves give the large-$\tlambda$ approximation in Eq.\eqref{eq:asymnew}, 
and those on the dashed curves give the approximation in Eq.\eqref{eq:asymold}.\  
}
\label{fig:fnl}
\end{figure}
\hspace*{-2mm}
The detailed derivations are presented in Eqs.\eqref{sm:uniformasym}-\eqref{sm:previousasym}
of Appendix\,\ref{App:F}.\ 
For comparison, we also show the large-$\tlambda$ limit of the nonlocal late-time result of 
Ref.\cite{You:2026CC} in the following:
\begin{equation}
\label{eq:asymold} 
f_{\rm{NL}}^{\rm{NLoc}}    
\simeq\frac{\,5\hp \pi^{\frac{5}{2}}}{\,288\sqrt{2\,}\,}\Pz\m^4\tlambda^{\frac{3}{2}}
e^{-9\pi\m^2/(4\tlambda)}\,.
\end{equation}
In Fig.\,\ref{fig:fnl}, we compare the two bispectrum aplitudes 
$\fNLCC$ (solid curves) of the present study and $f_{\rm{NL}}^{\rm{NLoc}}$ (dashed curves) 
of Ref.\cite{You:2026CC} at $\Lambda/H\!=\!30$ (red color) and $60$ (blue color), respectively.\ 
The two choices $\Lambda\!=\!(30,60)H$ correspond to $\tlambda\!\simeq\! (116, 58)$.\ 
The black dotted curves present the results by using the large-$\tlambda$ approximation 
in Eqs.\eqref{eq:asymnew} and \eqref{eq:asymold}.\  
For the benckmark choice of $\Lambda\!\!=\!30H$, the complete amplitude peaks at 
$\m\!\simeq\!3.7$ with $\fNLCC\!\simeq\!1.40\!\times\!10^{-2}$, 
whereas our previous result peaks at 
$\m\!\simeq\!5.8$ with $f_{\rm{NL}}^{\rm{NLoc}}\!\simeq\!8.35\!\times\!10^{-5}$\,\cite{You:2026CC}.\ 
At these two mass points, their results of the large-$\tlambda$ approximation 
agree with the corresponding precise results to the accuracies of about $0.2\%$ and $2.5\%$, respectively.\ 
At the same mass $\m\hsm\!=\hsm\!4$, the complete amplitude is about 
$2.5\!\times\!10^2$ times the previous result.\

\vs  

The distinct mass powers of the amplitude formulas \eqref{eq:asymnew} and \eqref{eq:asymold} 
explain both the separation at small $\m$ and the different peak positions.\ 
Before the exponential suppression becomes important, 
the complete result grows as $\m^2$, whereas our previous result grows as $\m^4$ 
and thus falls more rapidly toward small masses.\ 
For the fixed value of $\Pz$, reducing the input $\Lambda/H$ from $60$ to $30$ 
would increase the $\tlambda$ value by a factor 2{\hs}.\ 
The stronger $\tlambda^3$ prefactor and the weaker mass suppression at larger $\tlambda$ 
raise the complete amplitude and move its peak toward a larger $\m\hs$.\ 
For the comparison, we show the results for another choice $\Lambda\!=\!60H$
in Fig.\,\ref{fig:fnl}.\ 
 
\vs 

For fixed $\m$ and large $\tlambda\hs$, the hard factor has magnitude
$|\Ch|\!\propto\!\m^4\tlambda^3$ (according to Appendix\,\ref{sm:factorOrigins})
which is the same for the current result [contained in Eq.\eqref{eq:asymnew}] and for the
previous nonlocal late-time result [contained in Eq.\eqref{eq:asymold}].\ 
This means that the soft contributions to Eq.\eqref{eq:asymnew} and to Eq.\eqref{eq:asymold}
have different power-dependence in $\m$ and in $\tlambda\hs$.\  
By factoring $|\Ch|\!\propto\!\m^4\tlambda^3$ out of  
the bispectrum amplitudes \eqref{eq:asymnew} and \eqref{eq:asymold}, we obtain 
a residual soft term $|\Cs|$ proportional to  $1/\m^{2}\hs$ and $1/\tlambda^{3/2}\hs$ 
respectively, which differ from each other and can be understood as follows.\ 
In the complete soft integral,
the fermion modes contribute
$|\Gamma[-\ii(\tnu\!-\!\tlambda)]|\!\sim\! 1/(\tnu\!-\!\tlambda)$.
The soft normalization gives a factor $1/(\tnu\!+\!\tlambda)$, hence their product
scales as $1/\m^2$ [cf.\ Eq.\eqref{sm:completeSoftGapWeight}].\ 
In the previous nonlocal calculation, the spinor contraction cancels the
leading contribution from collinear loop momenta, generating a residual
soft prefactor proportional to $\tlambda^{-3/2}$
[cf.\ Eq.\eqref{sm:previousSoftGapWeight}].\ 
This explains that the different power-dependence of $\m$ and $\tlambda\hs$ 
in the bispectrum amplitudes \eqref{eq:asymnew} and \eqref{eq:asymold}
arise from the different soft contributions contained in them.

\vs

We can further understand the different exponential suppression factors 
in the bispectrum amplitudes \eqref{eq:asymnew} and \eqref{eq:asymold}.\ 
Denoting $\Delta\!=\!\tnu\!-\!\tlambda\simeq\m^2/(2\tlambda)$, 
we note that the hard factor contributes $e^{-2\pi\Delta}$ and 
the complete soft integral contributes $e^{-7\pi\Delta/2}$.\  
Hence, their product give the exponential factor $e^{-11\pi\Delta/2}$
in agreement with Eq.\eqref{eq:asymnew}.\
On the other hand, 
the nonlocal late-time result \eqref{eq:asymold} instead 
has total exponential suppression $e^{-9\pi\Delta/2}$,
and it becomes $e^{-5\pi\Delta/2}$ after removing the hard exponential factor $e^{-2\pi\Delta}$.\ 
Then, we can justify this soft suppression $e^{-5\pi\Delta/2}$
by using Eq.\,\eqref{sm:previousSoftGapWeight} under the late-time approximation.\ 
Hence, the above discussions have explained why the bispectrum amplitudes 
\eqref{eq:asymnew} and \eqref{eq:asymold} exhibit different dependence on the
fermion mass $\m$ and chemical potential $\tlambda$.\ 
Using the formulas \eqref{eq:asymnew} and \eqref{eq:asymold},
we see that as shown in Fig.\,\ref{fig:fnl} the present complete CC amplitude $\fNLCC$
(solid curves) is mostly much larger than the previous nonlocal late-time amplitude
$f_{\rm{NL}}^{\rm{NLoc}}$ because the former is enhanced by an extra large 
factor $\tlambda^{3/2}$ over the latter.\

\vspace*{4mm}
\noindent
{\bf 6.~Understanding Contributions from Mellin Poles}
\vspace*{0.5mm}

As discussed above, for the soft part, we retain the full Whittaker mode functions of 
both soft fermion lines at the vertex-3 to perform the loop and time integrals.\ 
This procedure could also be understood through the Mellin-Barnes 
representation\,\cite{Sleight:2019hfp,Qin:2022fbv}:
{\small 
\begin{align} 
\label{eq:Wmellin}
\hspace*{-4mm}
W_{\kappa,\ii\tnu}(X) =\, &
\frac{\,e^{-X/2}\,}{\,\Gamma(\frac{\,1\,}{\,2\,}\!-\!\kappa\!+\!\ii\tnu)
\Gamma(\frac{\,1\,}{\,2\,}\!-\!\kappa\!-\!\ii\tnu)\,}  
\\
\hspace*{-4mm}
& \times\!\!\int_{\mathcal C_{\rm {MB}}}\!\!
\frac{\,\di s\,}{\,\ii\hp 2\pi\,}
\Gamma(s\!+\!\ii\tnu)\Gamma(s\!-\!\ii\tnu)\Gamma(\tfrac{\,1\,}{\,2\,}\!-\!\kappa\!-\!s)X^{\frac{1}{2}-s} \,,
\nn 
\end{align}
}
\hspace*{-2.5mm}
where $X$ is the dimensionless argument of the Whittaker function 
(such as $X\!=\!2pz_3^{}/k_3^{}$ on the rotated $p$ line), 
$s$ is its Mellin integration variable, and $\mathcal C_{\rm{MB}}^{}$ 
is the contour in the complex $s$ plane.\ It initially separates the 
left poles $s\!=\!\pm\ii\tnu\!-\!n$ from the right poles 
$s\!=\!1/2\!-\!\kappa\!+\!n$, with $n\!=\!0,1,\cdots$ 
counting poles from the first in each family.\  
The local versus nonlocal distinction follows from the endpoint powers.\
For a soft line of momentum $p\hs$, let $\tau_h^{}$ denote its hard-end time 
and $\tau_3^{}$ its soft-end time.\
The Mellin variables at those endpoints are $s_h^{}$ and $s_s^{}$, respectively.\
Then, the Mellin power factor for the soft propagator is 
$(2p)^{-s_h\!-\!s_s}(-\tau_h)^{-s_h}(-\tau_3)^{-s_s}$.\  
With coefficients and exponentials factored out and 
by setting $s_h\!=\!-\ii\tnu$ for the positive external clock, 
we derive at the $n\!=\!0$ pole the different dependence on the time 
$\tau_3$ and momentum $p$ for 
the local pole ($s_s\!=\!+\ii\hs\tnu$) and nonlocal pole ($s_s\!=\!-\ii\hs\tnu$):
\begin{equation}\begin{aligned}[b]
s_s=+\ii\hs\tnu:\quad &(\tau_h/\tau_3)^{\ii\tnu}\,\,\,\,\,\,\quad(\mathrm{local}),
\\
s_s=-\ii\hs\tnu:\quad &(4p^2\tau_h\tau_3)^{\ii\tnu}\quad(\mathrm{nonlocal}).
\end{aligned}\label{eq:endpointpairings}\end{equation}
For the local contribution, the oscillatory powers of $p$ cancel, leaving a phase
that depends on the ratio of the endpoint times.\
For the nonlocal contribution, these powers combine into $p^{\ii\hs 2\tnu}$,
and the phase depends on the product of the endpoint times.\
The higher poles introduce additional powers $(-2p\tau_3)^n$,
giving the local family $s_s\!=\!+\ii\hp\tnu\!-\!n$ and the nonlocal family
$s_s\!=\!-\ii\hp\tnu\!-\!n$.\
Because the local contribution still carries a time-dependent phase, it could also generate
an oscillation in the external momentum ratio after the vertex integrations.\

\vs

\begin{figure}[!t]
\centering
\includegraphics[width=\columnwidth]{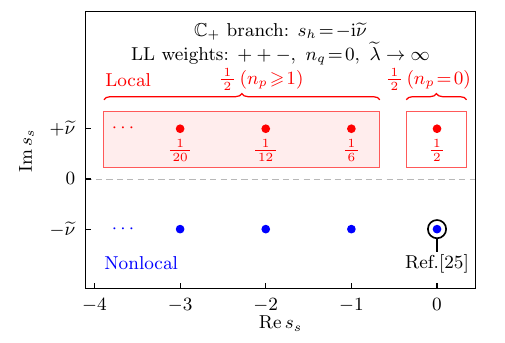}
\vspace*{-8mm}
\caption{Local and nonlocal poles at the soft endpoint of the $p$ line,
where local (nonlocal) poles are marked in red (blue) color.\ 
The indices $(n_p,n_q)$ label the poles on the $(p,q)$ lines respectively, 
starting from zero.\ The notation LL denotes local poles on both lines, 
and the SK signs refer to vertices\,1, 2 and 3 in order.\
For $\Cc$, Ref.\,\cite{You:2026CC} retains only the circled nonlocal pole $n_p\!=\!0\hs$, 
and the conjugate choice gives $\Cc^*$.\
The fractions are the integrated ratios $\mathbb C_+^{\rm{LL}(n_p,0);++-}\!/\Cc$
at fixed nonzero mass and large chemical potential, with the $q$ line at its first local pole
and both soft exponentials retained.\
The braces separate the first local pole's $\tfrac{\,1\,}{\,2\,}$ from the remaining
$\tfrac{\,1\,}{\,2\,}$ contributed by the higher local poles,
as given by Eq.\eqref{eq:ladder} and Appendix\,\ref{sm:poles}.}

    \label{fig:poles}

\end{figure}

Our previous late-time calculation\,\cite{You:2026CC} keeps the first nonlocal pole
for $\Cc$ and its complex conjugate for $\Cc^*$, and approximates the exponential factor $e^{-X/2}$ in Eq.\eqref{eq:Wmellin} by unity.\
In Fig.\ref{fig:poles}, we show the entire local family and the higher nonlocal poles
omitted by this approximation.\
We demonstrate that the previously omitted poles in Ref.\,\cite{You:2026CC} 
can provide the major contributions to the complete CC signals.

\vs 

We first evaluate the two first local poles, one on each soft line.\
Their dominant contribution to bispectrum amplitude $\Cc$ comes from the SK branch $(++-)$, 
i.e., $a\!=\!b\!=\!-c\!=\!+$ in Fig.\,\ref{fig:2}.\footnote{%
We present the detailed derivation in Appendix\,\ref{sm:SKchoice}.}\
From Eq.\eqref{sm:LLfinite}, retaining both soft exponentials,
we obtain the resultant $\Cc$ from only the first local poles 
and our complete result under the large chemical potential limit 
($\tlambda\!\rightarrow\! \infty$):
{\small 
\begin{equation}
\hspace*{-3mm}
\lim_{\tlambda\to\infty}\!\!\mathbb C_+^{\rm{LL}(0,0);++-}
\!\!=\!-\frac{\m^2}{\,32\pi\Lambda^3\,}\hs,
\hspace*{4mm}
\lim_{\tlambda\to\infty}\!\!\Cc\!=\!-\frac{\m^2}{\,16\pi\Lambda^3\,} \hs.
\label{eq:half}
\end{equation}
}
\hspace*{-2.5mm}
In the above, $\mathrm{LL}(0,0)$ denotes the first local pole on each soft line.\
This pair on the single $(++-)$ path gives half the complete coefficient
with the same asymptotic phase.\
Eqs.\,\eqref{sm:LLfinite}-\eqref{sm:LLhalf} give the finite Gamma-function expression
and its reduction to Eq.\eqref{sm:halfMain}.\

\vs 

To identify the remaining half, we keep the $q$ line at its first local pole
and include the higher local poles on the $p$ line.\
We denote these contributions by $\mathrm{LL}(n,0)$, 
where $n\!=\!n_p$ and $n_q\!=\!0\hs$.\
After the time and loop-momentum integrations, we obtain the following limit, 
for each given $n\!\geqslant\!0$ and $\m\hs$,
{\small
\begin{equation}
\hspace{-2mm}
\lim_{\tlambda\to\infty}^{}\!\!\frac{\,\mathbb C_+^{\rm{LL}(n,0);++-}\,}{\Cc}
=\int_0^1\!\!(1\!-\!x)x^n \di x
=\frac{1}{(n\!+\!1)(n\!+\!2)}\,,
\label{eq:ladder}
\end{equation}
}
\hspace*{-1.5mm}%
where $x\!=\!p/k_3$ is the momentum fraction at the collinear boundary
$p\!+\!q\!=\!k_3^{}\hs$.\
The first local pole gives $\tfrac{\,1\,}{\,2\,}$, whereas the higher poles give
$\tfrac{\,1\,}{\,6\,},\tfrac{\,1\,}{\,12\,},\ldots$, whose sum provides the other half.\

\vs 

In the above, the two soft lines play different roles because their spinor components
have different mode coefficients.\
The higher local poles on the $p$ line survive the large $\tlambda$ limit,
whereas those on the $q$ line are suppressed after integration.\
Keeping the $p$ line at its first local pole and summing the local family on the $q$ line
therefore leaves the first-pole result unchanged.\
Furthermore, the mixed local-nonlocal first-pole pairs give subleading corrections.\
We present the detailed derivation for this analysis in Appendix\,\ref{App:G}.

\vs 

In summary, the pole analysis thus identifies that the local poles in the soft propagator provide the leading contributions for the total CC signal, which is responsible for
the enhanced asymptotic amplitude compared with \cite{You:2026CC}.\
We note that in the above analysis, we first extract the relevant propagator residues and then evaluate the remaining time and loop-momentum integrals without further approximation. This differs from the procedure adopted in Refs.\,\cite{Qin:2022fbv,Qin,Qin:2024gtr,Aoki:2026vbc,Aoki:2026smw}, where these integrations are performed first, and poles arising from both the propagators and the integrations are subsequently taken into account. Despite this difference in the order of operations, both procedures can retain the full contribution of the time and loop-momentum integrals.



\vspace*{3mm}
\noindent 
{\bf 7.~Conclusions}
\vspace*{1.5mm}

In this work, we derived for the first time
a finite closed-form expression for the leading cosmological collider ( CC) signal 
as generated by right-handed Majorana neutrino loops with chemical-potential enhancement in the collinear squeezed bispectrum.\ We used a localized directional cut to isolate the contribution carrying the leading oscillatory coefficient, with which the loop diagram could be cut into the hard and soft parts.\ 
Retaining the complete mode functions at the soft vertex, we compute the complete loop integral, 
and sum over all the spinor, helicity and SK contributions.\ The resultant complex clock coefficient 
is expressed in terms of Euler Beta and ordinary Meijer $G$ functions and determines 
both the CC signal amplitude and phase at the finite mass and chemical potential.

\vs

For a given fermion mass, the large-chemical-potential expansion yields a power-law prefactor 
$\m^2\tlambda^3$, as compared with $\m^4\tlambda^{3/2}$ in the previous nonlocal 
late-time estimate\,\cite{You:2026CC}.\ Together with the exponential factors in Eqs.\eqref{eq:asymnew} and \eqref{eq:asymold}, these scaling behaviors account for the mass dependence in Fig.\,\ref{fig:fnl} 
and the significant enhancement relative to the previous estimate when the mass $\m$ decreases.

\vs

We analyzed the Mellin-pole structure and used it to explain the origin of the large difference
of our present complete bispectrum amplitude from the previous nonlocal 
late-time result\,\cite{You:2026CC}.\ 
In the same large-chemical-potential limit, the pair of first local poles on the two soft lines 
contributes half of the complete complex coefficient on the SK path $(++-)$, whereas 
higher local poles on the $p$ line provide the remaining half.\ 
These contributions were previously omitted by the nonlocal truncation\,\cite{You:2026CC}.\ 
They survive at leading squeezed order because the mode arguments at the soft vertex 
remain finite as $k_3/k_1\!\to\!0\hs$, 
requiring the inclusion of the complete soft-vertex mode functions.

\vspace*{3mm}
\noindent  
{\bf Acknowledgments}
\\[0.8mm]
We used GPT-6 Astra and Claude Opus\,5.5 to assist with and cross-check our analysis.\ 
We independently verified our methods and results for the closed-form correlation functions 
through machine-checked formal proofs in Lean.\ 
This research was supported in part by the National Natural Science Foundation of China (NSFC)
under Grants No.\,12435005 and No.\,12175136,
by Shenzhen Science and Technology Program
(Grant No.\,JCYJ2024 0813150911015),
by the State Key Laboratory of Dark Matter Physics,
by the Key Laboratory for Particle Astrophysics and Cosmology (MOE),
and by the Shanghai Key Laboratory for Particle Physics and Cosmology.\

\nocite{DLMF}
\bibliography{references}
\clearpage
\label{last:letter}
\onecolumngrid

\makeatletter
\let\set@footnotewidth\set@footnotewidth@one
\let\compose@footnotes\compose@footnotes@one
\makeatother

\clubpenalty=10000
\widowpenalty=10000
\displaywidowpenalty=10000
\raggedbottom
\appendix
\setcounter{secnumdepth}{2}
\renewcommand{\thesection}{\Alph{section}}
\renewcommand{\thesubsection}{\thesection.\arabic{subsection}}
\makeatletter
\def\p@subsection{}
\def\p@subsubsection{}
\makeatother
\setcounter{section}{0}
\setcounter{equation}{0}

\renewcommand{\theequation}{\ifnum\value{section}=0 S\else\thesection\fi\arabic{equation}}
\makeatletter\@addtoreset{equation}{section}\makeatother
\begin{center}
  {\large\bfseries Appendix}\\[3mm]
\end{center}
\noindent
In these appendices, we provide details of the calculation of the cosmological collider signal from the neutrino triangle loop and of its coefficient $\Cc$ multiplying $\rho^{2\ii\tnu}$ in the collinear squeezed limit.\ 
In Appendix\,\ref{sm:model}, we describe the dynamics of the Majorana fermion with chemical potential and use the Nambu field to combine the eight diagrams into a single trace. In Appendix\,\ref{sm:cut}, we prove the cutting relation and fix the finite part of the soft integral. In Appendices\,\ref{sm:hard}--\ref{sm:native}, we calculate the hard and soft parts separately and then evaluate their product. In Appendices\,\ref{sm:asym}--\ref{sm:checks}, we derive the large-chemical-potential limit of the CC signal and present checks based on Mellin poles and numerical analysis.

For clarity, we restate the parameter definitions used throughout the appendices. The dimensionless physical parameters, corresponding to Eq.\,\eqref{eq:physical} of the main text, are
\begin{equation}
  \m=\frac{M_R}{H},\qquad
  \tlambda=\frac{\lambda}{H},\qquad
  \tnu=\sqrt{\m^2+\tlambda^2}.
  \label{sm:overviewPhysical}
\end{equation}
The parameter combinations used in the closed-form expressions, corresponding to Eq.\,\eqref{eq:finitepars} of the main text, are
\begin{equation}
  \begin{aligned}
    a&=-\ii(\tnu-\tlambda),\qquad
    b=-\ii(\tnu+\tlambda),\\[2pt]
    A&=2+2\ii\tnu,\qquad
    \omega=\frac{\sinh(2\pi\tlambda)\sinh(\pi\tlambda)}{\sinh(\pi\tnu)}.
  \end{aligned}
  \label{sm:overviewFinitePars}
\end{equation}
These satisfy $a+b=2-A$ and $ab=-\m^2$. When used as Schwinger--Keldysh (SK) branch labels, $a,b,c=\pm1$ instead denote discrete contour indices and are distinct from the complex parameters $a,b$ defined above. Complex conjugation acts on the whole indicated expression with its stated boundary prescription.

\bigskip\noindent\textbf{Contents}\par\medskip
\begin{tabular}{@{}p{.88\textwidth}r@{}}
A.\  Fermion Loop at Finite Chemical Potential and Nambu Representation                       
   & \pageref{sm:model}  \\[5pt]
B.\  Factorization with Localized Directional Cutting 
& \pageref{sm:cut}    \\[5pt]
C.\  Computation of the Hard Part        
   & \pageref{sm:hard}   \\[5pt]
D.\  Computation of the Soft Part      & \pageref{sm:soft}   \\[5pt]
E.\  Derivation of the Complete Closed-Form Expression              
  & \pageref{sm:native} \\[5pt]
F.\  Inflaton Correlator in the Large-Chemical-Potential Limit      
  & \pageref{sm:asym}   \\[5pt]
G.\  Analyzing Inflaton Correlator through the Mellin Poles   
  & \pageref{sm:poles}  \\[5pt]
H.\  Benchmark Points and Numerical Analysis                 
  & \pageref{sm:checks}
\end{tabular}

\section{Fermion Loop at Finite Chemical Potential and Nambu Representation}
\label{sm:model}
\label{App:A}
We establish the field and propagator conventions for the Majorana loop. Starting from the inflaton--neutrino Lagrangian, we construct the canonical modes and express the three Weyl propagators of Ref.\,\cite{You:2026CC} as Nambu blocks. Their assembly reproduces all eight connected contractions in a single trace, with the Wick signs and normalization used in the main text.

\subsection{Lagrangian, Canonical Fields and Mode Functions}\label{sm:lagrangianmodes}
In this work, the relevant inflaton--right-handed-neutrino Lagrangian is
\begin{equation}
 \mathcal L=\sqrt{-g}\biggl[-\frac12\partial_\mu\phi\,\partial^\mu\phi-V(\phi)+\frac{\ii}{a(\tau)}N^\dagger\bar\sigma^\mu\mathrm D_\mu N-\frac{1}{a(\tau)\Lambda}\partial_\mu\phi\,N^\dagger\bar\sigma^\mu N-\frac{M_R}{2}\left(NN+\mathrm{h.c.}\right)\biggr].
 \label{sm:fulllagrangian}
\end{equation}
Here $g$ is the metric determinant and $V(\phi)$ is the inflaton potential. The field $N$ is the two-component Weyl field describing a right-handed neutrino with a real, positive Majorana mass $M_R$, and $NN=N^\alpha N_\alpha$. The scale $\Lambda$ suppresses the derivative coupling to the inflaton. We use $\di s^2=a^2(\tau)(-\di\tau^2+\di\vct x^2)$, $\sqrt{-g}=a^4$ and $a(\tau)=-1/(H\tau)$ at leading slow-roll order. The flat-frame matrices are $\sigma^\mu=(1,\vct\sigma)$ and $\bar\sigma^\mu=(1,-\vct\sigma)$; the factors $a^{-1}$ in Eq.\,\eqref{sm:fulllagrangian} come from the inverse vierbein. The spinor covariant derivative $\mathrm D_\mu$ includes the spin connection\,\cite{You:2026CC}.

The rescaling $N(x)=a^{-3/2}\Nt(x)$ removes the spin connection and canonically normalizes the fermion kinetic term. Splitting $\phi=\phi_0+\delta\phi$ then separates the derivative interaction into a background contribution and a coupling to the fluctuation. Since $\partial_\tau\phi_0=a\dot\phi_0$, the background contributes $-a\lambda\Nt^\dagger\bar\sigma^0\Nt$, where $\lambda\equiv\dot\phi_0/\Lambda$. The canonical quadratic Lagrangian and fluctuation interaction are therefore
\begin{subequations}\label{sm:canonicalL}
\begin{align}
 \widetilde{\mathcal L}_N^{(2)}&=\ii\Nt^\dagger\bar\sigma^\mu\partial_\mu\Nt
 -a\lambda\Nt^\dagger\bar\sigma^0\Nt-\frac{aM_R}{2}(\Nt\Nt+\mathrm{h.c.}),\label{sm:quadraticL}\\
 \mathcal L_{{\rm{int}},\delta\phi}^{(5)}&=-\frac{\sqrt{-g}}{a(\tau)\Lambda}\partial_\mu\delta\phi\,N^\dagger\bar\sigma^\mu N
 =-\frac{1}{\Lambda}\partial_\mu\delta\phi\,\Nt^\dagger\bar\sigma^\mu\Nt.
 \label{sm:cubicL}
\end{align}
\end{subequations}
The background chemical potential is included in the free propagators, while the fluctuation coupling is treated perturbatively. 

We first expand the canonical field in terms of the spinor mode functions $\xi_{s,\alpha}$ and $\chi_{s,\alpha}$:
\begin{equation}
 \begin{aligned}
 \Nt_\alpha(\tau,\vct x)
 &=\int\frac{\di^3\vct k}{(2\pi)^3}\sum_{s=\pm1}
 \left[\xi_{s,\alpha}(\tau,\vct k)b_s(\vct k)e^{\ii\vct k\cdot\vct x}
 +\chi_{s,\alpha}(\tau,\vct k)b_s^\dagger(\vct k)e^{-\ii\vct k\cdot\vct x}\right],\\
 \Nt^\dagger_{\dot\alpha}(\tau,\vct x)
 &=\int\frac{\di^3\vct k}{(2\pi)^3}\sum_{s=\pm1}
 \left[\xi^\dagger_{s,\dot\alpha}(\tau,\vct k)b_s^\dagger(\vct k)e^{-\ii\vct k\cdot\vct x}
 +\chi^\dagger_{s,\dot\alpha}(\tau,\vct k)b_s(\vct k)e^{\ii\vct k\cdot\vct x}\right].
 \end{aligned}
 \label{sm:fieldexpansion}
\end{equation}
The operators $b_s(\vct k)$ and $b_s^\dagger(\vct k)$ annihilate and create a Majorana excitation of comoving momentum $\vct k$, respectively. Undotted and dotted indices label the Weyl field and its adjoint. Here $s=\pm1$ is the helicity label, $k=|\vct k|$, and $\{b_s(\vct k),b_{s'}^\dagger(\vct k')\}=(2\pi)^3\delta_{ss'}\delta^3(\vct k-\vct k')$. The helicity decomposition is
\begin{equation}
 \xi_{s,\alpha}(\tau,\vct k)=u_s(\tau,k)h_{s,\alpha}(\vct k),\qquad
 \chi_s^{\dagger\dot\alpha}(\tau,\vct k)=v_s(\tau,k)\,s\,h_{-s}^{\dagger\dot\alpha}(\vct k).
 \label{sm:helicitymodes}
\end{equation}
The spinors $h_s$ carry the momentum-direction dependence, whereas $u_s$ and $v_s$ are scalar mode coefficients. We write $1_d$ for the $d$-dimensional identity matrix. The unit helicity eigenspinors satisfy
\begin{equation}
 (\vct\sigma\cdot\widehat{\vct k})h_s(\vct k)=s h_s(\vct k),\qquad
 h_s^\dagger(\vct k)h_{s'}(\vct k)=\delta_{ss'},\qquad
 \sum_{s=\pm1}h_s(\vct k)h_s^\dagger(\vct k)=1_2.
 \label{sm:helicitybasis}
\end{equation}

With primes denoting conformal-time derivatives of the mode functions, the equations derived from $\widetilde{\mathcal L}_N^{(2)}$ then take the form
\begin{equation}
 \begin{aligned}
 \ii u_\pm'\pm k u_\pm&=a\lambda u_\pm+aM_Rv_\pm,\\
 \ii v_\pm'\mp k v_\pm&=-a\lambda v_\pm+aM_Ru_\pm.
 \end{aligned}
 \label{sm:modeeq}
\end{equation}
Using the dimensionless parameters in Eq.\,\eqref{eq:physical}, the canonically normalized Bunch--Davies (BD) solutions are
\begin{subequations}\label{sm:modes}
  \begin{align}
    u_+(\tau,k)                                   & =\frac{\,\m e^{\pi\tlambda/2}\,}{\,\sqrt{-2k\tau}\,}
    W_{-1/2\!-\!\ii\tlambda,\ii\tnu}(\ii\hp 2 k\tau), &
    u_-(\tau,k)                                   & =\frac{\,\ii e^{-\pi\tlambda/2}\,}{\,\sqrt{-2k\tau}\,}
    W_{1/2\!+\!\ii\tlambda,\ii\tnu}(\ii\hp 2 k\tau)\,,\label{sm:umodes}                                     \\
    v_+(\tau,k)                                   & =\frac{\,\ii e^{\pi\tlambda/2}\,}{\,\sqrt{-2k\tau}\,}
    W_{1/2\!-\!\ii\tlambda,\ii\tnu}(\ii\hp 2 k\tau),  &
    v_-(\tau,k)                                   & =\frac{\,\m e^{-\pi\tlambda/2}\,}{\,\sqrt{-2k\tau}\,}
    W_{-1/2\!+\!\ii\tlambda,\ii\tnu}(\ii\hp 2 k\tau)\,.
    \label{sm:modes:last}
  \end{align}
\end{subequations}

\subsection{Weyl Propagators and Their Nambu Representation}\label{sm:weylnambu}
We now introduce the fermion propagators in the Schwinger--Keldysh (SK) formalism\,\cite{SK} and express them in the Nambu representation.
The normal Weyl propagator $D_{ab\,\alpha\dot\beta}$ contracts $\Nt$ with $\Nt^\dagger$, whereas the anomalous propagators $\hat D_{ab\,\alpha\beta}$ and $\check D_{ab\,\dot\alpha\dot\beta}$ contract two $\Nt$ fields or two $\Nt^\dagger$ fields, respectively. The labels $a,b=\oplus,\ominus$ specify the forward and backward SK branches. In the conventions of Ref.\,\cite{You:2026CC}, their $\ominus\oplus$ components are
\begin{subequations}\label{sm:weylpropagators}
\begin{align}
 D_{\ominus\oplus\,\alpha\dot\beta}(\vct k;\tau_1,\tau_2)
 &=\sum_s\xi_{s,\alpha}(\tau_1,\vct k)\xi^\dagger_{s,\dot\beta}(\tau_2,\vct k),
 &&\text{type-1},\label{sm:normalweyl}\\
 \hat D_{\ominus\oplus\,\alpha\beta}(\vct k;\tau_1,\tau_2)
 &=\sum_s\xi_{s,\alpha}(\tau_1,\vct k)\chi_{s,\beta}(\tau_2,\vct k),
 &&\text{type-2},\label{sm:anomalousweyl}\\
 \check D_{\ominus\oplus\,\dot\alpha\dot\beta}(\vct k;\tau_1,\tau_2)
 &=\sum_s\chi^\dagger_{s,\dot\alpha}(\tau_1,\vct k)\xi^\dagger_{s,\dot\beta}(\tau_2,\vct k),
 &&\text{type-3}.\label{sm:conjugateweyl}
\end{align}
\end{subequations}
When spinor indices are suppressed, $D_{ab}$ denotes the corresponding two-component spinor matrix. For a line connecting endpoints $i$ and $j$, we use the following abbreviated notation for the propagators:
\begin{equation}
  D_{ab}\llbracket ij\rrbracket\equiv D_{ab}(\vct p_{ij};\tau_i,\tau_j),
  \label{sm:weylShorthand}
\end{equation}
and analogously for $\hat D$ and $\check D$. Here $a,b$ specify the SK branches at the two endpoints, while $\vct p_{ij}$ is the oriented momentum assigned to that line. The double brackets thus abbreviate its momentum and two time arguments; they do not denote an additional operation.
We denote the greater and lesser Wightman propagators by $D^>\equiv D_{\ominus\oplus}$ and $D^<\equiv D_{\oplus\ominus}$, respectively, with the same convention for $\hat D$ and $\check D$. More components are given in Ref.\,\cite{You:2026CC}. 

For the Nambu formulas and contour sums below, we identify $\oplus,\ominus$ with $+,-$, or numerically $+1,-1$. The momentum $\vct k$ in these definitions is replaced by the routed momentum of each internal line.

We now collect these propagators into Nambu form. We use $\epsilon\equiv\ii\sigma_2=\bigl(\begin{smallmatrix}0&1\\-1&0\end{smallmatrix}\bigr)$ for the antisymmetric spinor matrix. Its indexed forms satisfy $[\epsilon^{\alpha\beta}]=[\epsilon^{\dot\alpha\dot\beta}]=\epsilon$ and $[\epsilon_{\alpha\beta}]=[\epsilon_{\dot\alpha\dot\beta}]=-\epsilon$, so that $\epsilon^{12}=\epsilon^{\dot1\dot2}=1$ and $\epsilon_{12}=\epsilon_{\dot1\dot2}=-1$. In matrix notation, $\epsilon^T=\epsilon^\dagger=-\epsilon$ and $\epsilon^2=-1_2$. With these conventions, define
\begin{equation}
 \Psi=\begin{pmatrix}\Nt\\\epsilon\Nt^{\dagger T}\end{pmatrix},\qquad
 \Psi^\dagger=(\Nt^\dagger,-\Nt^T \epsilon),\qquad
 S_{ab}(\vct p_{ij};\tau_i,\tau_j)=\langle T_C\Psi_a(i)\Psi_b^\dagger(j)\rangle_{\vct p}.
 \label{sm:nambudef}
\end{equation}
We use a similar abbreviation for the Nambu propagator:
\begin{equation}
  S_{ab}\llbracket ij\rrbracket\equiv S_{ab}(\vct p_{ij};\tau_i,\tau_j).
  \label{sm:nambuShorthand}
\end{equation}
In Eq.\,\eqref{sm:nambudef}, $i=(\tau_i,\vct x)$, $j=(\tau_j,\vct0)$, $T_C$ denotes ordering along the SK contour, and $\langle\cdots\rangle_{\vct p}$ includes the Fourier transform $\int\di^3x\,e^{-\ii\vct p\cdot\vct x}$, with $\vct p=\vct p_{ij}$ on the chosen line. For convenience, we leave the contour assignment implicit and write $S_{ij}\equiv S_{a_i a_j}\llbracket ij\rrbracket$ and $D_{ij}\equiv D_{a_i a_j}\llbracket ij\rrbracket$, where $a_i$ is the SK branch at endpoint $i$. The two Nambu entries distinguish the field from its charge conjugate, independently of these SK labels. The propagator and vertex matrices are
\begin{equation}
 S_{ab}\llbracket ij\rrbracket=\begin{pmatrix}D_{ab}\llbracket ij\rrbracket&\hat D_{ab}\llbracket ij\rrbracket\\\check D_{ab}\llbracket ij\rrbracket&Q_{ab}\llbracket ij\rrbracket\end{pmatrix},\qquad
 \Gamma^\mu=\begin{pmatrix}\bar\sigma^\mu&0\\0&-\sigma^\mu\end{pmatrix}.
 \label{sm:blocks}
\end{equation}
The normal block has components $(D_{ab})_{\alpha\dot\beta}=D_{ab\,\alpha\dot\beta}$. The lower diagonal block $Q_{ab}$ propagates the charge-conjugate field; its endpoint-reversal formula is given below. The anomalous blocks use the raised-index propagators of Ref.\,\cite{You:2026CC}. Using the spinor-index conventions stated above, their definitions give
\begin{equation}
 \begin{aligned}
 (\hat D_{ab})_\alpha{}^\beta
 &\equiv\hat D_{ab\,\alpha}{}^\beta
 =\epsilon^{\beta\gamma}\hat D_{ab\,\alpha\gamma}
 =-\hat D_{ab\,\alpha\gamma}\epsilon^{\gamma\beta},\\
 (\check D_{ab})^{\dot\alpha}{}_{\dot\beta}
 &\equiv\check D_{ab}{}^{\dot\alpha}{}_{\dot\beta}
 =\epsilon^{\dot\alpha\dot\gamma}\check D_{ab\,\dot\gamma\dot\beta}
 =-\epsilon^{\dot\gamma\dot\alpha}\check D_{ab\,\dot\gamma\dot\beta},\\
 Q_{ab}(\vct p;\tau_i,\tau_j)
 &=\epsilon D_{ba}^{T}(-\vct p;\tau_j,\tau_i)\epsilon.
 \end{aligned}
 \label{sm:weylnambumap}
\end{equation}

Using Eqs.\,\eqref{sm:helicitymodes} and \eqref{sm:weylnambumap}, introduce the mode columns
\begin{equation}
  \Pi_s(\vct p)=\frac{\,1\!+\!s\widehat{\vct p}\cdot\vct\sigma\,}{\,2\,},\qquad
  \mathbf u_s=\begin{pmatrix}u_s\\v_s\end{pmatrix},\qquad
  \mathbf v_s=\begin{pmatrix}-v_s^*\\u_s^*\end{pmatrix}\,.
  \label{sm:columns}
\end{equation}
Writing $S^>_{ij}\equiv S_{-+}\llbracket ij\rrbracket$ and $S^<_{ij}\equiv S_{+-}\llbracket ij\rrbracket$, the resulting Nambu Wightman matrices are
\begin{equation}
  S^>_{ij}=\sum_s\!\mathbf u_s(\tau_i)\mathbf u_s^\dagger(\tau_j)\otimes\Pi_s(\vct p),\qquad
  S^<_{ij}=-\sum_s\!\mathbf v_s(\tau_i)\mathbf v_s^\dagger(\tau_j)\otimes\Pi_s(\vct p)\,.
  \label{sm:wightman}
\end{equation}
The superscript $\dagger$ on the scalar functions $u_s,v_s$ in Ref.\,\cite{You:2026CC} denotes complex conjugation; we use $*$ for these scalars in the Nambu columns.

We then define the spectral matrix through the anticommutator:
\begin{equation}
  \varrho_{ij}=S^>_{ij}\!-\!S^<_{ij}=\langle\{\Psi(i),\Psi^\dagger(j)\}\rangle,
  \qquad \varrho(\tau,\tau)=1_4\,.
  \label{sm:spectral}
\end{equation}
The equal-time identity follows from the canonical anticommutation relations. Unlike the individual, occupation-dependent Wightman matrices, $\varrho$ is an evolution kernel fixed by the free equation and this initial condition. Its smooth momentum dependence is the key property used for the time-ordering remainders in Appendix\,\ref{sm:remainderproof}.

\subsection{Eight Connected Contractions and One Trace}\label{sm:contractions}

\begin{figure*}[t]
  \centering
  \includegraphics[width=0.90\linewidth]{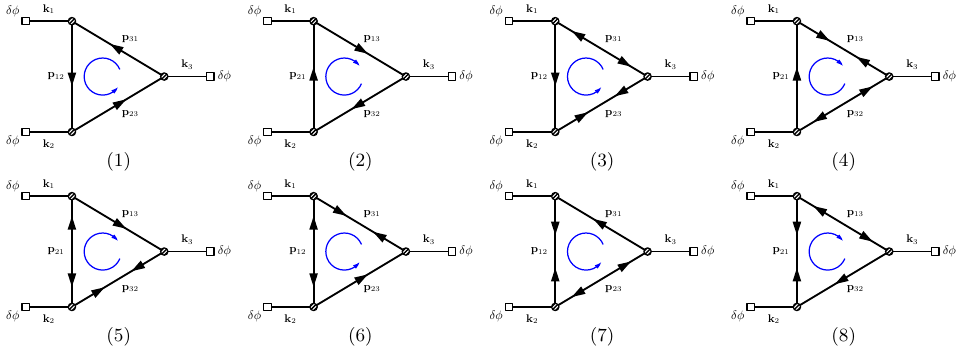}
  \caption{The eight contractions of the right-handed Majorana neutrino loop, numbered as in Ref.\cite{You:2026CC}. Arrowed internal lines denote neutrino propagators; lines joining a vertex to a square are bulk-to-boundary inflaton propagators. Each shaded vertex carries an independent SK label ($\oplus$ or $\ominus$), giving eight contour assignments for each diagram.}
  \label{fig:eight}
\end{figure*}

Fig.\ref{fig:eight} displays the original eight contractions of Ref.\cite{You:2026CC}. Using the Weyl-to-Nambu map in Eq.\,\eqref{sm:weylnambumap}, we express the contractions through the blocks in Eq.\,\eqref{sm:blocks}. Their vertex arguments are
\begin{equation}
  D_{ij}=D_{a_i a_j}(\vct p_{ij},\tau_i,\tau_j),\qquad
  \hat D_{ij}=\hat D_{a_i a_j}(\vct p_{ij},\tau_i,\tau_j),\qquad
  \check D_{ij}=\check D_{a_i a_j}(\vct p_{ij},\tau_i,\tau_j),\qquad
  (a_1,a_2,a_3)=(a,b,c)\,.
  \label{sm:contractionargs}
\end{equation}
The routed momenta, including the reversed lines, are
\begin{equation}
  \vct p_{12}=\vct q\!+\!\vct k_1,\qquad
  \vct p_{23}=\vct q\!-\!\vct k_3,\qquad
  \vct p_{31}=\vct q,\qquad \vct p_{ji}=-\vct p_{ij}\,.
  \label{sm:contractionmomenta}
\end{equation}
Thus $D_{21}$, for example, means $D_{ba}(-\vct p_{12},\tau_2,\tau_1)$; endpoint reversal changes times, SK labels and oriented momentum together.\  In the following list $\mathcal Y(j)$ denotes $\mathcal Y_{abc}^{\mu_1\mu_2\mu_3}(j)$ in Eq.\,\eqref{eq:parent}, and ``$\operatorname{tr}$'' contracts the two-component spinor indices.\  The eight functions are
\begin{subequations}\label{sm:fourcontractions}
\begin{align}
\mathcal Y(1) &=-\operatorname{tr}[\bar\sigma^{\mu_1}D_{12}\bar\sigma^{\mu_2}D_{23}\bar\sigma^{\mu_3}D_{31}]\,,\qquad & \mathcal Y(2) &=-\operatorname{tr}[\bar\sigma^{\mu_2}D_{21}\bar\sigma^{\mu_1}D_{13}\bar\sigma^{\mu_3}D_{32}]\,,\label{sm:Y12}\\
\mathcal Y(3) &=+\operatorname{tr}[\bar\sigma^{\mu_1}D_{12}\bar\sigma^{\mu_2}\hat D_{23}\sigma^{\mu_3}\check D_{31}]\,,\qquad & \mathcal Y(4) &=+\operatorname{tr}[\bar\sigma^{\mu_2}D_{21}\bar\sigma^{\mu_1}\hat D_{13}\sigma^{\mu_3}\check D_{32}]\,,\label{sm:Y34}\\
\mathcal Y(5) &=+\operatorname{tr}[\sigma^{\mu_2}\check D_{21}\bar\sigma^{\mu_1}D_{13}\bar\sigma^{\mu_3}\hat D_{32}]\,,\qquad & \mathcal Y(6) &=+\operatorname{tr}[\sigma^{\mu_1}\check D_{12}\bar\sigma^{\mu_2}D_{23}\bar\sigma^{\mu_3}\hat D_{31}]\,,\label{sm:Y56}\\
\mathcal Y(7) &=+\operatorname{tr}[\bar\sigma^{\mu_1}\hat D_{12}\sigma^{\mu_2}\check D_{23}\bar\sigma^{\mu_3}D_{31}]\,,\qquad & \mathcal Y(8) &=+\operatorname{tr}[\bar\sigma^{\mu_2}\hat D_{21}\sigma^{\mu_1}\check D_{13}\bar\sigma^{\mu_3}D_{32}]\,.\label{sm:Y78}
\end{align}
\end{subequations}
Simultaneous exchange of the complete labels $1\leftrightarrow2$ pairs diagrams $(1,2)$, $(3,4)$, $(6,5)$ and $(7,8)$. The signs in Eq.\,\eqref{sm:fourcontractions} are the Wick signs; the SK weight $abc$ and vertex factors remain outside this list in Eq.\,\eqref{eq:parent}. Since the vertices are diagonal in Nambu space, the trace sums over two block choices at each of three vertices. Its $2^3=8$ terms reproduce the eight contractions:
\begin{equation}
  \sum_{j=1}^{8}\!\mathcal Y(j)=-\operatorname{Tr}[\Gamma^{\mu_1}S_{12}\Gamma^{\mu_2}S_{23}\Gamma^{\mu_3}S_{31}],
  \label{sm:eighttrace}
\end{equation}
Here ``$\operatorname{Tr}$'' runs over both Nambu and two-component spinor indices, whereas ``$\operatorname{tr}$'' in Eq.\,\eqref{sm:fourcontractions} runs only over the latter.

At vertex $i$ on SK branch $a_i$, we contract $\Gamma^{\mu_i}$ with the differentiated external inflaton line [Eq.\,\eqref{eq:externalSource}]:
\begin{equation}
  \mathcal J_i^{a_i}\equiv\mathcal F_{\mu_i a_i}(\vct k_i,\tau_i)\,\Gamma^{\mu_i}\,,
  \qquad (a_1,a_2,a_3)=(a,b,c)\,.
  \label{eq:source}
\end{equation}
Using Eq.\,\eqref{sm:eighttrace} for the sum over the eight contractions in Eq.\,\eqref{eq:parent} and collecting one $\mathcal J_i^{a_i}$ at each vertex gives
{\small
\begin{equation}
  \Bphi={}-(-\ii/\Lambda)^3\sum_{a,b,c=\pm1}\!abc \int_{-\infty}^{0}\di\tau_1\di\tau_2\di\tau_3 \int\frac{\,\di^3q\,}{\,(2\pi)^3\,}\operatorname{Tr}[\mathcal J_1^a S_{ab}\llbracket 12\rrbracket\mathcal J_2^b S_{bc}\llbracket 23\rrbracket \mathcal J_3^c S_{ca}\llbracket 31\rrbracket] \,. \label{sm:parenttrace}
\end{equation}
}
{\par\parindent=2.5mm
\hspace*{-2.5mm}%
Here $S_{ab}\llbracket 12\rrbracket\!=\!S_{ab}(\vct p_{12},\tau_1,\tau_2)$, with the analogous definitions on the other two edges.\  The overall minus sign follows from Eq.\,\eqref{sm:eighttrace}, which collects the Wick signs contained in the eight functions $\mathcal Y(j)$.\
\par}

\section{Factorization with Localized Directional Cutting}\label{sm:cut}\label{App:B}
We derive Eq.\,\eqref{eq:cut} in two steps, at fixed positive mass and chemical potential in the collinear squeezed limit. First, localized directional cutting isolates the leading oscillatory coefficient of the original loop. Second, a controlled hard-endpoint projection factorizes that coefficient as $\Cc=\Ch\Cs$. The momentum boundary inherited from the original integral fixes the finite part of $\Cs$; it is not an independent subtraction prescription.

\subsection{The Signal to Be Isolated and the Two Cuts}\label{sm:cutsetup}
Throughout this appendix, $\mathcal B\equiv\Bphi$ denotes the full triangle-loop contribution in Eq.\,\eqref{sm:parenttrace}, including the external legs, and $\widehat{\mathcal B}=k_1^6\mathcal B/H^6$. We take the collinear squeezed limit $\rho=k_3/k_1\to0$ at fixed $k_1$, with $\vct k_1$ and $\vct k_3$ pointing in the same direction. Our target is the coefficient of the leading oscillations $\rho^{\pm2\ii\tnu}=e^{\pm2\ii\tnu\ln\rho}$ in $\widehat{\mathcal B}$. The factorization $\Cc=\Ch\Cs$ concerns this coefficient; the full correlator also contains nonoscillatory background and subleading terms.

Two distinct operations enter the argument. A \emph{momentum separation} partitions the loop integral into a soft region and its complement, retaining both. A \emph{propagator cut} decomposes each of the two soft-line propagators into a chosen Wightman function and an explicit time-ordering correction within each SK assignment. We first write this algebraic decomposition over the full momentum domain, then apply it only inside the soft region. No term is discarded at the decomposition stage.

The two cut lines, 23 and 31, share vertex~3. The adjacent-cut argument of Ref.\,\cite{Qin:2023bjk} assumes at least two external lines at the shared vertex, allowing their total spatial momentum to vanish while the sum of their energy magnitudes remains finite. Here vertex~3 has only one external line, so its external energy also vanishes as $k_3\to0$. The soft-time integral can then introduce additional nonanalytic dependence on $k_3$, and the cited argument cannot be applied directly. We therefore establish the required remainder bounds for this three-point limit below.

We use $S_{-+}=S^>$, $S_{+-}=S^<$ and the spectral matrix $\varrho=S^>-S^<$ defined in Appendix\,\ref{sm:weylnambu}. With $\theta_{ij}=\theta(\tau_i-\tau_j)$, the time-ordered propagators satisfy
\begin{equation}
 S_{++}\llbracket ij\rrbracket=S^<_{ij}+\theta_{ij}\varrho_{ij},\qquad
 S_{--}\llbracket ij\rrbracket=S^>_{ij}-\theta_{ij}\varrho_{ij}.
  \label{sm:orderedSKpropagators}
\end{equation}
Thus the two soft lines in the parent trace, Eq.\,\eqref{sm:parenttrace}, obey the exact identities
\begin{equation}
  S_{bc}\llbracket 23\rrbracket=S_{-c,c}\llbracket 23\rrbracket\!-\!c\delta_{bc}\theta_{32}\varrho_{23}\,,\quad S_{ca}\llbracket 31\rrbracket=S_{c,-c}\llbracket 31\rrbracket\!+\!c\delta_{ac}\theta_{31}\varrho_{31}\,. \label{sm:cutidentity}
\end{equation}
The Kronecker deltas $\delta_{bc},\delta_{ac}$ select equal branches at the endpoints of the corrected line. The first terms form the Wightman pair $(S^>,S^<)$ for $c=+$ and $(S^<,S^>)$ for $c=-$, with no step function linking $\tau_3$ to either hard time. Choosing a correction on one or both lines instead introduces $\theta_{31}$ or $\theta_{32}$. These corrections vanish when the soft vertex is earliest, but are generally nonzero elsewhere. The claim that the Wightman pair in the soft region alone captures the leading CC coefficient requires the integrated remainder estimates below; it does not follow from Eq.\,\eqref{sm:cutidentity} alone.

Substituting Eq.\,\eqref{sm:cutidentity} into the original triangle trace and expanding gives four terms, according to whether neither, one, or both soft lines contribute a time-ordering correction:
\begin{equation}
  \mathcal B=\mathcal B_{\rm{cut}}\!+\!\mathcal R_1\!+\!\mathcal R_2\!+\!\mathcal R_{12}\,.
  \label{sm:cutexpand}
\end{equation}
where the four contributions are
\begin{subequations}\label{sm:cutFourTerms}
\begin{align}
 \mathcal B_{\rm{cut}}
 &=-\left(\frac{-\ii}{\Lambda}\right)^3\sum_{a,b,c=\pm1}abc
 \int_{-\infty}^{0}\di\tau_1\di\tau_2\di\tau_3
 \int\frac{\di^3q}{(2\pi)^3}\notag\\*[-2pt]
 &\qquad\times\operatorname{Tr}\!\left[
 \mathcal J_1^a S_{ab}\llbracket 12\rrbracket\mathcal J_2^b S_{-c,c}\llbracket 23\rrbracket
 \mathcal J_3^c S_{c,-c}\llbracket 31\rrbracket\right],\label{sm:cutWightmanTerm}\\[5pt]
 \mathcal R_1
 &=-\left(\frac{-\ii}{\Lambda}\right)^3\sum_{a,b,c=\pm1}abc\,c\delta_{ac}
 \int_{-\infty}^{0}\di\tau_1\di\tau_2\di\tau_3\,\theta_{31}
 \int\frac{\di^3q}{(2\pi)^3}\notag\\*[-2pt]
 &\qquad\times\operatorname{Tr}\!\left[
 \mathcal J_1^a S_{ab}\llbracket 12\rrbracket\mathcal J_2^b S_{-c,c}\llbracket 23\rrbracket
 \mathcal J_3^c\varrho_{31}\right],\label{sm:cutRoneTerm}\\[5pt]
 \mathcal R_2
 &=+\left(\frac{-\ii}{\Lambda}\right)^3\sum_{a,b,c=\pm1}abc\,c\delta_{bc}
 \int_{-\infty}^{0}\di\tau_1\di\tau_2\di\tau_3\,\theta_{32}
 \int\frac{\di^3q}{(2\pi)^3}\notag\\*[-2pt]
 &\qquad\times\operatorname{Tr}\!\left[
 \mathcal J_1^a S_{ab}\llbracket 12\rrbracket\mathcal J_2^b\varrho_{23}
 \mathcal J_3^c S_{c,-c}\llbracket 31\rrbracket\right],\label{sm:cutRtwoTerm}\\[5pt]
 \mathcal R_{12}
 &=+\left(\frac{-\ii}{\Lambda}\right)^3\sum_{a,b,c=\pm1}abc\,\delta_{ac}\delta_{bc}
 \int_{-\infty}^{0}\di\tau_1\di\tau_2\di\tau_3\,\theta_{31}\theta_{32}
 \int\frac{\di^3q}{(2\pi)^3}\notag\\*[-2pt]
 &\qquad\times\operatorname{Tr}\!\left[
 \mathcal J_1^a S_{ab}\llbracket 12\rrbracket\mathcal J_2^b\varrho_{23}
 \mathcal J_3^c\varrho_{31}\right].\label{sm:cutRbothTerm}
\end{align}
\end{subequations}
Here $\mathcal B_{\rm{cut}}$ uses the Wightman term on both lines, while $\mathcal R_1$, $\mathcal R_2$ and $\mathcal R_{12}$ contain corrections on line 31, line 23 and both lines, respectively.

We now localize the cut. Choose a positive dimensionless separator $\varepsilon_{\rm{sep}}$ independently of $\rho$, and take $\rho\to0$ with $\varepsilon_{\rm{sep}}$ fixed, so that $k_3\ll\varepsilon_{\rm{sep}}k_1\ll k_1$. For example, $\varepsilon_{\rm{sep}}=1/16$ is a convenient choice \footnote{The upper bound $1/16$ is a conservative convenience, not a physical threshold; any fixed, sufficiently small bound satisfying the estimates below is admissible. This separator partitions the momentum domain and is distinct from the ultraviolet regulator.}. Introduce a smooth cutoff function $\chi_{\varepsilon_{\rm{sep}}}(\vct q)$ that equals unity for $q<\varepsilon_{\rm{sep}}k_1/2$ and vanishes for $q>\varepsilon_{\rm{sep}}k_1$, with a smooth transition in between. The precise form of the transition is unimportant, as long as it is smooth enough to allow the estimates below. The separator and cutoff satisfy
\begin{equation}
 \begin{gathered}
  0<k_3<\varepsilon_{\rm{sep}} k_1/4,\quad
  \chi_{\varepsilon_{\rm{sep}}}(\vct q)=
  \begin{cases}
   1, & q<\varepsilon_{\rm{sep}} k_1/2,\\[2pt]
   0, & q>\varepsilon_{\rm{sep}} k_1.
  \end{cases}
 \end{gathered}
 \label{sm:separator}
\end{equation}
This choice places both potentially soft points, $\vct q=0$ and $\vct q=\vct k_3$, inside the region where $\chi_{\varepsilon_{\rm{sep}}}=1$. On its support, $q<\varepsilon_{\rm{sep}}k_1$ and $p=|\vct q-\vct k_3|<5\varepsilon_{\rm{sep}}k_1/4$, while the hard-line momentum is at least $(1-\varepsilon_{\rm{sep}})k_1$. These bounds allow hard external damping to control the time-ordering corrections. 

We split the momentum integral into two weighted pieces: $\chi_{\varepsilon_{\rm{sep}}}$ selects the soft region, while $1-\chi_{\varepsilon_{\rm{sep}}}$ selects its complement. Explicitly, for any contribution $X$, we write
\begin{equation}
 X[w]\equiv\int\frac{\di^3q}{(2\pi)^3}\,w(\vct q)\,\mathcal K_X(\vct q),
  \label{sm:weightedMomentumIntegral}
\end{equation}
where $w(\vct q)$ is the chosen momentum weight and $\mathcal K_X$ contains the time integrals, propagator trace, vertices and SK sum defining $X$. We apply the propagator decomposition of Eq.\,\eqref{sm:cutexpand} only to the soft piece and leave the original propagators in the complement. This gives
\begin{equation}
 \begin{aligned}
 \mathcal B
 &=\underbrace{\mathcal B[\chi_{\varepsilon_{\rm{sep}}}]}_{\text{soft region}}
   +\underbrace{\mathcal B[1-\chi_{\varepsilon_{\rm{sep}}}]}_{\text{complementary region}}\\[5pt]
 &=\underbrace{\mathcal B_{\rm{cut}}[\chi_{\varepsilon_{\rm{sep}}}]}_{\text{soft-region Wightman pair}}
   +\underbrace{\mathcal R_1[\chi_{\varepsilon_{\rm{sep}}}]+\mathcal R_2[\chi_{\varepsilon_{\rm{sep}}}]+\mathcal R_{12}[\chi_{\varepsilon_{\rm{sep}}}]}_{\text{soft-region time-ordering remainders}}
   +\underbrace{\mathcal B[1-\chi_{\varepsilon_{\rm{sep}}}]}_{\text{original propagators}}\,.
 \end{aligned}
 \label{sm:localizedcutdifference}
\end{equation}
Our task is to show that the last two groups in the second line---the three soft-region remainders and the complementary parent integral---have zero leading clock coefficient, while their full contributions need not vanish. The remaining term, $\mathcal B_{\rm{cut}}[\chi_{\varepsilon_{\rm{sep}}}]$, then reproduces the leading oscillatory coefficient of $\mathcal B$. Appendices\,\ref{sm:outsideproof}--\ref{sm:UVproof} treat the complement, first at finite ultraviolet cutoff and then with that cutoff removed; Appendix\,\ref{sm:remainderproof} treats the three remainders. Appendices\,\ref{sm:matchingproof} and \ref{sm:hardprojectionproof} establish separator independence, fix the soft finite part and justify the projection yielding $\Ch\Cs$.

To show that neither the complementary contribution nor the localized time-ordering remainders contain the leading CC oscillation, we examine their smoothness at $k_3=0$ after stripping the external $k_3^{-3}$ factor. After stripping the external $k_3^{-3}$ factor, the target term is proportional to $k_3^3\rho^{\pm2\ii\tnu}$ and has an unbounded fourth derivative. Four continuous derivatives therefore suffice to exclude this leading oscillation. Equivalently, its coefficient can be selected by a Mellin pole after subtracting the integer-power background:
\begin{equation}
  \mathcal M[\widehat X](s)=\int_0^{\rho_0}\rho^{s-1}\widehat X(\rho)\di\rho,
  \qquad
  \int_0^{\rho_0}\rho^{s-1+\ii\hp 2\tnu}\di\rho
  =\frac{\,\rho_0^{s+\ii\hp 2\tnu}\,}{\,s\!+\!\ii\hp 2\tnu\,}\,.
  \label{sm:Mellinselect}
\end{equation}
The Mellin transform $\mathcal M$ maps the momentum ratio $\rho$ to the complex variable $s$. Here $\widehat X=k_1^6X/H^6$ and $\rho_0>0$ is a fixed upper endpoint within the small-$\rho$ domain. The residue at $s=-2\ii\tnu$ selects the coefficient of $\rho^{2\ii\tnu}$ in $\widehat X$; for the parent correlator this is $\Cc$. Integer powers instead give real-axis poles. This criterion separates the oscillatory coefficient from the background without introducing a further approximation.

We estimate the integrals on the Bunch--Davies rotated contours $x_i=-\tau_i=e^{-\ii a_i\vartheta}Y_i$, with $Y_i>0$, $(a_1,a_2,a_3)=(a,b,c)$ and $0<\vartheta\le\pi/2$. Here $Y_i=|x_i|$ is the radial time on the ray and $\vartheta$ is its rotation angle. Oscillatory time factors then become damping factors. Same-branch ordered times rotate together; conjugate modes, square roots and measures are continued from the original real-time prescription. At $\vartheta=\pi/2$, $\tau_3=\ii cY_3$.

\subsection{Why the Outside Region Has No Leading Clock Signal}\label{sm:outsideproof}
We first consider $\mathcal B[1-\chi_{\varepsilon_{\rm{sep}}}]$ at finite ultraviolet cutoff. On this domain, $q\ge\varepsilon_{\rm{sep}}k_1/2$ and $p=|\vct q-\vct k_3|>\varepsilon_{\rm{sep}}k_1/4$, so both potentially soft momenta stay bounded away from zero as $k_3\to0$. Choosing the unit vector $\hat{\vct z}$ along $\vct k_3$ gives
\begin{equation}
  p=|\vct q\!-\!k_3\hat{\vct z}|=q\!-\!k_3\hat{\vct q}\cdot\hat{\vct z}
  +\frac{k_3^2}{2q}\bigl[1\!-\!(\hat{\vct q}\cdot\hat{\vct z})^2\bigr]
  +O(k_3^3/q^2)\,.
  \label{sm:hardmomentumexpansion}
\end{equation}
The two soft-line modes are therefore smooth in the external momentum at fixed times. To carry this property through the time integrals, we need damping that remains uniform as $k_3\to0$. The Euclidean propagator bounds are written using a matrix norm $\|\cdot\|$ and a positive constant $C$ independent of the integration times:
\begin{equation}
  \|S_{aa}(p;Y_i,Y_j)\|\leqslant C e^{-p|Y_i-Y_j|},\qquad
  \|S_{a,-a}(p;Y_i,Y_j)\|\leqslant C e^{-p(Y_i+Y_j)}\,.
  \label{sm:linebound}
\end{equation}
Keeping polynomial source factors separate, the total damping exponent is at least
\begin{equation}
  k_1Y_1\!+\!(k_1\!+\!k_3)Y_2\!+\!k_3Y_3\!+\!q|Y_3\!-\!Y_1|\!+\!p|Y_3\!-\!Y_2|\,.
  \label{sm:fullexponent}
\end{equation}
Using $Y_i+|Y_i-Y_3|\ge Y_3$ and the lower bounds on $p,q$ gives
\begin{equation}
  \text{Eq.~\eqref{sm:fullexponent}}\geqslant
  \frac{\,k_1\,}{\,4\,}(Y_1\!+\!Y_2)\!+\!\frac{\,\varepsilon_{\rm{sep}} k_1\,}{\,2\,}Y_3\,.
  \label{sm:gap}
\end{equation}
The region $Y_3\sim1/k_3$ is thus suppressed by $e^{-\varepsilon_{\rm{sep}}/(2\rho)}$: the complementary integral has no independent long soft-time region. Equation\,\eqref{sm:gap} holds for all time orderings, including those in which a hard vertex supplies the damping of $Y_3$.

After stripping the external $k_3^{-3}$ factor, the time factors generated by $k_3$ derivatives remain integrable under this uniform damping. The hard-line zero at $\vct q=-k_1\hat{\vct z}$ is fixed as $k_3$ varies and is locally integrable. This controls the finite momentum domain. To conclude that the full complementary contribution has no leading clock coefficient, we must still control the ultraviolet tail, as follows.

\subsection{Removing the Ultraviolet Cutoff without Losing Smoothness}\label{sm:UVproof}
To complete the complementary-region estimate, we seek an integrable large-$q$ bound uniform in $k_3$ through four derivatives. The complement retains the original ordered propagators, so the cancellations below are applied to the parent kernel, rather than to separately cut terms. We first bound the exact modes, then use the Nambu trace and source structure to obtain an integrable tail.

Set $a_\kappa(z)=e^{z/2}z^{-1/2}W_{\kappa,\ii\tnu}(z)$ and $D_z=z\partial_z$. The exact connection formula is
\begin{equation}
  a_\kappa(z)=\sum_{\epsilon=\pm1}\!
  \frac{\,\Gamma(-\ii\hp 2\epsilon\tnu)\,}{\,\Gamma(\frac{\,1\,}{\,2\,}\!-\!\ii\epsilon\tnu\!-\!\kappa)\,}
  z^{\ii\epsilon\tnu}
    {}_1F_1(\tfrac{\,1\,}{\,2\,}\!+\!\ii\epsilon\tnu\!-\!\kappa;1\!+\!\ii\hp 2\epsilon\tnu;z)\,.
  \label{sm:fullconnection}
\end{equation}
For each fixed nonnegative integer $r$, the logarithmic derivative $D_z^r a_\kappa$ is bounded at zero. At large argument,
\begin{equation}
  a_\kappa(z)=z^{\kappa-\frac{1}{2}}\left[1\!-\!
    \frac{\,(\frac{\,1\,}{\,2\,}\!-\!\kappa)^2\!+\!\tnu^2\,}{\,z\,}\!+\!O(z^{-2})\right],
  \qquad |D_z^rO(z^{-2})|\leqslant C_r|z|^{-2}\,.
  \label{sm:Wasym}
\end{equation}
Since $\Re\kappa=\pm1/2$, joining the small- and large-argument bounds gives, for rotation angles $\vartheta_0\le\vartheta\le\pi/2$ with $c\!=\!\sin\vartheta_0>0$,
\begin{equation}
  \|\partial_{\vct p}^{\alpha}S_{ab}(\vct p;Y_i,Y_j)\|
  \leqslant C_\alpha p^{-|\alpha|}(1\!+\!pD)^{|\alpha|}e^{-cpD}\,.
  \label{sm:symbolbound}
\end{equation}
Here $\alpha$ is a multi-index for Cartesian momentum derivatives, $|\alpha|$ their total order, and $D$ is $|Y_i-Y_j|$ on equal branches or $Y_i+Y_j$ on opposite branches; $c$ in the damping bounds is a positive constant, distinct from the SK label used elsewhere. The constants depend on fixed physical parameters and contour angle, not on $k_3$ or the ultraviolet cutoff. Using exact modes at $pY\lesssim1$ avoids extending the large-argument expansion into the future endpoint.

To identify the first ultraviolet cancellation, introduce reference \emph{internal fermion propagators} with the mass set to zero while keeping the chemical potential fixed \footnote{The physical loop still has finite mass; this reference isolates the part that cancels before estimating the mass-dependent corrections. }. With $x=-\tau$, $p=|\vct p|$ and $\vct n=\vct p/p$, define
\begin{equation}
  G(x)=x^{\ii\tlambda\mathsf t_3}\otimes1_2,\qquad
  P_\epsilon(\vct n)=\frac{\,1_4\!+\!\epsilon\mathsf t_3\otimes(\vct\sigma\cdot\vct n)\,}{\,2\,}\,.
  \label{sm:freeprojectors}
\end{equation}
Both matrices act in Nambu $\otimes$ spin space: $\mathsf t_i$ are Pauli matrices acting on the two Nambu components, whereas $\sigma_i$ act on the two spin components. In the ordered Nambu basis $\Psi=(\Nt,\epsilon\Nt^{\dagger T})^T$ of Eq.\,\eqref{sm:nambudef}, $\mathsf t_3=\operatorname{diag}(1,-1)$, so that $\mathsf t_3\otimes1_2=\operatorname{diag}(1_2,-1_2)$ multiplies the field and charge-conjugate components by $+1$ and $-1$, respectively, without mixing them. The matrix $G(x)$ supplies the opposite chemical-potential phases $x^{\pm\ii\tlambda}$ of the Nambu components. The matrices $P_\epsilon$ project onto the eigenvalue-$\epsilon$ subspaces of $\mathsf t_3\otimes(\vct\sigma\cdot\vct n)$; they obey $P_\epsilon^2=P_\epsilon$, $P_+P_-=0$ and $P_++P_-=1_4$. Here $\epsilon=\pm1$ labels the kinetic projector, independently of the SK branch.

Combining these matrices with the free propagation phases gives the two reference Wightman kernels,
\begin{equation}
 \begin{aligned}
 S^{0,>}_{ij}(\vct p)&=G(x_i)e^{\ii p(x_i-x_j)}P_-(\vct n)G(x_j)^{-1},\\
 S^{0,<}_{ij}(\vct p)&=-G(x_i)e^{-\ii p(x_i-x_j)}P_+(\vct n)G(x_j)^{-1}.
 \end{aligned}
  \label{sm:referenceWightman}
\end{equation}
The superscript $0$ denotes the massless reference, and the ordered propagators follow from the same SK combinations as in Eq.\,\eqref{sm:orderedSKpropagators}. External inflaton lines remain in the source matrices $\mathcal J_i$ of Eq.\,\eqref{eq:source}.

Since $G(x_i)$ commutes with $\mathcal J_i$, its endpoint factors cancel around the closed trace. The remaining matrix structure contains only the projectors and vertices. In the collinear configuration, the external momenta and all routed loop momenta lie in a common plane, which we choose as the $xz$ plane. Conjugation by $\mathsf R=\mathsf t_1\otimes\sigma_y$ preserves each projector but reverses each vertex:
\begin{equation}
  \mathsf RP_\epsilon\mathsf R=P_\epsilon,\qquad
  \mathsf R\mathcal J_i\mathsf R=-\mathcal J_i,
  \qquad
  \operatorname{Tr}[\mathcal J_1S^0_{12}\mathcal J_2S^0_{23}\mathcal J_3S^0_{31}]=0\,.
  \label{sm:masslesszero}
\end{equation}
The three vertex sign reversals multiply to $-1$, whereas the trace is invariant under conjugation because $\mathsf R^2=1_4$. The reference trace therefore equals its negative and vanishes for each choice of Wightman kernels, and hence for their ordered combinations. This cancellation uses the complete Nambu trace, not an individual Weyl contraction. It removes the massless reference contribution from the ultraviolet estimate; the surviving finite-mass terms and their momentum suppression are evaluated next.

To resolve the large-$q$ time scales, write each equal-branch ordering in terms of its minimum time $T$ and the two successive gaps $u/q$ and $v/q$. The permutation $\pi$ orders the three $Y_i$ from smallest to largest:
\begin{equation}
  Y_{\pi(1)}=T,\qquad Y_{\pi(2)}=T\!+\!u/q,\qquad
  Y_{\pi(3)}=T\!+\!(u\!+\!v)/q,\qquad T,u,v>0\,.
  \label{sm:gaps}
\end{equation}
In the bulk $qT\ge1$, the first mass correction is off-diagonal at order $\m/(qT)$, while its diagonal phase enters a line as
\begin{equation}
  \frac{\,\ii\m^2\,}{\,2p\,}\left(\frac{\,1\,}{\,x_j\,}\!-\!\frac{\,1\,}{\,x_i\,}\right),\qquad
  \left|\frac{\,1\,}{\,x_j\,}\!-\!\frac{\,1\,}{\,x_i\,}\right|
  \leqslant\frac{\,|h_i\!-\!h_j|\,}{\,qT^2\,},\qquad Y_i=T\!+\!h_i/q\,.
  \label{sm:phasecancel}
\end{equation}
The time difference supplies a further inverse power of $q$, so this phase is second order in $1/q$. A single off-diagonal insertion has zero trace against Nambu-diagonal lines and vertices. The first surviving terms therefore contain two insertions or a second-order remainder, bounded by $O((qT)^{-2})$ times a gap polynomial and $e^{-c(u+v)}$.

The endpoint region $qT\le1$ requires the exact modes. Here a temporal external source supplies the suppressing factor $k_i^2Y_i$. If all three sources are spatial, a second cancellation replaces this suppression: we average the spin trace over a common unit momentum direction $\vct n$. In this trace $s,t,u$ are helicity signs, $i,j,l$ are spatial indices and $\epsilon_{ijl}$ is the three-dimensional Levi-Civita symbol:
\begin{equation}
  \left\langle\operatorname{tr}(\sigma_i\Pi_s\sigma_j\Pi_t\sigma_l\Pi_u)\right\rangle_{\vct n}
  =\frac{\,\ii\epsilon_{ijl}\,}{\,12\,}(3\!-\!st\!-\!su\!-\!tu)\,.
  \label{sm:spatialzero}
\end{equation}
Its contraction with three conserved external momenta vanishes. Comparing the routed momenta with the common-momentum value requires one momentum derivative, which supplies the extra $q^{-1}$ in Eq.\,\eqref{sm:symbolbound}.

Let $F$ include the angular average, sources, time-gap Jacobian $q^{-2}$ and radial measure $q^2$. Suppressing a common polynomial times $e^{-ck_1T-c(u+v)}$, these two cancellations give
\begin{equation}
 \begin{array}{c|cc}
 &qT\le1&qT\ge1\\\hline
 \text{at least one temporal source}&T+q^{-1}&(q^2T)^{-1}\\
 \text{three spatial sources}&q^{-1}&(q^3T^2)^{-1}
 \end{array}
 \label{sm:regionbounds}
\end{equation}
A temporal source retains its factor $Y_i$ under $k_3$ derivatives. Spatial-source routing derivatives retain the improvement above; further derivatives are bounded by Eq.\,\eqref{sm:symbolbound}. Through four derivatives,
\begin{equation}
  |\partial_{k_3}^nF(q,k_3;T,u,v)|
  \leqslant\frac{\,C_n\mathcal P_n(k_1T,u,v)e^{-ck_1T-c(u+v)}\,}{\,q(1\!+\!qT)\,},
  \qquad 0\leqslant n\le4\,.
  \label{sm:majorant}
\end{equation}
The polynomial $\mathcal P_n$ bounds factors of the rescaled times generated by $n$ derivatives. It can be absorbed into slightly weaker exponentials. The remaining time integral obeys
\begin{equation}
  \int_0^\infty\frac{\,e^{-ck_1T}\,}{\,q(1\!+\!qT)\,}\di T
  =\frac{\,e^{ck_1/q}E_1(ck_1/q)\,}{\,q^2\,}
  \leqslant C\frac{\,1\!+\!\log(q/Q_{\rm{UV}})\,}{\,q^2\,},\qquad q>Q_{\rm{UV}}\,,
  \label{sm:UVtail}
\end{equation}
Here $E_1(z)=\int_z^\infty e^{-t}\di t/t$ and $Q_{\rm{UV}}$ is a fixed scale above which the bound applies. The resulting $(1+\log q)/q^2$ tail is integrable. Mixed SK assignments contain two opposite-branch lines; rescaling all times by $q^{-1}$ and using the same source cancellations gives the same bound. The two all-equal branch assignments each have six time orderings, while each of the six mixed assignments has two. The bound thus covers all $2\times6+6\times2=24$ nonempty ordering domains.

To connect these rotated-contour estimates to the original Bunch--Davies prescription, we also bound the contour arcs. A common-time future arc of radius $\varepsilon$ is bounded by
\begin{equation}
  C\varepsilon\int_{Q_{\rm{UV}}}^\infty\frac{\,\di q\,}{\,q(1\!+\!q\varepsilon)\,}
  =C\varepsilon\log\left(1\!+\!\frac{\,1\,}{\,Q_{\rm{UV}}\varepsilon\,}\right)\longrightarrow0\,.
  \label{sm:arc}
\end{equation}
Ordered-gap and mixed-branch arcs are respectively $O(\varepsilon[1+|\log\varepsilon|^2])$ and $O(\varepsilon[1+|\log\varepsilon|])$, also after four derivatives; past arcs vanish by damping. We first work in a closed BD sector, remove the ultraviolet cutoff using these bounds, and then take the contour boundary value.

Together with the finite-domain estimate in Appendix\,\ref{sm:outsideproof}, the integrable tail permits dominated convergence through four derivatives. Writing $\widehat{\mathcal B}_{\rm{outside}}=k_1^6\mathcal B[1-\chi_{\varepsilon_{\rm{sep}}}]/H^6$, we obtain the required $C^4$ regularity of the stripped complement:
\begin{equation}
  k_3^3\widehat{\mathcal B}_{\rm{outside}}(k_3)
  =b_0\!+\!b_1k_3\!+\!b_2k_3^2\!+\!b_3k_3^3\!+\!O(k_3^4)\,.
  \label{sm:hardTaylor}
\end{equation}
The $b_j$ are Taylor coefficients at $k_3\!=\!0$ with the separator held fixed. Dividing by $k_3^3$ leaves integer Laurent terms through order $k_3^0$, plus $O(k_3)$. These are background terms at the order of the leading clock. In contrast,
\begin{equation}
  \frac{\di^4}{\di k_3^4}\bigl[k_3^3\rho^{\ii\hp 2\tnu}\bigr]
  =\frac{\rho^{\ii\hp 2\tnu}}{k_3}
  (\ii\hp 2\tnu)(1+\ii\hp 2\tnu)(2+\ii\hp 2\tnu)(3+\ii\hp 2\tnu)\,.
  \label{sm:clockderivativetest}
\end{equation}
This derivative is unbounded for $\tnu>0$, excluding the leading clock term and its conjugate. Equivalently, the $O(k_3)$ remainder after Laurent-background subtraction has a Mellin transform holomorphic for $\Re s>-1$. Neither target pole is present, so the complementary term $\mathcal B[1-\chi_{\varepsilon_{\rm{sep}}}]$ in Eq.\,\eqref{sm:localizedcutdifference} has no leading clock coefficient.

\subsection{Why the Time-Ordering Remainders Have No Leading Clock Signal}\label{sm:remainderproof}
We now show that the three localized remainders $\mathcal R_j[\chi_{\varepsilon_{\rm{sep}}}]$, $j=1,2,12$, cannot contribute the leading clock coefficient. As in Appendix\,\ref{sm:UVproof}, the required statement is that $k_3^3\mathcal R_j[\chi_{\varepsilon_{\rm{sep}}}]$ has four continuous derivatives at $k_3=0$, with $k_1$, the physical parameters and the separator held fixed. The factor $k_3^3$ removes the external soft-leg normalization; we do not require the unstripped remainders themselves to be regular. Their momentum domains include the soft points, so the proof needs three ingredients: smooth spectral kernels, uniform control of the time integrals, and treatment of the remaining Wightman line.

First, consider the spectral kernel $\varrho=S^>-S^<$ that enters a corrected line. Its momentum dependence follows from the Nambu field equation, written in terms of $x=-\tau>0$:
\begin{equation}
  \ii\partial_x\Psi_{\vct p}=\left[\mathsf t_3\otimes\vct\sigma\cdot\vct p
    \!-\!\frac{\,\tlambda\mathsf t_3\!+\!\m\mathsf t_1\,}{\,x\,}\otimes1_2\right]\Psi_{\vct p}\,.
  \label{sm:canonical}
\end{equation}
Denote the matrix in brackets by $\mathsf H_{\vct p}(x)$. The chemical-potential term is diagonal in Nambu space, whereas the mass term mixes the two components through $\mathsf t_1=\bigl(\begin{smallmatrix}0&1\\1&0\end{smallmatrix}\bigr)$. Here the physical mass is retained. Let $U_{\vct p}(x_i,x_j)$ evolve the field from $x_j$ to $x_i$, so that $\Psi_{\vct p}(x_i)=U_{\vct p}(x_i,x_j)\Psi_{\vct p}(x_j)$. The anticommutator definition in Eq.\,\eqref{sm:spectral} and its equal-time normalization then give, with the common momentum delta function suppressed,
\begin{equation}
 \varrho_{\vct p}(x_i,x_j)
 =U_{\vct p}(x_i,x_j)
 \bigl\langle\{\Psi_{\vct p}(x_j),\Psi_{\vct p}^\dagger(x_j)\}\bigr\rangle
 =U_{\vct p}(x_i,x_j).
  \label{sm:spectralEvolutionIdentity}
\end{equation}
Thus the spectral matrix is fixed by the same evolution equation and a unit-matrix initial condition:
\begin{equation}
 \ii\partial_{x_i}\varrho_{\vct p}(x_i,x_j)
 =\mathsf H_{\vct p}(x_i)\varrho_{\vct p}(x_i,x_j),
 \qquad \varrho_{\vct p}(x_j,x_j)=1_4.
  \label{sm:spectralInitialValue}
\end{equation}
Continue these identities along a common BD ray, $x_i=e^{-\ii a\vartheta}Y_i$, where $a=\pm1$ is the shared SK branch and $0<\vartheta\le\pi/2$ is the contour rotation angle. The $1/Y$ term generates a norm-preserving rotation, while the momentum term can grow at rate $p\sin\vartheta$. Differentiating the evolution equation with respect to momentum therefore gives
\begin{equation}
  \|\partial_{\vct p}^{\alpha}\varrho(Y_i,Y_j)\|
  \leqslant C_\alpha|Y_i\!-\!Y_j|^{|\alpha|}
  e^{p\sin\vartheta|Y_i-Y_j|}\,.
  \label{sm:causalbound}
\end{equation}
Because $\mathsf H_{\vct p}$ is linear in vector momentum and the initial condition is momentum independent, the full spectral matrix, including both helicities, is entire in $\vct p$ at fixed nonzero times. The fractional momentum powers of the individual Wightman functions cancel in their difference. Retaining the complete spectral matrix is essential for this cancellation. Equation\,\eqref{sm:causalbound} controls its derivatives, but their time-dependent bound must still be integrated.

Second, use the step functions to control the time integrations. In $\mathcal R_1$, $\delta_{ac}$ requires $a=c$, and $\theta_{31}$ imposes $Y_3\le Y_1$ on their common rotated ray. At $\vartheta=\pi/2$, combining the spectral growth with the damping from the remaining lines and external legs gives
\begin{equation}
  e^{-(k_1-q)Y_1-(k_1+k_3-p)Y_2-(k_3+q+p)Y_3},\qquad Y_3\leqslant Y_1\,.
  \label{sm:remgap}
\end{equation}
Within $\chi_{\varepsilon_{\rm{sep}}}$, $q<\varepsilon_{\rm{sep}} k_1$ and $p<5\varepsilon_{\rm{sep}} k_1/4$. The hard external damping dominates the spectral growth. Using $Y_1\ge Y_3$ to assign part of that damping to the soft time yields
\begin{equation}
  e^{-(k_1-q)Y_1-(k_1+k_3-p)Y_2-(k_3+q+p)Y_3}
  \le e^{-k_1(Y_1+Y_2+Y_3)/4}\,.
  \label{sm:remgapexplicit}
\end{equation}
For $\mathcal R_2$, interchange vertices-1 and 2; for $\mathcal R_{12}$, both constraints hold, so $Y_3\le\min(Y_1,Y_2)$. At any fixed smaller rotation angle, the exponent acquires the positive factor $\sin\vartheta$. Thus the hard scale controls all three time integrations uniformly as $k_3\to0$. The time polynomials generated by momentum derivatives remain integrable against this damping.

Finally, handle the Wightman line that remains in each single-cut correction. In $\mathcal R_1$, this line has a moving nonsmooth point at $\vct q=\vct k_3$, so derivatives cannot simply be taken through it. Fix that point by the exact change of variable $\vct r=\vct q-\vct k_3$:
\begin{equation}
  \int\di^3q\,A(\vct q,\vct k_3;Y)B(\vct q\!-\!\vct k_3;Y)
  =\int\di^3r\,A(\vct r\!+\!\vct k_3,\vct k_3;Y)B(\vct r;Y)\,.
  \label{sm:shift}
\end{equation}
Here $Y$ stands for $(Y_1,Y_2,Y_3)$. The bounded, locally integrable Wightman factor $B(\vct r;Y)$ is now independent of $\vct k_3$. All external-momentum derivatives act on $A$, which contains the shifted separator, spectral line, nonzero hard line and vertices. Its support stays in a fixed compact momentum domain; Eqs.\,\eqref{sm:causalbound} and \eqref{sm:remgapexplicit} control the derivative factors and their time integrals. In $\mathcal R_2$, the remaining Wightman line's nonsmooth point is already fixed at $\vct q=0$; in $\mathcal R_{12}$, both soft lines are smooth spectral kernels. These bounds permit differentiation under the stripped localized integrals through fourth order and establish the required $C^4$ regularity.

A leading clock term in an unstripped remainder would give $k_3^3\rho^{\pm2\ii\tnu}$ after stripping, whose fourth derivative is unbounded by Eq.\,\eqref{sm:clockderivativetest}. The regularity just established excludes this term from all three localized remainders; it does not make the remainders vanish or exclude every subleading nonanalytic term. Combining this result with the complementary-region estimate yields $\mathbb C_+^{\rm{cut}}(\varepsilon_{\rm{sep}})=\Cc$, where $\mathbb C_+^{\rm{cut}}$ is extracted from $\mathcal B_{\rm{cut}}[\chi_{\varepsilon_{\rm{sep}}}]$. The localized cut therefore preserves the target coefficient. Fixing the soft finite part and justifying the hard-endpoint projection remain separate steps.

\subsection{Separator Independence and the Finite Part of the Soft Integral}\label{sm:matchingproof}
This subsection has two aims: to show that the leading CC coefficient is independent of the auxiliary momentum separator, and to determine the finite part of the soft integral from the original loop boundary. The second step fixes the finite endpoint contributions that must be retained in $\Cs$ to preserve the clock coefficient. We address these aims in turn, keeping each separator fixed before taking $\rho\to0$.

Choose two admissible smooth profiles $\chi_1$ and $\chi_2$, which may have different separator values or transition shapes. By linearity,
\begin{equation}
 \mathcal B_{\rm{cut}}[\chi_1]-\mathcal B_{\rm{cut}}[\chi_2]
 =\mathcal B_{\rm{cut}}[\chi_1-\chi_2].
 \label{sm:separatorDifference}
\end{equation}
For sufficiently small $\rho$, both profiles equal one near $\vct q=0$ and $\vct q=\vct k_3$. Their difference therefore vanishes near both soft points and has bounded support away from them. On this support, the momentum dependence is smooth and $p+q$ stays bounded below independently of $k_3$, giving uniform soft-time damping. After removing the external $k_3^{-3}$ factor, these bounds allow four continuous derivatives through the difference integral. The criterion of Appendix\,\ref{sm:remainderproof} then excludes a leading clock term from Eq.\,\eqref{sm:separatorDifference}. Hence both profiles yield the same complex CC coefficient, although their background contributions and full integral values can differ. Holding the separators fixed is essential to this argument: it keeps the difference region away from the soft points as $k_3\to0$.

We can therefore choose a boundary adapted to the soft integral, $p+q<d$, with $d=\varepsilon_{\rm{sep}}k_1$. Its radial boundary is $q=(d^2-k_3^2)/[2(d-k_3\cos\theta)]$, where $\cos\theta=\widehat{\vct q}\cdot\widehat{\vct k}_3$. For $k_3<d/4$, it lies between $3d/8$ and $5d/8$ and varies smoothly with $k_3$. Smooth approximations to this boundary differ from the original profile only in a bounded region away from both soft points. The preceding estimate therefore preserves the clock coefficient when this domain is used.

To express the inherited boundary in dimensionless variables, set
\begin{equation}
  z=\frac{\,k_3\,}{\,p\!+\!q\,},\qquad w=\frac{\,q\!-\!p\,}{\,k_3\,},\qquad u=(p\!+\!q)Y_3,\qquad k_3Y_3=z_3=zu\,.
  \label{sm:geometry}
\end{equation}
The triangle inequality and azimuthal integration give
\begin{equation}
  0<z\le1,\quad -1\leqslant w\le1,\quad
  \frac{\,p\,}{\,k_3\,}=\frac{\,1\!-\!wz\,}{\,2z\,},\quad\frac{\,q\,}{\,k_3\,}=\frac{\,1\!+\!wz\,}{\,2z\,},\qquad
  \frac{\,\di^3q\,}{\,(2\pi)^3k_3^3\,}=\frac{\,1\!-\!w^2z^2\,}{\,32\pi^2z^4\,}\di z\di w\,.
  \label{sm:Jacobian}
\end{equation}
The chosen domain becomes $\rho/\varepsilon_{\rm{sep}}<z\le1$. Its upper endpoint is $p+q=k_3$, while its moving lower endpoint is $z_{\min}=\rho/\varepsilon_{\rm{sep}}$. As $\rho\to0$, this lower endpoint approaches zero even though $p+q=d$ remains fixed. Thus small $z$ means large momentum relative to $k_3$, not necessarily an ultraviolet momentum of the original loop. The boundary lies in the overlap $k_3\ll d\ll k_1$, where the hard-endpoint expansion in $(p+q)/k_1$ and the fixed-momentum expansion in $k_3/(p+q)$ both apply.

Now consider the projected soft kernel, whose preservation of the target residue will be established in Appendix\,\ref{sm:hardprojectionproof}. For its combined channel bracket $F(z)$ in Eq.\,\eqref{sm:Csoftintegral}, set $f_n=F^{(n)}(0)/n!$. Angular parity gives $f_0=0$, so $F(z)=f_1z+f_2z^2+O(z^3)$. These derivatives follow under the damped $u,w$ integrals of Eq.\,\eqref{sm:fourresponse}, using $\partial_xU(c,d,x)=-cU(c+1,d+1,x)$. Dividing by $z^{3+2\ii\tnu}$ makes the first two terms nonintegrable at zero. We must therefore integrate them with the inherited lower limit before extending the remainder to zero.

Including the overall positive-clock power, the first term gives
\begin{equation}
  \rho^{\ii\hp 2\tnu}f_1\int_{\rho/\varepsilon_{\rm{sep}}}^{1}z^{-2-\ii\hp 2\tnu}\di z
  =\frac{\,f_1\varepsilon_{\rm{sep}}^{1+\ii\hp 2\tnu}\,}{\,1\!+\!\ii\hp 2\tnu\,}\rho^{-1}
  \!-\!\frac{\,f_1\,}{\,1\!+\!\ii\hp 2\tnu\,}\rho^{\ii\hp 2\tnu}\,.
  \label{sm:overlapexample}
\end{equation}
Its separator-dependent contribution is proportional to $\rho^{-1}$, whereas its clock coefficient is $-f_1/(1+2\ii\tnu)$. The second term similarly gives
\begin{equation}
 \rho^{2\ii\tnu}f_2\int_{\rho/\varepsilon_{\rm{sep}}}^{1}z^{-1-2\ii\tnu}\di z
 =\frac{f_2\varepsilon_{\rm{sep}}^{2\ii\tnu}}{2\ii\tnu}
 -\frac{f_2}{2\ii\tnu}\rho^{2\ii\tnu}.
 \label{sm:overlapSecondTerm}
\end{equation}
Here the separator-dependent term is constant in $\rho$, and the clock coefficient is $-f_2/(2\ii\tnu)$. Although $\varepsilon_{\rm{sep}}^{2\ii\tnu}$ is complex, it does not oscillate with $\rho$ at fixed separator. For the remaining $O(z^3)$ term, the integral from zero to $z_{\min}$ is $O(z_{\min})$ in magnitude, so extending its lower limit to zero changes only subleading terms.

Combining the convergent remainder with the two clock coefficients just obtained defines
\begin{equation}
  \FP\int_0^1\frac{\,F(z)\,}{\,z^{3+\ii\hp 2\tnu}\,}\di z
  =\int_0^1\frac{\,F(z)\!-\!f_1z\!-\!f_2z^2\,}{\,z^{3+\ii\hp 2\tnu}\,}\di z
  \!-\!\frac{\,f_1\,}{\,1\!+\!\ii\hp 2\tnu\,}\!-\!\frac{\,f_2\,}{\,\ii\hp 2\tnu\,}\,.
  \label{sm:finitepart}
\end{equation}
The finite constants in this expression are thus fixed by the original moving boundary. Discarding the two subtracted terms altogether would lose these clock contributions and change the desired coefficient. This finite part implements soft-region boundary matching; it is distinct from the ultraviolet convergence of the parent loop in Appendix\,\ref{sm:UVproof}. The remaining step is to justify the hard-endpoint projection used above.

\subsection{From the Retained Cut to the Factorized Coefficient}\label{sm:hardprojectionproof}
In the previous subsection, we showed that the leading clock coefficient is independent of the auxiliary momentum separator. We now show that it is also preserved by the hard-endpoint projection. 

Removing the step functions that link the soft and hard times makes their integration domains independent, but the integrand still couples them through the loop momenta. We therefore need one further operation: replace the two soft propagators' \emph{hard-endpoint} mode functions by their leading small-momentum powers. This subsection shows why that replacement preserves the leading clock coefficient. The soft-vertex modes and the hard propagator remain complete throughout.

\noindent\textit{Step 1: keep the exact dependence that prevents factorization.}
Use the momentum domain of Appendix\,\ref{sm:matchingproof} and introduce
\begin{equation}
 r=\frac{p+q}{k_1}=\frac{\rho}{z},\qquad
 z=\frac{k_3}{p+q},\qquad
 \frac{\rho}{\varepsilon_{\rm{sep}}}<z\le1.
 \label{sm:hardProjectionVariables}
\end{equation}
The variable $r$ measures the loop momenta relative to the hard external scale. At the hard times, $|\tau_{1,2}|\sim k_1^{-1}$, the mode arguments $p\tau_h$ and $q\tau_h$ are of order $r$. In contrast, at the soft time $|\tau_3|\sim k_3^{-1}$ they need not be small. This is why the endpoint expansion is applied only at the hard vertices.

After extracting the powers associated with the positive clock, retain all higher hard-endpoint terms in a kernel $\mathcal Q(r,z)$:
\begin{equation}
 \widehat{\mathcal B}_{\rm{cut},+}(\rho)
 =\rho^{2\ii\tnu}\int_{\rho/\varepsilon_{\rm{sep}}}^1
 z^{-3-2\ii\tnu}\mathcal Q(\rho/z,z)\di z.
 \label{sm:exacthardprofile}
\end{equation}
Here $\mathcal Q$ includes the time and angular integrations. The hat denotes the normalization used in Eq.\,\eqref{sm:Mellinselect}. The subscript $+$ selects the sector whose leading oscillation is $\rho^{2\ii\tnu}$; the conjugate sector is treated in the same way.

\noindent\textit{Step 2: identify the factorized leading kernel.}
Setting $r=0$ in this kernel means retaining the leading hard-endpoint powers, not setting the soft momenta to zero everywhere. The corresponding Nambu spinors have fixed component ratios and form rank-one matrices. Equation\,\eqref{sm:rankone} then separates the trace into a hard overlap and a soft overlap. After the remaining contractions, this gives
\begin{equation}
 \mathcal Q(0,z)=\Ch\,\frac{2^{-1-2\ii\tnu}}{8\pi^2\Lambda}F(z).
 \label{sm:projectedHardProfile}
\end{equation}
The function $F(z)$ is the soft bracket in Eq.\,\eqref{sm:Csoftintegral}, with the full soft-vertex modes retained. Thus the factorized expression is obtained if we may replace $\mathcal Q(\rho/z,z)$ by $\mathcal Q(0,z)$ when extracting the leading clock coefficient.

This replacement needs an integral-level argument. At every fixed $z>0$, $r=\rho/z\to0$, so the omitted hard-endpoint terms are suppressed. However, at the moving lower endpoint,
\begin{equation}
 z_{\min}=\frac{\rho}{\varepsilon_{\rm{sep}}}
 \quad\Longrightarrow\quad
 r\big|_{z=z_{\min}}=\varepsilon_{\rm{sep}},
 \label{sm:hardProjectionBoundary}
\end{equation}
which stays fixed as $\rho\to0$. Pointwise suppression therefore does not justify dropping the correction throughout the integral. We must check explicitly whether the region near this boundary changes the clock coefficient.

\noindent\textit{Step 3: separate the terms that require endpoint matching.}
As in Appendix\,\ref{sm:matchingproof}, isolate the first three powers of $z$ before taking any limit:
\begin{equation}
 \mathcal Q(r,z)=\sum_{n=0}^2g_n(r)z^n+G(r,z),\qquad
 g_n(r)=\frac{\partial_z^n\mathcal Q(r,0)}{n!}.
 \label{sm:hardProfileSubtraction}
\end{equation}
The required bounds are $g_n\in C^1[0,\varepsilon_{\rm{sep}}]$ and $G,\partial_rG=O(z^3)$ uniformly on that interval. Consequently,
\begin{equation}
 g_n(r)-g_n(0)=O(r),\qquad
 G(r,z)-G(0,z)=O(rz^3).
 \label{sm:hardProfileDifferenceBounds}
\end{equation}
We keep the $n=0$ term because the identity $f_0=0$ used in Appendix\,\ref{sm:matchingproof} applies at $r=0$; it does not set $g_0(r)$ to zero for finite $r$.

The bounds above use the exact confluent hypergeometric factors ${}_1F_1$ in the hard-endpoint mode functions. These factors and their first $r$ derivatives are bounded by a time polynomial times $e^{3\varepsilon_{\rm{sep}}k_1(Y_1+Y_2)}$, which is dominated by the hard external damping. The three $z$ derivatives needed for the Taylor remainder have an integrable soft-time bound near $z=0$; the exact modes and geometric measure control the remaining endpoint degeneracies. These bounds control the expansion after time integration, including the region near the moving momentum boundary.

\noindent\textit{Step 4: check which terms can have the clock pole.}
The leading positive-clock coefficient is the residue at $s=-2\ii\tnu$ of the Mellin transform in $\rho$. For this calculation, use the convenient upper endpoint $\rho_0=\varepsilon_{\rm{sep}}$. Changing the original fixed upper endpoint adds only an interval away from zero, whose Mellin transform is entire, so it cannot change this residue. This choice does not extend the squeezed-limit claim beyond Eq.\,\eqref{sm:separator}.

For $\Re s>2$, the integrals converge absolutely. The substitution $\rho=rz$ changes the triangular domain $0<\rho<\varepsilon_{\rm{sep}}$, $\rho/\varepsilon_{\rm{sep}}<z<1$ into the rectangle $0<r<\varepsilon_{\rm{sep}}$, $0<z<1$, giving
\begin{equation}
 \mathcal M[\widehat{\mathcal B}_{\rm{cut},+}](s)
 =\int_0^{\varepsilon_{\rm{sep}}}r^{s+2\ii\tnu-1}\di r
   \int_0^1z^{s-3}\mathcal Q(r,z)\di z.
 \label{sm:hardProfileDoubleMellin}
\end{equation}
This form makes the origin of the clock pole explicit: the $r$-independent part $\mathcal Q(0,z)$ multiplies
\begin{equation}
 \int_0^{\varepsilon_{\rm{sep}}}r^{s+2\ii\tnu-1}\di r
 =\frac{\varepsilon_{\rm{sep}}^{s+2\ii\tnu}}{s+2\ii\tnu}.
 \label{sm:hardProfilePoleOrigin}
\end{equation}
In contrast, each difference in Eq.\,\eqref{sm:hardProfileDifferenceBounds} supplies an extra power of $r$. Its $r$ integral is therefore regular near $s=-2\ii\tnu$. The endpoint subtraction also controls the $z$ integral: the polynomial terms give $1/(s+n-2)$, with poles only at the real positions $s=2,1,0$, while the $O(z^3)$ remainder is integrable for $\Re s>-1$. For fixed $\tnu>0$, none of these produces a pole at the target point.

Combining these observations and analytically continuing from $\Re s>2$ yields
\begin{equation}
 \begin{aligned}
 \mathcal M[\widehat{\mathcal B}_{\rm{cut},+}](s)
 ={}&\frac{\varepsilon_{\rm{sep}}^{s+2\ii\tnu}}{s+2\ii\tnu}
 \left[\int_0^1z^{s-3}G(0,z)\di z
       +\sum_{n=0}^2\frac{g_n(0)}{s+n-2}\right]\\
 &+\text{a function holomorphic near }s=-2\ii\tnu.
 \end{aligned}
 \label{sm:hardprofileMellin}
\end{equation}
Thus only the projected kernel $\mathcal Q(0,z)$ contributes to the leading clock residue. The higher hard-endpoint terms can change other parts of the integral, but cannot change this coefficient.

\noindent\textit{Step 5: read off the hard--soft product.}
At the target pole the residue of the prefactor in Eq.\,\eqref{sm:hardprofileMellin} is one. Substituting Eq.\,\eqref{sm:projectedHardProfile}, the bracket becomes exactly the soft finite part fixed in Appendix\,\ref{sm:matchingproof}, including its two finite endpoint contributions. Hence
\begin{equation}
 \res_{s=-2\ii\tnu}\mathcal M[\widehat{\mathcal B}_{\rm{cut},+}](s)
 =\Ch\,\frac{2^{-1-2\ii\tnu}}{8\pi^2\Lambda}\FP\int_0^1z^{-3-2\ii\tnu}F(z)\di z
 =\Ch\Cs.
 \label{sm:cutMellinResidue}
\end{equation}
Together with the earlier exclusion of the complementary momentum region and the time-ordering remainders, this establishes
\begin{equation}
 \boxed{\mathbb C_+^{\rm{cut}}(\varepsilon_{\rm{sep}})=\Cc=\Ch\Cs\,.}
 \label{sm:factorizedresult}
\end{equation}
The equality concerns the leading oscillatory coefficient, not the full correlator: background and subleading terms need not be preserved by the hard-endpoint projection. No late-time expansion of the soft-vertex modes or exchange of an infinite descendant sum with the integrals is used. The result holds at fixed positive physical parameters and is independent of the admissible separator; the zero-frequency limit requires a separate treatment of logarithmic terms.

\section{Computation of the Hard Part}
\label{sm:hard}
\label{App:C}

\begin{figure}[!ht]
\centering
\includegraphics[width=0.68\linewidth]{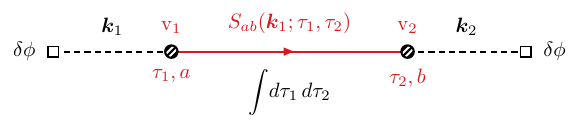}
\caption{The hard subgraph defining $\Ch$, obtained by rotating the red hard block of Fig.\,\ref{fig:time}. Dashed lines are external inflaton legs, hatched circles are interaction vertices, and the red arrowed line is the complete fermion propagator at leading hard momentum $\vct k_1$. The soft lines' hard-endpoint projections are implicit. Both time orderings and the full SK sum over $a,b$ are retained; horizontal placement does not impose a time ordering.}
\label{fig6}
\end{figure}

The factor $\Ch$ corresponds to the red hard block in the middle of Fig.\,\ref{fig:time}, redrawn horizontally in Fig.\,\ref{fig6}. It contains the two external inflaton legs at vertices 1 and 2, the fermion propagator joining them, and the two time integrations over $\tau_1$ and $\tau_2$. The leading hard-endpoint factors of the two soft lines are included through the endpoint projections defined below; their soft-vertex modes and the loop-momentum integral belong to $\Cs$. We retain the complete hard propagator and both hard-time orderings. After fixing the projection and normalization of Eq.\,\eqref{eq:hardIntegral}, we reduce the spinor and SK sums to two basic time integrals, whose combination gives the compact Beta-function result in Eq.\,\eqref{eq:hardfinite}.

\subsection{Endpoint Spinors and Trace Factorization}
We begin with the endpoint states that separate the Nambu trace into hard and soft factors. Only the soft lines' hard endpoints are projected; the propagator joining the hard vertices remains exact. Define the positive gaps
\begin{equation}
  \Delta=\tnu\!-\!\tlambda,\qquad \Sigma=\tnu\!+\!\tlambda,\qquad
  \Delta\Sigma=\m^2,\qquad 2\ii\tnu=A\!-\!2\,.
  \label{sm:gapsdef}
\end{equation}
These quantities are determined by Eq.\,\eqref{eq:physical}. Multiplying Eq.\,\eqref{sm:modeeq} by $\tau$ and keeping the leading hard-endpoint power yields the matrix $\tlambda\mathsf t_3\!+\!\m\mathsf t_1$. Here $\mathsf t_i$ are Pauli matrices acting on the two mode components $(u,v)$, distinct from the spin matrices $\sigma_i$. Its normalized eigenvectors are
\begin{equation}
  e_+=\frac{\,(\m,\Delta)^T\,}{\,\sqrt{2\tnu\Delta}\,},\qquad
  e_-=\frac{\,(-\Delta,\m)^T\,}{\,\sqrt{2\tnu\Delta}\,},\qquad
  (\tlambda\mathsf t_3\!+\!\m\mathsf t_1)e_\epsilon=\epsilon\tnu e_\epsilon\,.
  \label{sm:eigenvectors}
\end{equation}
The power $(-p\tau)^{\ii\epsilon\tnu}$ obeys $\ii\tau\partial_\tau=-\epsilon\tnu$, fixing the exact endpoint ratio $v/u=(\epsilon\tnu-\tlambda)/\m$.\

For the positive clock, line 23 supplies $e_+$ at its hard end, while line 31 supplies the row $e_-^T$. For each fixed helicity and source component, their two-dimensional Nambu factors are $e_+r_s$ and $c_s e_-^T$, where the row $r_s$ and column $c_s$ retain the complete soft-time dependence. In the following identity the source and propagator symbols denote these Nambu factors; the helicity projectors and their spin trace are still to be contracted. The Nambu trace gives
\begin{equation}
  \operatorname{tr}_N[\mathcal J_1S_{12}\mathcal J_2(e_+r_s)\mathcal J_3(c_se_-^T)]
  =(e_-^T\mathcal J_1S_{12}\mathcal J_2e_+)(r_s\mathcal J_3c_s)\,.
  \label{sm:rankone}
\end{equation}
The first bracket in Eq.\,\eqref{sm:rankone} contains the hard propagator and vertices; the second contains the complete soft-vertex modes. The remaining helicity contraction determines whether a common scalar hard response can be extracted from all soft channels.

After taking the helicity projector trace, exchange the two soft endpoints in the momentum integral. The temporal bilinear is symmetric and the spatial bilinear antisymmetric under this exchange, while their accompanying angular factors have the opposite parity. The soft numerator multiplying the part odd in the hard helicity therefore integrates to zero. The surviving terms allow the two hard-helicity responses to be combined into a common scalar factor.

To write this factor explicitly, use $x_i\!=\!-k_1\tau_i$, strip the external factors $H^2/(2k_1^2)$ from the two hard vertices and define
\begin{subequations}\label{sm:Hproject}
  \begin{align}
    V_1^a(x_1)   & =-x_1\mathsf t_3\!+\!\ii s_{12}(1\!+\!\ii ax_1)1_N\,,\label{sm:hardVertexOne} \\
    V_2^b(x_2)   & =-x_2\mathsf t_3\!-\!\ii s_{12}(1\!+\!\ii bx_2)1_N\,,\label{sm:hardVertexTwo} \\
    H_{\rm{proj}} & =\sum_{s_{12},a,b=\pm1}\!ab\int_0^\infty\di x_1\di x_2
    (x_1x_2)^{\ii\tnu}e^{-\ii(ax_1+bx_2)}
    e_-^TV_1^aS_{ab,s_{12}}(1;x_1,x_2)V_2^be_+\,.
    \label{sm:Hproject:last}
  \end{align}
\end{subequations}
In Eqs.\,\eqref{sm:Hproject} and \eqref{sm:hardsksum}, $a,b\!=\!\pm1$ label the SK branches; the special-function parameters are defined separately in Eq.\,\eqref{eq:finitepars}; $s_{12}\!=\!\pm1$ is the hard helicity and $1_N$ the Nambu identity. In $S_{ab,s_{12}}(1;x_1,x_2)$ the first argument is the hard momentum in units of $k_1$, and each term follows its own BD contour.\  Each frequency sign fixes its damped rotation ray.\

To identify the schematic quantities in the main text, restore the two external sources and the dimensionful times:
\begin{equation}
 \Ch=\mathcal N_h\sum_{a,b=\pm1}ab\int_{-\infty}^{0}\di\tau_1\di\tau_2\,
 \mathcal F_{\mu a}(\vct k_1,\tau_1)\mathcal F_{\nu b}(-\vct k_1,\tau_2)
 \mathcal P_h(\tau_1,\tau_2)
 [\Gamma^\mu S_{ab}(\vct k_1;\tau_1,\tau_2)\Gamma^\nu].
 \label{sm:hardIntegralExplicit}
\end{equation}
The symbol $\mathcal P_h$ denotes a linear contraction with the selected endpoint states, rather than another propagator. For a matrix $M$ acting on Nambu and spinor indices, it is
\begin{equation}
 \mathcal P_h(\tau_1,\tau_2)[M]
 =(x_1x_2)^{\ii\tnu}\sum_{s=\pm1}
 (e_-^T\otimes h_s^\dagger)M(e_+\otimes h_s),
 \qquad h_s=h_s(\vct k_1),\quad x_i=-k_1\tau_i.
 \label{sm:hardProjectionExplicit}
\end{equation}
The helicity spinors are defined in Eq.\,\eqref{sm:helicitybasis}, and $(x_1x_2)^{\ii\tnu}$ supplies the selected hard-endpoint powers. To fix the split between $\Ch$ and $\Cs$, we use channel $C$ as the reference: its endpoint coefficient is included in $\Ch$, while the relative channel weights in Eq.\,\eqref{sm:hardratios} are included in $\Cs$. The normalization factors $\mathcal N_x$ and $\mathcal N_h$ refer to dimensionless and dimensionful time variables, respectively:
\begin{equation}
 \Ch=\mathcal N_x H_{\rm{proj}},\qquad
 \mathcal N_h=\frac{4k_1^6}{H^4}\mathcal N_x,\qquad
 \mathcal N_x=-\frac{\ii\,2^{1+2\ii\tnu}\tnu\m^3\sinh(\pi\tnu)}{16\Lambda^2}
 R_\kappa^+R_{\bar\kappa}^+,\quad \kappa=-\tfrac12-\ii\tlambda.
 \label{sm:hardIntegralNormalization}
\end{equation}
Here $R_\kappa^+=R_{\kappa,+1}$ is the connection coefficient in Eq.\,\eqref{eq:Wdefs}. The two external normalization factors and the change of time measures together supply $H^4/(4k_1^6)$, explaining the relation between $\mathcal N_h$ and $\mathcal N_x$. The external phases are already present in $H_{\rm{proj}}$. Equations\,\eqref{sm:hardIntegralExplicit}--\eqref{sm:hardIntegralNormalization} thus specify the schematic hard integral in the main text, including its coupling factors and phase convention.

\subsection{Both Orderings and the Exact Time Seeds}
The next task is to evaluate $H_{\rm{proj}}$. Both hard-time orderings must be retained. Before contour rotation, summing the four SK assignments at fixed ordering gives
\begin{equation}
  \sum_{a,b=\pm1}\!ab\,\mathcal J_1^aS_{ab}\llbracket 12\rrbracket\mathcal J_2^b
  =\begin{cases}
    (\mathcal J_1^+\!-\!\mathcal J_1^-)(S^>\mathcal J_2^+\!-\!S^<\mathcal J_2^-), & \tau_1>\tau_2\,, \\
    (\mathcal J_1^+S^<\!-\!\mathcal J_1^-S^>)(\mathcal J_2^+\!-\!\mathcal J_2^-), & \tau_2>\tau_1\,.
  \end{cases}
  \label{sm:hardsksum}
\end{equation}
The matrix order in Eq.\,\eqref{sm:hardsksum} is essential. Equivalently, one may organize the four SK assignments into opposite-branch terms, whose time integrals separate, and equal-branch terms, whose propagators impose a time ordering. This organization reduces the calculation to two elementary integrals, denoted $L$ and $J$ below.

On opposite hard branches each time integral has the form $L_\kappa(\ell)$ below. Here $y\!=\!k_1|\tau|$ is the rotated hard time and $\ell$ is a complex moment index; the soft moments in Eq.\,\eqref{eq:Wseed} have their own index $\ell\!=\!0,1$:
\begin{equation}
  L_\kappa(\ell)=\int_0^\infty y^{\ell-\frac{1}{2}}e^{-y}W_{\kappa,\ii\tnu}(2y)\di y
  =\frac{\,2^{-\ell-\frac{1}{2}}\Gamma(\ell\!+\!1\!+\!\ii\tnu)\Gamma(\ell\!+\!1\!-\!\ii\tnu)\,}{\,\Gamma(\ell\!+\!3/2\!-\!\kappa)\,}\,.
  \label{sm:Lseed}
\end{equation}
To obtain Eq.\,\eqref{sm:Lseed}, write $W=e^{-y}(2y)^{1/2+\ii\tnu}U(1/2+\ii\tnu-\kappa,1+2\ii\tnu,2y)$ and insert the Euler representation of Tricomi's confluent hypergeometric function $U$. The $y$ integral becomes a Gamma integral, and the auxiliary Euler integral becomes a Beta integral. The derivation starts at $\Re\ell>-1$, $\Re\kappa<1/2$ and is continued to the required indices $\ell=n+\ii\epsilon\tnu$, $n=0,1$, $\epsilon=\pm1$. The original Laplace integral remains convergent there.

For equal branches, the constraint between the two times requires the ordered integral
\begin{equation}
  J(\ell_1,\ell_2;\kappa_1,\kappa_2)=
  \int_{0<y_1<y_2}\di y_1\di y_2\,
  y_1^{\ell_1-\frac{1}{2}}y_2^{\ell_2-\frac{1}{2}}e^{-y_1-y_2}
  W_{\kappa_1,\ii\tnu}(-2y_1\!-\!\ii0)W_{\kappa_2,\ii\tnu}(2y_2)\,.
  \label{sm:Jseed}
\end{equation}
The negative argument is approached from below.\  Let $\sigma\!=\!\pm1$ label the two branches of the first Whittaker function. For this seed define
\begin{equation}
  u_\sigma=\tfrac{\,1\,}{\,2\,}\!-\!\kappa_1\!-\!\ii\sigma\tnu,\quad
  v_\sigma=1\!-\!\ii\hp 2\sigma\tnu,\quad
  d_\sigma=1\!+\!\ell_1\!-\!\ii\sigma\tnu,\quad
  L_\sigma=\ell_2\!+\!d_\sigma,\quad
  q_\sigma=-\ii e^{-\sigma\pi\tnu}2^{\frac{1}{2}-\ii\sigma\tnu}
  \frac{\,\Gamma(\ii\hp 2\sigma\tnu)\,}{\,\Gamma(\frac{\,1\,}{\,2\,}\!-\!\kappa_1\!+\!\ii\sigma\tnu)\,}\,.
  \label{sm:Jparameters}
\end{equation}
The symbols $u_\sigma,v_\sigma,q_\sigma$ are parameters of this seed formula, distinct from the mode functions and loop momentum. Expanding the first Whittaker function by the exact connection formula and integrating over the first time variable gives
\begin{equation}
  \int_0^{y_2}y_1^{d_\sigma-1}{}_1F_1(u_\sigma;v_\sigma;-2y_1)\di y_1
  =\frac{\,y_2^{d_\sigma}\,}{\,d_\sigma\,}
  {}_2F_2(u_\sigma,d_\sigma;v_\sigma,d_\sigma\!+\!1;-2y_2)\,.
  \label{sm:innerseed}
\end{equation}
Applying Eq.\,\eqref{sm:Lseed} to each remaining power gives
\begin{equation}
  J=\sum_{\sigma=\pm1}\!\frac{\,q_\sigma L_{\kappa_2}(L_\sigma)\,}{\,d_\sigma\,}
  {}_4F_3\!\left[\begin{matrix}u_\sigma,d_\sigma,L_\sigma\!+\!1\!+\!\ii\tnu,L_\sigma\!+\!1\!-\!\ii\tnu \\
      v_\sigma,d_\sigma\!+\!1,L_\sigma\!+\!3/2\!-\!\kappa_2\end{matrix}\middle|-1\right]\,.
  \label{sm:Jclosed}
\end{equation}
To justify integrating the series term by term, first multiply the argument $-2y_1$ of ${}_1F_1$ by an Abel parameter of magnitude less than one. The resulting hypergeometric series is continued along the negative real axis to the value at $-1$ in Eq.\,\eqref{sm:Jclosed}. Reversing the time ordering exchanges all endpoint data, including powers and Whittaker indices. The $--$ contribution uses the conjugate upper-lip value, with the corresponding conjugated endpoint data and reversed frequency.

\subsection{Reduction of 45 Component Terms to Two Functions}
We now assemble the elementary integrals into the projected hard response. Expanding Eq.\,\eqref{sm:Hproject} produces 18 ordered $J$ terms for the positive endpoint sign, 18 conjugated ordered terms for the negative sign, and nine products of $L$ integrals after conjugate products are identified. These $36+9$ terms are components of the same Nambu expression, rather than additional diagrams. We first factor out their common channel dependence, then give a bilinear rule that generates their coefficients, and finally reduce the sum to two hypergeometric values.

Gamma recurrence for adjacent mode indices reduces all hard endpoint coefficients to two:
\begin{equation}
  \kappa=-\tfrac{\,1\,}{\,2\,}\!-\!\ii\tlambda,\quad\eta=\kappa\!+\!1,\qquad
  R_\eta^m=\ii(\tlambda\!-\!m\tnu)R_\kappa^m,\qquad
  R_{\overline\eta}^m=-\ii(\tlambda\!+\!m\tnu)R_{\overline\kappa}^m\,,
  \label{sm:adjacent}
\end{equation}
where $R_\kappa^m\equiv R_{\kappa,m}$ was defined in Eq.\,\eqref{eq:Wdefs}. Terms with the same seed are collected before any special function is evaluated.\  Let $a_j$ denote the hard coefficient multiplying soft channel $j\!=\!A,B,C,D$, whose mode content is given in Eq.\,\eqref{sm:channelTable}. These coefficients have the ratios
\begin{equation}
  a_A=\mathcal R_Aa_C,\quad a_B=\mathcal R_Ba_C,\quad \m^2a_D=a_C,
  \quad
  \mathcal R_A=-\ii\frac{\,R_\kappa^+\,}{\,R_{\bar\kappa}^+\,}
  \frac{\,\sinh[\pi(\tnu\!+\!2\tlambda)]\,}{\,\Sigma\sinh(\pi\tnu)\,},\quad
  \mathcal R_B=-\ii\frac{\,R_{\bar\kappa}^+\,}{\,R_\kappa^+\,}
  \frac{\,\sinh[\pi(\tnu\!-\!2\tlambda)]\,}{\,\Delta\sinh(\pi\tnu)\,}\,.
  \label{sm:hardratios}
\end{equation}
The identities hold coefficient by coefficient, including at zeros of the hard response, and $\Ch=a_C/\Lambda^2$. All 45 coefficients are generated by the formal bilinear rule

\begin{equation}
 \mathfrak J_\epsilon[y^r w_{\kappa_1},y^n w_{\kappa_2}]
 :=J(r+\ii\epsilon\tnu,n+\ii\epsilon\tnu;\kappa_1,\kappa_2),\qquad
 \alpha_\epsilon=\ii(\epsilon\tnu-\tlambda),\quad\beta_\epsilon=\ii(\epsilon\tnu+\tlambda),
 \label{sm:hardassemblyrule}
\end{equation}
Here $\epsilon=\pm1$ is the endpoint-power sign in the seed, and $r,n=0,1$ are integer powers, not loop variables. Extend $\mathfrak J_\epsilon$ linearly in both arguments. The symbols $y,w_\kappa$ are bookkeeping markers for a power and a Whittaker index; the two arguments specify the earlier and later endpoints of $J$. With $X=e^{\pi\tlambda}$ and $Q=e^{\pi\tnu}$, define an ordered block $B_\epsilon^{\rm{hard}}$ and the complete seed sum $H_{\rm{hard}}$ by

\begin{align}
 B_\epsilon^{\rm{hard}}(\tlambda)
 &=\mathfrak J_\epsilon[w_\eta+\alpha_\epsilon(1+2y)w_\kappa,
                    (1+2y)w_{\bar\eta}+\beta_\epsilon w_{\bar\kappa}],\label{sm:hardassemblyblock}\\
 H_{\rm{hard}}
 &=X\overline{B_-^{\rm{hard}}(\tlambda)}-X^{-1}\overline{B_-^{\rm{hard}}(-\tlambda)}
   -Q^2[X B_+^{\rm{hard}}(\tlambda)-X^{-1}B_+^{\rm{hard}}(-\tlambda)]\nonumber\\
 &\quad+4\ii Q(X-X^{-1})(1+2\ii\tnu)^2L_\kappa(\ii\tnu)L_{\bar\kappa}(\ii\tnu).
 \label{sm:hardassembly}
\end{align}
In $B_\epsilon^{\rm{hard}}(-\tlambda)$ interchange barred and unbarred indices and reverse $\tlambda$ in $\alpha_\epsilon,\beta_\epsilon$, at fixed $\tnu$. Bars conjugate coefficients and seeds. Each block generates nine ordered terms; the $\kappa,\bar\kappa$ term with $r=n=0$ has coefficient $\alpha_\epsilon\beta_\epsilon=-\m^2$. The nine Laplace products are $\ii Q(X-X^{-1})T_{\tlambda}T_{-\tlambda}$, where

\begin{equation}
 T_{\tlambda}=L_\eta(\ii\tnu)+2L_\eta(1+\ii\tnu)+\ii\Delta L_\kappa(\ii\tnu)
        =2(1+2\ii\tnu)L_\kappa(\ii\tnu).
 \label{sm:laplaceassembly}
\end{equation}
Equation\,\eqref{sm:Lseed} gives the last equality and hence the last line of Eq.\,\eqref{sm:hardassembly}. This accounts for all $36+9$ terms. The phase and normalization relating this Euclidean seed sum to the projected integral are
\begin{equation}
 H_{\rm{proj}}=\frac{\ii\m}{2\tnu e^{\pi\tnu}}H_{\rm{hard}}.
 \label{sm:hardProjectionBridge}
\end{equation}
Thus $H_{\rm{hard}}$ is a rescaled intermediate form of $H_{\rm{proj}}$; the physical factor is still $\Ch=\mathcal N_xH_{\rm{proj}}$.

To simplify the special functions in this sum, set $\alpha=\ii\Delta=-a$ and $\beta=\ii\Sigma=-b$, so that $\alpha+\beta=2\ii\tnu$ and $\alpha\beta=-\m^2$. Substituting Eq.\,\eqref{sm:Jclosed} into Eq.\,\eqref{sm:hardassembly} and collecting terms leaves only two regularized hypergeometric values, $F_h(\alpha)$ and $F_h(\beta)$. It is convenient to remove a common prefactor by defining
\begin{equation}
 H_{\rm{hard}}=\frac{e^{\pi\tnu}\Gamma(2\ii\tnu+2)}{2^{1+2\ii\tnu}\sinh(2\pi\tnu)}\mathcal H_h.
 \label{sm:hardReducedNormalization}
\end{equation}
The reduced sum and its auxiliary functions are
\begin{subequations}
  \begin{align}
    \mathcal H_h={} & -2\pi(2\ii\tnu\!+\!1)\sinh(\pi\tlambda)\left[
      \frac{\,T(\alpha)F_h(\alpha)\,}{\,\Gamma(\beta)\,}\!+\!
      \frac{\,T(\beta)F_h(\beta)\,}{\,\Gamma(\alpha)\,}\right]
    \!+\!\frac{\,\ii\Gamma(2\ii\tnu\!+\!2)\sinh(\pi\tlambda)\sinh(2\pi\tnu)\,}{\,(2\ii\tnu\!+\!1)\Gamma(2\!+\!\alpha)\Gamma(2\!+\!\beta)\,}P_3\nonumber \\
                    & +\frac{\,\ii(2\ii\tnu\!+\!2)(2\ii\tnu\!-\!1\!-\!\m^2)\sinh(\pi\tnu)\sinh(2\pi\tlambda)
    \!+\!4\tlambda[\cosh(\pi\tnu)\cosh(2\pi\tlambda)\!-\!\cosh(3\pi\tnu)]\,}{\,2\ii\tnu\!+\!1\!-\!\m^2\,}\,,\label{sm:Htwo}                          \\
    T(c)            & =(2\ii\tnu\!+\!2)c\!+\!2\ii\tnu\!-\!2,\qquad
    P_3=(2\ii\tnu\!+\!2)((2\ii\tnu)^2\!+\!2\ii\tnu\!+\!4\!+\!2\m^2)\!+\!8(2\ii\tnu\!+\!1)\,,\label{sm:hardPolynomialDefs}                                           \\
    F_h(c)            & ={}_3\widetilde F_2\!\left[\begin{matrix}2\!+\!2\ii\tnu,-c,1\\3\!+\!c,1\!-\!2\ii\tnu\end{matrix}\middle|-1\right]\,.
    \label{sm:twoF}
  \end{align}
\end{subequations}
The tilde in ${}_3\widetilde F_2$ denotes division by the Gamma functions of both lower parameters. The next subsection evaluates $F_h(\alpha)$ and $F_h(\beta)$ exactly; the rest of $\mathcal H_h$ is already elementary or expressed in Gamma functions.

\subsection{Exact Gamma Summation and the Independent Beta Value}
The two remaining functions are special complementary-parameter sums.\  The identity
\begin{equation}
  {}_3F_2(1,u,v;1\!-\!u,1\!-\!v;-1)
  =\frac{\,1\,}{\,2\,}\left[1\!+\!\frac{\,\Gamma(1\!-\!u)\Gamma(1\!-\!v)\,}{\,\Gamma(1\!-\!u\!-\!v)\,}\right]
  \label{sm:Whipple}
\end{equation}
follows initially for $0<\Re u,\Re v<1/4$ by applying Parseval's identity to $[2\cos(\theta/2)]^{-2u}$ and its shifted $v$ counterpart on $(-\pi,\pi)$. Their overlap evaluates to a trigonometric Beta integral. Meromorphic continuation then gives the complex-parameter identity used here.

Separate the first term of Eq.\,\eqref{sm:Whipple} at $u\!=\!1\!+\!2\ii\tnu$, $v\!=\!-1\!-\!\alpha$, shift the summation index once, and divide by the lower-parameter Gamma factors.\  The result is
\begin{equation}
  F_h(\alpha)=\frac{\,1\,}{\,2(2\ii\tnu\!+\!1)(1\!+\!\alpha)\,}
  \left[\frac{\,1\,}{\,\Gamma(1\!-\!\beta)\,}\!-\!
    \frac{\,1\,}{\,\Gamma(-2\ii\tnu)\Gamma(2\!+\!\alpha)\,}\right],\qquad\beta=2\ii\tnu\!-\!\alpha\,.
  \label{sm:Fgamma}
\end{equation}
Exchange $\alpha$ and $\beta$ for the second function.\  Substitution into Eq.\,\eqref{sm:Htwo} uses the polynomial identity
\begin{equation}
  P_3\!-\!\frac{\,\beta(1\!+\!\beta)T(\alpha)\,}{\,1\!+\!\alpha\,}
  \!-\!\frac{\,\alpha(1\!+\!\alpha)T(\beta)\,}{\,1\!+\!\beta\,}
  =\frac{\,4(2\ii\tnu\!+\!1)(2\ii\tnu\!+\!2)^2\,}{\,(1\!+\!\alpha)(1\!+\!\beta)\,}\,.
  \label{sm:polycancel}
\end{equation}
Reflection, $\Gamma(z)\Gamma(1\!-\!z)\!=\!\pi/\sin\pi z$, cancels the remaining imaginary elementary term.\  Writing $\mathcal R$ for the remaining Gamma-function combination gives the final hard seed sum
\begin{subequations}
  \begin{align}
    H_{\rm{hard}} & =\frac{\,(e^{2\pi\tnu}\!-\!1)\Gamma(2\ii\tnu\!+\!2)\,}{\,2^{2\ii\tnu}(2\ii\tnu\!+\!1\!-\!\m^2)\,}\mathcal R,
    \qquad
    \mathcal R=\frac{\,\ii(2\ii\tnu\!+\!2)^2\Gamma(2\ii\tnu\!+\!2)\sinh(\pi\tlambda)\,}{\,\Gamma(2\!+\!\alpha)\Gamma(2\!+\!\beta)\sinh(\pi\tnu)\,}\!-\!2\tlambda\,,\label{sm:Hgamma} \\
    \Ch          & =\frac{\,2^{1+\ii\hp 2\tnu}\,}{\,64\Lambda^2\,}R_\kappa^+R_{\bar\kappa}^+\m^4
    (1\!-\!e^{-2\pi\tnu})H_{\rm{hard}}\,.
    \label{sm:Hnormal}
  \end{align}
\end{subequations}
Combining Eq.\,\eqref{sm:hardProjectionBridge} with Eq.\,\eqref{sm:Hnormal} gives exactly $\mathcal N_x$ in Eq.\,\eqref{sm:hardIntegralNormalization}, since $e^{\pi\tnu}(1-e^{-2\pi\tnu})=2\sinh(\pi\tnu)$. This fixes the phase and normalization connecting the original projected integral to the seed sum.

Finally, $a\!+\!b\!=\!2\!-\!A$ and $ab\!=\!-\m^2$ imply
\begin{equation}
  (1\!-\!a)(1\!-\!b)=A\!-\!1\!-\!\m^2,\qquad
  \mathrm B(1\!-\!a,1\!-\!b)=-\frac{\,\m^2\,}{\,(A\!-\!2)(A\!-\!1)\,}\overline{\mathrm B(a,b)}\,.
  \label{sm:Betarecur}
\end{equation}
To make the normalization in Eq.\,\eqref{eq:hardfinite} explicit, first combine the endpoint coefficients and reflect the remaining Gamma pair:
\begin{subequations}\label{sm:hardBetaBridge}
  \begin{align}
    R_\kappa^+R_{\bar\kappa}^+ & =-\frac{\Gamma(-2\ii\tnu)}{\m^2\mathrm B(a,b)},\qquad
    \Gamma(-2\ii\tnu)\Gamma(A)=\frac{\ii\pi(A\!-\!1)}{\sinh(2\pi\tnu)},\label{sm:hardBetaEndpoint}                                                            \\
    \Ch                        & =-\frac{\ii\pi\m^2(A\!-\!1)\tanh(\pi\tnu)}{16\Lambda^2(A\!-\!1\!-\!\m^2)\mathrm B(a,b)}\mathcal R,\label{sm:hardBetaPrefactor} \\
    \mathcal R                 & =\frac{2\tnu}{\m^2\overline{\mathrm B(a,b)}}
    \left[\frac{A^2(A\!-\!1)\sinh(\pi\tlambda)}{(A\!-\!1\!-\!\m^2)\sinh(\pi\tnu)}
      \!-\!\frac{\tlambda\m^2}{\tnu}\overline{\mathrm B(a,b)}\right].\label{sm:hardBetaBracket}
  \end{align}
\end{subequations}
The first two lines follow from Eq.\,\eqref{sm:Hnormal}; the last uses Eq.\,\eqref{sm:Betarecur} in $\mathcal R$ from Eq.\,\eqref{sm:Hgamma}. Multiplying the last two lines and using $\mathrm B(a,b)\overline{\mathrm B(a,b)}=|\mathrm B(a,b)|^2$ gives
\begin{equation}
 \Ch=-\frac{\ii\pi\tnu(A-1)\tanh(\pi\tnu)}{8\Lambda^2(A-1-\m^2)|\mathrm B(a,b)|^2}
 \left[\frac{A^2(A-1)\sinh(\pi\tlambda)}{(A-1-\m^2)\sinh(\pi\tnu)}
 -\frac{\tlambda\m^2}{\tnu}\overline{\mathrm B(a,b)}\right].
 \label{sm:hardFinal}
\end{equation}
This is exactly the hard factor used in Eq.\,\eqref{eq:hardfinite}, with $a,b,A$ defined in Eq.\,\eqref{eq:finitepars}. The only independent complex special-function value is $\mathrm B(a,b)$; its conjugate follows directly for real mass and chemical potential. Both hard time orderings and the normalization of Eq.\,\eqref{sm:hardIntegralExplicit} have been retained.

\section{Computation of the Soft Part}
\label{sm:soft}
\label{App:D}
\begin{figure}[!ht]
\centering
\includegraphics[width=0.55\linewidth]{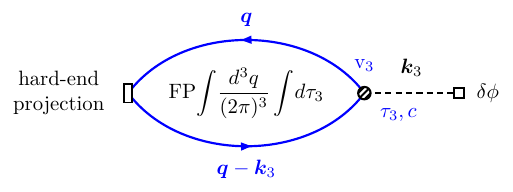}
\caption{The soft subgraph defining $\Cs$, extracted from the blue block of Fig.\,\ref{fig:time}. The two blue arrowed fermion lines carry $\vct q$ and $\vct q-\vct k_3$ and retain their complete mode functions at the hatched vertex $\mathrm v_3$. The dashed line is the external inflaton leg. The open rectangle denotes the projection of the two hard endpoints. The soft factor includes the soft-time and loop-momentum integrations, with $\FP$ fixed by Appendix\,\ref{sm:matchingproof}, and the full spinor, helicity and SK sums.}
\label{fig7}
\end{figure}

The factor $\Cs$ corresponds to the blue soft block on the right of Fig.\,\ref{fig:2}, shown separately in Fig.\,\ref{fig7}. It contains the two soft fermion lines, the external inflaton leg at vertex 3, and the soft-time and loop-momentum integrations. The fermion modes at $\tau_3$ are kept complete. The open rectangle represents the hard-endpoint projection established in Appendix\,\ref{sm:hardprojectionproof}; it introduces no additional interaction vertex or time integral.

We first evaluate the time integral at fixed loop momenta and specify its continuation to the physical triangle. We then assemble the four spinor channels with the relative weights obtained in Appendix\,\ref{sm:hard}. Performing the time, angular and radial integrations reduces their sum to three moments built from two Euler kernels. The radial finite part is fixed by Appendix\,\ref{sm:matchingproof}; the auxiliary Euler moments require a separate continuation described below. Appendix\,\ref{sm:native} evaluates those moments in terms of Meijer $G$ functions, and we collect the normalized result at the end of this appendix.

\subsection{The Product of Two Whittaker Functions}\label{sm:Wseed}
At fixed comoving magnitudes $p=|\vct q-\vct k_3|$ and $q=|\vct q|$, the soft-time integral contains one complete Whittaker function from each soft line. The physical momenta satisfy $|p-q|\le k_3\le p+q$. The external inflaton source in Eq.\,\eqref{eq:externalSource} supplies a constant and a term linear in time, so only the moments $\mathcal I_0$ and $\mathcal I_1$ of Eq.\,\eqref{eq:Wseed} are needed. Using $Y_3=z_3/k_3$ gives $\di Y_3\,Y_3^{\ell-1}=k_3^{-\ell}\di z_3\,z_3^{\ell-1}$. Canonical square roots and spinor weights are restored when the channels are assembled in Appendix\,\ref{app:softchannels}.

The integrals themselves are convergent: near $Y_3=0$, the product of the two Whittaker functions has real power one, giving a time-integrand envelope proportional to $Y_3^{\Re\ell}$; at large $Y_3$, the damping is $e^{-(k_3+p+q)Y_3}$. For fixed positive $p,q$, this defines a holomorphic function for $\Re\ell>-1$ and $\Re k_3>-p-q$, locally away from endpoint degeneracies. The auxiliary domain used next is needed to integrate the hypergeometric series term by term, rather than to make the original time integral converge.

For the analytic evaluation, we use the exact two-branch identity\,\cite{DLMF}
\begin{equation}
  W_{\kappa,\ii\tnu}(z)=e^{-z/2}\sum_{m=\pm1}\!
  R_{\kappa,m}z^{\frac{1}{2}+\ii m\tnu}
  {}_1F_1(\tfrac{\,1\,}{\,2\,}\!+\!\ii m\tnu\!-\!\kappa;1\!+\!\ii\hp 2m\tnu;z)\,,
  \label{sm:Wconnection}
\end{equation}
where $R_{\kappa,m}=\Gamma(-2\ii m\tnu)/\Gamma(1/2-\ii m\tnu-\kappa)$. Each branch retains the full confluent hypergeometric function ${}_1F_1$. In the auxiliary domain $k_3>p+q$, insert both branches into the time integral and expand the two confluent hypergeometric functions ${}_1F_1$.\  Each pair of series orders $n_p,n_q\ge0$ is integrated using
\begin{equation}
 \int_0^\infty Y_3^{d_{mn}+n_p+n_q-1}e^{-(k_3+p+q)Y_3}\di Y_3
 =\frac{\Gamma(d_{mn})(d_{mn})_{n_p+n_q}}{(k_3+p+q)^{d_{mn}+n_p+n_q}},
 \qquad d_{mn}=\ell+1+\ii(m+n)\tnu.
 \label{sm:Wseriesproof}
\end{equation}
The remaining double series is Appell's $F_2$, giving Eq.\,\eqref{eq:Wseed}, in the convention
\begin{equation}
  F_2(d;u,v;c,e;x,y)=\sum_{n_p,n_q\ge0}\!
  \frac{\,(d)_{n_p\!+\!n_q}(u)_{n_p}(v)_{n_q}\,}{\,(c)_{n_p}(e)_{n_q}\,n_p!n_q!\,}x^{n_p}y^{n_q},
  \qquad |x|\!+\!|y|<1\,.
  \label{sm:F2def}
\end{equation}
Here $(c)_n=\Gamma(c+n)/\Gamma(c)$. Substitution of the series parameters therefore gives
\begin{subequations}\label{sm:WresultDerived}
\begin{align}
 \mathcal I_\ell(p,q;k_3)
 &=k_3^{-\ell}\int_0^\infty\di z_3\,z_3^{\ell-1}e^{-z_3}
 W_{\kappa_1,\ii\tnu}\!\left(\frac{2p}{k_3}z_3\right)
 W_{\kappa_2,\ii\tnu}\!\left(\frac{2q}{k_3}z_3\right)\nonumber\\
 &=\sum_{m,n=\pm1}\frac{R_{\kappa_1,m}R_{\kappa_2,n}\Gamma(d_{mn})
 (2p)^{\frac12+\ii m\tnu}(2q)^{\frac12+\ii n\tnu}}
 {(k_3+p+q)^{d_{mn}}}\nonumber\\
 &\quad\times F_2\!\left(d_{mn};\tfrac12+\ii m\tnu-\kappa_1,\tfrac12+\ii n\tnu-\kappa_2;
 1+2\ii m\tnu,1+2\ii n\tnu;\frac{2p}{k_3+p+q},\frac{2q}{k_3+p+q}\right),
 \label{sm:WseedDerived}\\
 R_{\kappa,m}&=\frac{\Gamma(-2\ii m\tnu)}{\Gamma(\frac12-\ii m\tnu-\kappa)},
 \qquad d_{mn}=\ell+1+\ii(m+n)\tnu.
 \label{sm:WdefsDerived}
\end{align}
\end{subequations}
These are exactly Eqs.\,\eqref{eq:Wseed} and \eqref{eq:Wdefs}. The sum of the two $F_2$ arguments is $2(p+q)/(k_3+p+q)<1$ in the auxiliary domain. The next step specifies how this derived expression is used in the physical triangle, where its defining double series cannot be used directly.

\subsection{Continuation into the Physical Triangle Region}\label{sm:Wcontinuation}
In the physical triangle $k_3\le p+q$, the defining $F_2$ series cannot be used directly. The original Whittaker integral nevertheless remains convergent: the growing contributions from the confluent hypergeometric functions ${}_1F_1$ cancel when the branches are combined. We therefore continue the \emph{whole four-term expression} from real $k_{3,0}>p+q$ through the upper half $k_3$ plane to $k_3+\ii0$, keeping the real logarithms of $2p,2q$ fixed. The identity theorem identifies this sum with the original integral throughout $\Re k_3>-p-q$. Lower-half-plane continuation gives the same combined value. At $k_3=p+q$, the branch terms must likewise be combined before taking the limit.

Two useful checks of this continued time integral follow directly from its fixed-$Y_3$ representation:
\begin{equation}
  -\partial_{k_3}\mathcal I_0(p,q;k_3)=\mathcal I_1(p,q;k_3),\qquad
  \mathcal I_\ell(cp,cq;ck_3)=c^{-\ell}\mathcal I_\ell(p,q;k_3),\quad c>0\,.
  \label{sm:Wchecks}
\end{equation}
In the first identity $p,q$ are held fixed, and differentiation acts on $e^{-k_3Y_3}$ under the integral. The second follows by rescaling the time variable. These identities check the moment powers and normalization; the analytic continuation into the physical triangle is fixed by the combined four-term expression above.

\subsection{Exact Mapping to the Four Spinor Channels}
\label{app:softchannels}

We next restore the spinor weights and identify which time integral enters each channel. The four labels $A,B,C,D$ refer to the Whittaker-index pairs selected by the Nambu contraction. They are separate from the four branch choices $(m,n)$ within each $\mathcal I_\ell$ and from the SK labels. With $\kappa=-1/2-\ii\tlambda$ and $\eta=1/2-\ii\tlambda$ as in Eq.\,\eqref{sm:adjacent}, the pairs are listed in the order $(p\text{ line},q\text{ line})$:
\begin{equation}
  A:(\bar\kappa,\bar\eta),\qquad B:(\kappa,\eta),\qquad
  C:(\bar\kappa,\kappa),\qquad D:(\eta,\bar\eta)\,.
  \label{sm:pairs}
\end{equation}
Reuse the variables of Appendix\,\ref{sm:matchingproof}; they separate the momentum ratio $z$, the asymmetry $w$ between the two soft lines, and the rescaled time $u$:
\begin{equation}
  z=\frac{\,k_3\,}{\,p\!+\!q\,},\qquad w=\frac{\,q\!-\!p\,}{\,k_3\,},\qquad u=\frac{\,(p\!+\!q)z_3\,}{\,k_3\,}=\frac{\,z_3\,}{\,z\,},
  \qquad \frac{\,2pz_3\,}{\,k_3\,}=(1\!-\!wz)u,\quad\frac{\,2qz_3\,}{\,k_3\,}=(1\!+\!wz)u\,.
  \label{sm:softvars}
\end{equation}
The domain is $0<z\le1$, $-1\leqslant w\le1$, $u>0$. In channel $j$, $\mathcal W_{j,n}$ is the rescaled time moment of order $n\!=\!0,1$; $\kappa_{1j},\kappa_{2j}$ are the ordered pair in Eq.\,\eqref{sm:pairs}:
\begin{equation}
  \mathcal W_{j,n}(z,w)=\int_0^\infty\di u\,u^{n-1}e^{-zu}
  W_{\kappa_{1j},\ii\tnu}((1\!-\!wz)u)W_{\kappa_{2j},\ii\tnu}((1\!+\!wz)u)
  =(p\!+\!q)^n\mathcal I_n^{(j)}(p,q;k_3)\,.
  \label{sm:Wrescaled}
\end{equation}
The superscript $(j)$ on $\mathcal I_n^{(j)}$ specifies the index pair in Eq.\,\eqref{sm:pairs}; $n$ labels the time moment. Thus the same two moments derived above supply every channel. The range $0<z\le1$ is used with the finite-part prescription of Appendix\,\ref{sm:matchingproof}, inherited from the original lower boundary $z_{\min}=\rho/\varepsilon_{\rm{sep}}$.

To expose the cancellation of canonical powers, use
\begin{equation}
  W_{\varkappa,\ii\tnu}(x)=e^{-x/2}x^{(1-2\ii\tnu)/2}
  U\!\left(\frac{\,1\!-\!2\ii\tnu\,}{\,2\,}\!-\!\varkappa,1\!-\!2\ii\tnu,x\right)\,.
  \label{sm:WU}
\end{equation}
The product of powers is $(1\!-\!w^2z^2)^{(1-2\ii\tnu)/2}u^{1-2\ii\tnu}$ and the product of exponentials is $e^{-u}$. Combining these powers with the spinor weights and the Jacobian in Eq.\,\eqref{sm:Jacobian} cancels the common angular denominator. The resulting response for each channel is
\begin{equation}
  \mathcal A_j(z)=\int_{-1}^{1}\di w\int_0^\infty\di u\,
  u^{-2\ii\tnu}e^{-(1+z)u}\widetilde K_j(z,w,u)
  U(\alpha_j,1\!-\!2\ii\tnu,(1\!-\!wz)u)U(\beta_j,1\!-\!2\ii\tnu,(1\!+\!wz)u)\,,
  \label{sm:fourresponse}
\end{equation}
where $\alpha_j,\beta_j$ are the first parameters of the $p$- and $q$-line Tricomi functions, respectively, and $\widetilde K_j$ is their spinor weight:
\begin{equation}
  \begin{array}{c|cc|l}
    j & \alpha_j & \beta_j & \widetilde K_j(z,w,u)             \\\hline
    A & 1\!+\!b  & b       & 2(1\!-\!z^2)[w\!+\!uz(w\!+\!1)]   \\
    B & 1\!+\!a  & a       & 2(1\!-\!z^2)[w\!+\!uz(w\!+\!1)]   \\
    C & 1\!+\!b  & 1\!+\!a & -2z(1\!-\!w^2)[1\!+\!uz(1\!-\!z)] \\
    D & a        & b       & -2z(1\!-\!w^2)[1\!+\!uz(1\!+\!z)]
  \end{array}\,.
  \label{sm:channelTable}
\end{equation}
Every weight in the last column is linear in $u$. Write it as $\widetilde K_j=k_{j0}+u k_{j1}$; the inner time integral in Eq.\,\eqref{sm:fourresponse} then equals
\begin{equation}
  (1\!-\!w^2z^2)^{-(1-2\ii\tnu)/2}
  \left[k_{j0}\mathcal W_{j,0}(z,w)\!+\!k_{j1}\mathcal W_{j,1}(z,w)\right]\,.
  \label{sm:timebridge}
\end{equation}
Here $k_{j0}$ and $k_{j1}$ are the coefficients of $1$ and $u$ in the last column of Eq.\,\eqref{sm:channelTable}; thus no additional time moments occur. The factor multiplying the bracket removes the canonical Whittaker powers in Eq.\,\eqref{sm:WU}, and the remaining integrations are over angle and loop momentum.

With $\mathcal R_A,\mathcal R_B$ in Eq.\,\eqref{sm:hardratios}, the soft factor, including the loop normalization, is
\begin{equation}
  \Cs=\frac{\,2^{-1-2\ii\tnu}\,}{\,8\pi^2\Lambda\,}\FP\int_0^1\di z\,z^{-3-2\ii\tnu}F(z),\qquad
  F(z)=\mathcal R_A\mathcal A_A(z)+\mathcal R_B\mathcal A_B(z)
    +\mathcal A_C(z)+\m^{-2}\mathcal A_D(z).
  \label{sm:Csoftintegral}
\end{equation}
The factor $2^{-1-2\ii\tnu}/(8\pi^2)$ follows from the azimuthal integral, the Jacobian in Eq.\,\eqref{sm:Jacobian} and the two hard-endpoint powers. This explicitly identifies the combined bracket $F(z)$ used in Appendices\,\ref{sm:matchingproof} and \ref{sm:hardprojectionproof}. Its finite part is fixed by Eq.\,\eqref{sm:finitepart}; the four channels are summed before the endpoint terms are subtracted.

For direct comparison with the main text, substitute Eq.\,\eqref{sm:timebridge} and $\mathcal W_{j,\ell}=(p+q)^\ell\mathcal I_\ell^{(j)}$ into Eq.\,\eqref{sm:Csoftintegral}, then undo the Jacobian. This gives
\begin{equation}
 \Cs=\FP\int\frac{\di^3\vct q}{(2\pi)^3}
 \sum_{j=A,B,C,D}\sum_{\ell=0}^{1}
 \mathcal K_{j\ell}(\vct q,\vct k_3)\,\mathcal I_\ell^{(j)}(p,q;k_3),
 \label{sm:softIbridgeExplicit}
\end{equation}
where $\gamma_j$ are the relative channel weights and $\mathcal K_{j\ell}$ are the azimuthally averaged kernels:
\begin{equation}
 \mathcal K_{j\ell}
 =\frac{2^{1-2\ii\tnu}}{\Lambda k_3^3}\,
 z^{1-2\ii\tnu}(1-w^2z^2)^{(2\ii\tnu-3)/2}\,
 \gamma_j k_{j\ell}(z,w)(p+q)^\ell,
 \qquad (\gamma_A,\gamma_B,\gamma_C,\gamma_D)
 =(\mathcal R_A,\mathcal R_B,1,\m^{-2}).
 \label{sm:softKexplicit}
\end{equation}
All quantities on the right are specified in Eqs.\,\eqref{sm:hardratios}, \eqref{sm:softvars} and \eqref{sm:channelTable}; the Whittaker-index pair for $\mathcal I_\ell^{(j)}$ is Eq.\,\eqref{sm:pairs}. In particular, multiplying $\mathcal K_{j\ell}$ by the measure in Eq.\,\eqref{sm:Jacobian} restores the prefactor $2^{-1-2\ii\tnu}/(8\pi^2\Lambda)$ and $z^{-3-2\ii\tnu}(1-w^2z^2)^{-(1-2\ii\tnu)/2}$ of the time-moment representation. Equation\,\eqref{sm:softIbridgeExplicit} is the result used in Eq.\,\eqref{eq:softIbridge}, now with its weights and normalization fixed. Its finite part is taken after the coherent channel sum, not separately for each term.

\subsection{Time and Momentum Integrations}
The $F_2$ form gives the soft-time integral at fixed $p,q$, but directly integrating its continued branch terms over momentum would obscure the cancellations between channels. We instead return to the equivalent Tricomi representation in Eq.\,\eqref{sm:fourresponse}. Introducing one auxiliary Euler variable for each soft line makes the time integral a Gamma integral and the radial and angular integrations elementary. This is an alternative evaluation of the same soft factor, not an additional approximation. For $\Re c>0$ and $\Re x>0$, use
\begin{equation}
  U(c,1\!-\!2\ii\tnu,x)=\frac{\,1\,}{\,\Gamma(c)\,}\int_0^1\di R\,
  (1\!-\!R)^{c-1}R^{2\ii\tnu-1}e^{-x(1-R)/R}\,.
  \label{sm:EulerU}
\end{equation}
Use $0<R,S<1$ as the Euler integration variables of the $p$ and $q$ lines, respectively.\  The combined exponential is $e^{-uQ/(RS)}$, with
\begin{equation}
  Q=R\!+\!S\!-\!RS\!+\!zRS\!+\!wz(R\!-\!S),\qquad
  \int_0^\infty u^{n-2\ii\tnu}e^{-uQ/(RS)}\di u
  =\Gamma(n\!+\!1\!-\!2\ii\tnu)\left(\!\frac{\,RS\,}{\,Q\,}\!\right)^{\!\!n+1-2\ii\tnu}\,.
  \label{sm:EulerTime}
\end{equation}
The time integration initially requires $\Re(n+1-2\ii\tnu)>0$ and $\Re[Q/(RS)]>0$. Combining the factor $(RS)^{n+1-2\ii\tnu}$ with the Euler measures leaves the polynomial factor $(RS)^n$, for $n=0,1$. All steps begin with positive real endpoint exponents and convergent moments; the combined expression is then continued to physical $a,b,2\ii\tnu$. This evaluates the same time integral as the $F_2$ route.\

In particular, writing $\widetilde K_j=k_{j0}+u k_{j1}$ as above gives the channel response after the time integration explicitly:
\begin{equation}
 \begin{aligned}
 \mathcal A_j(z)={}&\frac{1}{\Gamma(\alpha_j)\Gamma(\beta_j)}
 \int_0^1\di R\int_0^1\di S\,
 (1-R)^{\alpha_j-1}(1-S)^{\beta_j-1}\\
 &\times\int_{-1}^1\di w\,
 \left[\Gamma(1-2\ii\tnu)k_{j0}(z,w)Q^{2\ii\tnu-1}
       +\Gamma(2-2\ii\tnu)RS\,k_{j1}(z,w)Q^{2\ii\tnu-2}\right].
 \end{aligned}
 \label{sm:EulerChannelResponse}
\end{equation}
This expression retains the channel weights of Eq.\,\eqref{sm:channelTable} and makes the remaining $z,w$ integrations explicit.

After the $u$ integration, perform the radial integral and then the angular integral. The substitution $y=1/z$ turns the finite-part radial integral into integrals of powers of an affine function of $y$; the angular integral has the same structure. The finite boundary terms are fixed by Eq.\,\eqref{sm:finitepart}. Channels $A$ and $B$ lead to equal-parameter Euler measures, with exponents $(b,b)$ and $(a,a)$, and share a kernel $w_U$. Channels $C$ and $D$ combine into a mixed-parameter measure with exponents $(a,b)$ and kernel $w_V$. Thus there are two distinct kernels but three square moments. For their explicit expressions, abbreviate $D=R+S-RS$ and $d=R-S$:
\begin{subequations}
  \begin{align}
    w_U(R,S) & =\frac{\,1\,}{\,d\,}\left[-\frac{\,2(R^{2\ii\tnu+2}\!-\!S^{2\ii\tnu+2})\,}{\,(2\ii\tnu\!+\!2)D\,}
    \!+\!\frac{\,2(R^{2\ii\tnu+3}\!-\!S^{2\ii\tnu+3})\,}{\,(2\ii\tnu\!+\!1)D^2\,}\right.\nonumber\\
    &\qquad\left.\!-\!\frac{\,4(R^{2\ii\tnu+4}\!-\!S^{2\ii\tnu+4})\,}{\,(2\ii\tnu\!+\!1)(2\ii\tnu\!+\!4)D^3\,}\right]\,,\label{sm:wU} \\
    w_V(R,S) & =\frac{\,8\,}{\,(2\ii\tnu\!+\!1)(2\ii\tnu\!+\!2)Dd^3\,}
    \left[-(2\ii\tnu\!+\!1)(R^{2\ii\tnu+2}S^2\!-\!R^2S^{2\ii\tnu+2})\right.\nonumber\\
    &\qquad\left.\!+\!2\ii\tnu(R^{2\ii\tnu+3}S\!-\!RS^{2\ii\tnu+3})
    \!-\!\frac{\,2\ii\tnu\,}{\,2\ii\tnu\!+\!4\,}(R^{2\ii\tnu+4}\!-\!S^{2\ii\tnu+4})\right]\,.
    \label{sm:wV}
  \end{align}
\end{subequations}
The Euler variables satisfy $0<R,S<1$, so the powers use real logarithms of positive bases. The denominators $d$ and $d^3$ do not give genuine singularities at $R=S$: the numerators vanish to the corresponding orders. On the diagonal, with $r$ denoting the common Euler coordinate, the finite values are
\begin{equation}
  w_U(r,r)=r^{2\ii\tnu}\left[-\frac{\,2\,}{\,2\!-\!r\,}\!+\!\frac{\,2(2\ii\tnu\!+\!3)\,}{\,(2\ii\tnu\!+\!1)(2\!-\!r)^2\,}
    \!-\!\frac{\,4\,}{\,(2\ii\tnu\!+\!1)(2\!-\!r)^3\,}\right],\qquad
  w_V(r,r)=\frac{\,8\ii\tnu(2\ii\tnu\!-\!1)r^{2\ii\tnu}\,}{\,3(2\ii\tnu\!+\!1)(2\!-\!r)\,}\,.
  \label{sm:diagonals}
\end{equation}
The values $w_U(1,1)\!=\!0$ and $w_V(1,1)\!=\!8\ii\tnu(2\ii\tnu\!-\!1)/[3(2\ii\tnu\!+\!1)]$ are useful checks of the endpoint terms.\

Define the analytically continued moment over the unit square by
\begin{equation}
  \mathcal J_{\alpha,\beta}[w]=\AC\int_0^1\di R\int_0^1\di S\,
  (1\!-\!R)^{\alpha-1}(1\!-\!S)^{\beta-1}w(R,S)\,,
  \label{sm:squaremoment}
\end{equation}
Here $\alpha,\beta$ are generic edge exponents, and $\AC$ denotes meromorphic continuation from a convergent domain with positive real edge exponents. The physical exponents are assigned separately to each of the three moments. Including their channel weights, the full reduced combination is
\begin{equation}
  \mathcal S=\mathcal J_{a,b}[w_V]\!-\!2s_A\mathcal J_{b,b}[w_U]
  \!-\!2s_B\mathcal J_{a,a}[w_U],\qquad
  s_A=\frac{\,\sinh[\pi(\tnu\!+\!2\tlambda)]\,}{\,\sinh(\pi\tnu)\,},\quad
  s_B=\frac{\,\sinh[\pi(\tnu\!-\!2\tlambda)]\,}{\,\sinh(\pi\tnu)\,}\,.
  \label{sm:Stwokernels}
\end{equation}
The first square moment contains the combined $C,D$ contribution, while the second and third contain channels $A$ and $B$. The coefficients $s_A,s_B$ are the real hyperbolic weights remaining after the connection coefficients in $\mathcal R_A,\mathcal R_B$ are combined with the Euler Gamma factors. Restoring those factors gives the physical normalization
\begin{equation}
  \Cs=-\frac{\,\Gamma(-2\ii\tnu)\,}{\,8\pi^2\Lambda\Gamma(1\!+\!a)\Gamma(1\!+\!b)\,}\mathcal S=\frac{\mathcal S}{8\pi^2\Lambda\m^2\mathrm B(a,b)}\,.
  \label{sm:Snormal}
\end{equation}
The second equality uses $a+b\!=\!-2\ii\tnu$ and $ab\!=\!-\m^2$. Equation\,\eqref{sm:Snormal} expresses the same normalized $\Cs$ as Eq.\,\eqref{sm:Csoftintegral}, now in terms of the three square moments.\

\subsection{Endpoint Subtraction and Analytic Continuation}
The radial finite part has already been included in $w_U,w_V$. A separate issue remains: at the physical values of $a,b$, the factors $(1-R)^{a-1}$ and $(1-S)^{b-1}$ do not define ordinary convergent integrals at the upper edges. We evaluate their analytic continuation by subtracting the edge values and adding back their integrals in closed form. Both kernels are symmetric in $R,S$. For either kernel $w$, define the remainder $D_w$, which vanishes on both upper edges, and the one-edge moment $E_c[w]$:
\begin{equation}
  D_w(R,S)=w(R,S)\!-\!w(R,1)\!-\!w(1,S)\!+\!w(1,1),\qquad
  E_c[w]=\AC\int_0^1(1\!-\!R)^{c-1}w(R,1)\di R\,.
  \label{sm:edgeSub}
\end{equation}
Adding and subtracting the edges gives
\begin{equation}
  \mathcal J_{\alpha,\beta}[w]
  =\int_0^1\di R\int_0^1\di S\,(1\!-\!R)^{\alpha-1}(1\!-\!S)^{\beta-1}D_w(R,S)
  \!+\!\frac{\,E_\alpha[w]\,}{\,\beta\,}\!+\!\frac{\,E_\beta[w]\,}{\,\alpha\,}
  \!-\!\frac{\,w(1,1)\,}{\,\alpha\beta\,}\,.
  \label{sm:edges}
\end{equation}
The double integral now has extra factors $(1\!-\!R)(1\!-\!S)$ at the upper edges and is integrable at the physical imaginary exponents.\  The remaining one-dimensional edge values are continued by the same subtract-and-add rule.\  For example
\begin{equation}
  E_c[w]=\int_0^1(1\!-\!R)^{c-1}[w(R,1)\!-\!w(1,1)]\di R\!+\!\frac{\,w(1,1)\,}{\,c\,}\,.
  \label{sm:edgefinite}
\end{equation}
The lower corner also needs control. On $R=r,S=rv$, $0\le v\le1$, either explicit kernel obeys $|w(r,rv)|\le C r^{\Re(2\ii\tnu)}$ locally for $\Re(2\ii\tnu)>-2$. After factoring out $r^{2\ii\tnu}$, $1+v-rv$ stays nonzero, the divided differences remove the apparent pole at $v=1$, and the least remaining complex power at $v=0$ is $v^{2\ii\tnu+2}$. The measure is $r\di r\di v$; hence the corner is integrable, also on the exchanged triangle. Together with the edge subtraction, this continues Eq.\,\eqref{sm:edges} to $\Re\alpha,\Re\beta>-1$, $\Re(2\ii\tnu)>-2$, away from parameter poles. The physical values have zero real parts. The finite edge terms are retained throughout.\

The two endpoint operations therefore have different roles. The radial finite part $\FP$ handles $z\to0$, where the loop momentum becomes large relative to $k_3$ in the expanded soft integral; Appendix\,\ref{sm:matchingproof} fixes its finite terms. Those terms are already incorporated in $w_U,w_V$. The continuation $\AC$ then handles the upper edges of the auxiliary Euler square and retains the edge and corner terms in Eq.\,\eqref{sm:edges}. It does not introduce a second freedom to subtract the physical loop integral.

\subsection{The Closed Form in the Main-Text Notation}\label{sm:softmainresult}
We have now reduced $\Cs$ to the three continued moments in Eq.\,\eqref{sm:Stwokernels}, with both endpoint prescriptions fixed. Appendix\,\ref{sm:native} performs the remaining special-function reduction using a common Barnes representation and its boundary residues. We state that result here so that the soft calculation ends with the same normalization as the main text. In the notation of Eq.\,\eqref{eq:finitepars}, the reduced combination is
\begin{equation}
 \mathcal S=-\frac{2^{A-1}\pi\m^2|\mathrm B(a,b)|^2}
 {\tnu A(A-1)(A+2)\sinh(2\pi\tnu)}
 \left[(A-2)(4+A-A^2)-16A\tlambda^2
 +\frac{8\ii(\mathcal T_a+\mathcal T_b)}{\pi^3(A+1)}\right].
 \label{sm:softSquareFinal}
\end{equation}
This is Eq.\,\eqref{sm:Srednormal} with the reduced bracket evaluated; $\mathcal T_c$ is defined below. Substitution into the independently normalized Eq.\,\eqref{sm:Snormal}, using $|\mathrm B|^2/\mathrm B=\overline{\mathrm B}$, gives
\begin{subequations}\label{sm:softFinalAll}
\begin{align}
 \Cs&=-\frac{2^{A-4}\overline{\mathrm B(a,b)}}
 {\pi\Lambda\tnu A(A-1)(A+2)\sinh(2\pi\tnu)}
 \left[(A-2)(4+A-A^2)-16A\tlambda^2
 +\frac{8\ii(\mathcal T_a+\mathcal T_b)}{\pi^3(A+1)}\right],
 \label{sm:softFinal}\\
 \mathcal T_c&=\sin\pi c\left[(\omega-\cos\pi c)\mathcal H(A,c)
 +(\omega+\cos\pi c)\overline{\mathcal H(\bar A,\bar c)}\right],\qquad c=a,b,
 \label{sm:Tfinal}\\
 \mathcal H(A,c)&=\sum_{r=0}^{1}\left[\frac{A(1-A)}{4}\right]^{1-r}
 \Gn[\!]{-1}{-\frac A2,\frac{1-A}{2},-r,1-c}{\varnothing}{0,1,-\frac12}{1-A-c}.
 \label{sm:Hfinal}
\end{align}
\end{subequations}
These are exactly Eqs.\,\eqref{eq:softfinite}--\eqref{eq:Hfinite}. The branch is $\operatorname{Log}(-1)=+\ii\pi$ in $\mathcal H$; conjugating the parameters and then the function value supplies the opposite boundary value as in Eq.\,\eqref{sm:Glips}. The polynomial in Eq.\,\eqref{sm:softFinal} includes the finite residue terms and cannot be dropped. Thus the complete soft modes, the finite terms retained in Eq.\,\eqref{sm:finitepart} and the Euler continuation all enter the same normalized $\Cs$ that multiplies Eq.\,\eqref{sm:hardFinal} in the factorization formula, Eq.\,\eqref{sm:factorizedresult}.

\section{Derivation of the Complete Closed-Form Expression}
\label{sm:native}
\label{App:E}
We evaluate the two Euler integrals of Appendix\,\ref{sm:soft} through a common Mellin--Barnes representation. Keeping the finite endpoint terms and the residues from poles crossed during contour shifts yields both the polynomial contribution and the ordinary Meijer $G$ functions in Eq.\,\eqref{eq:softfinite}. The two boundary values at $z=-1$ retain the branch prescription inherited from the soft integral.

\subsection{A Common Barnes Kernel}

The equal moments have identical Euler exponents, $\mathcal J_{a,a}[w_U]$ or $\mathcal J_{b,b}[w_U]$, while the mixed moment is $\mathcal J_{a,b}[w_V]$ in Eq.\,\eqref{sm:Stwokernels}. Split the Euler square into $S<R$ and $R<S$ and set $S=vR$ in the first triangle. The Jacobian is $R$, and $R+S-RS=R(1+v-vR)$ turns the radial integral into a Beta integral with a Gauss function. With $r\!=\!(1-v)/(1+v)$, the following odd-power identity (for a complex exponent $h$)
\begin{equation}
  \frac{\,(1\!+\!r)^h\!-\!(1\!-\!r)^h\,}{\,2hr\,}
  ={}_2F_1\!\left(\frac{\,1\!-\!h\,}{\,2\,},\frac{\,2\!-\!h\,}{\,2\,};\frac{\,3\,}{\,2\,};r^2\right)
  \label{sm:oddbinomial}
\end{equation}
puts the equal and mixed moments on related Barnes contours.\  Here $s$ is the Mellin variable for the Euler moment, distinct from the single-endpoint Mellin variables in Eq.\,\eqref{eq:Wmellin}. Write all Barnes measures as $\di s/(\ii\hp 2\pi)$. With $A,a,b$ from Eq.\,\eqref{eq:finitepars}, the equal-channel kernel is
\begin{equation}
  e_c(s)=\frac{\,\Gamma(-s)^2\Gamma(A/2\!+\!s)\Gamma((A\!+\!1)/2\!+\!s)
    \Gamma(1\!+\!s)\Gamma(c\!+\!s)\,}{\,\Gamma(3/2\!+\!s)\Gamma(A\!+\!c\!+\!s)\,},\qquad c=a,b\,.
  \label{sm:commonkernel}
\end{equation}
At physical parameters a convenient fixed upward line is $L:\Re s\!=\!-3/4$. Moving the Euler exponents to $\Re c\!=\!0$ crosses the pole at $s\!=\!-c$. Accordingly the canonical continuation is
\begin{subequations}
  \begin{align}
    \AC\int w(s)e_c(s)M(s)\frac{\,\di s\,}{\,\ii\hp 2\pi\,}
    ={}                  & \int_Lw(s)e_c(s)M(s)\frac{\,\di s\,}{\,\ii\hp 2\pi\,}
    \!+\!w(-c)M(-c)\res_{s=-c}e_c(s)\,,\label{sm:BarnesAC}                       \\
    \res_{s=-c}e_c(s)={} &
    \frac{\,\Gamma(c)^2\Gamma(A/2\!-\!c)\Gamma((A\!+\!1)/2\!-\!c)\Gamma(1\!-\!c)\,}{\,\Gamma(3/2\!-\!c)\Gamma(A)\,}\,.
    \label{sm:movingpole}
  \end{align}
\end{subequations}
Here $w$ is the polynomial weight introduced below, and $M$ is $1$, $\tan\pi s$ or the indicated linear combination. With $L$ held fixed, the explicit residue compensates for the moving pole crossing the contour as the parameters are continued. The tangent pole at $s=-1/2$ stays to the right of $L$ and contributes no crossed residue. Thus the line integral and the residue together represent the continued Euler moment.

\subsection{Contour Shifts and Boundary Residues}
The common cubic and the polynomial appearing in the final soft bracket are
\begin{equation}
  P_A(s)=s(s\!+\!A/2)(s\!+\!h),\qquad h=\frac{\,4\!+\!A\!-\!A^2\,}{\,4\,},\qquad
  \mathcal N=(A\!-\!2)(4\!+\!A\!-\!A^2)\!-\!16A\tlambda^2\,.
  \label{sm:cubic}
\end{equation}
These polynomials are used only in the reduction.\  The shift ratio of the kernel in Eq.\,\eqref{sm:commonkernel} is
\begin{equation}
  \frac{\,e_c(s\!+\!1)\,}{\,e_c(s)\,}=\frac{\,n_c(s)\,}{\,r_c(s\!+\!1)\,},\qquad
  n_c(s)=(s\!+\!A/2)(s\!+\!(A\!+\!1)/2)(s\!+\!c),\quad
  r_c(s)=s(s\!+\!1/2)(s\!+\!A\!+\!c\!-\!1)\,.
  \label{sm:shiftRatio}
\end{equation}
Write $p_c$ for the equal-channel linear weight. Subtracting the unit-shift difference $d_c$ generated by $f_c$ leaves a multiple of the common cubic $P_A$:
\begin{subequations}\label{sm:certificate}
  \begin{align}
    p_c(s) & =\frac{\,\mathcal N\,}{\,4\,}(s\!+\!A/2)
    \!-\!\frac{\,A(A\!-\!1)\,}{\,8\,}(A\!+\!2c\!-\!2)(3A\!+\!4c\!-\!4)\,,\label{sm:pcDefinition} \\
    f_c(s) & =3A\!+\!4c\!-\!4\!-\!\frac{\,4(s\!+\!A/2)\,}{\,A\!+\!1\,}\,,\label{sm:fcDefinition} \\
    d_c(s) & =n_c(s)f_c(s\!+\!1)\!-\!r_c(s)f_c(s),\qquad
    p_c(s)\!-\!d_c(s)=\frac{\,8\,}{\,A\!+\!1\,}P_A(s)\,.
    \label{sm:certificate:last}
  \end{align}
\end{subequations}
Expanding the cubic polynomials and using both $c(c\!+\!A\!-\!2)\!=\!\m^2$ and $4\tlambda^2\!=\!-(A+2c\!-\!2)^2$, valid for $c\!=\!a,b$, verifies the last equality.\  Multiplication by $e_c$ turns $d_c$ into a unit contour shift, the Barnes version of integration by parts, with the boundary residue displayed below.\

In an initial domain near $A\!=\!2.1,c\!=\!0.8$ with $\Re s\!=\!-0.65$, the shift crosses zero.\  Since
\begin{equation}
  e_c(s)=2^{2-A}\mathrm B(A,c)s^{-2}\!+\!O(s^{-1}),\qquad
  \AC\int d_c(s)e_c(s)\frac{\,\di s\,}{\,\ii\hp 2\pi\,}
  =f_c(0)(A\!+\!c\!-\!1)2^{1-A}\mathrm B(A,c)\,,
  \label{sm:shiftresidue}
\end{equation}
the contour shift contributes a finite residue that cancels the equal-channel radial boundary term in Eq.\,\eqref{sm:edges}. Continuation of the complete identity through Eq.\,\eqref{sm:BarnesAC} retains the moving pole.\  The equal and mixed channels can then use the same cubic weight.\

\subsection{The Mixed Remainder Is an Elementary Residue}
The two mixed channels are first combined using the inverse-argument Gauss connection formula.\  At $a\!+\!b\!=\!2\!-\!A$ the reflected Barnes kernel can be written in terms of $e_b$ and one remainder.\  Denote the reflected kernel by $K_b$, and write its remainder $M_b$ as a Gamma factor $\ell_b$ times a trigonometric factor:
\begin{subequations}\label{sm:mixedkernels}
  \begin{align}
    K_b(s)    & =e_b(s)\frac{\,\sin\pi(b\!-\!s)\sin\pi(b\!+\!s)\,}{\,\cos\pi s\sin\pi(a\!-\!s)\,}\,,\label{sm:KbDefinition} \\
    \ell_b(s) & =\frac{\,\Gamma(-s)\Gamma(a\!-\!1\!-\!s)\Gamma(A/2\!+\!s)
      \Gamma((A\!+\!1)/2\!+\!s)\Gamma(b\!+\!s)\,}{\,\Gamma(3/2\!+\!s)\,},\qquad
    M_b(s)=\ell_b(s)\frac{\,\sin\pi(s\!-\!b)\,}{\,\cos\pi s\,}\,.
    \label{sm:mixedkernels:last}
  \end{align}
\end{subequations}
Gamma reflection gives the pointwise identity
\begin{equation}
  K_b=\frac{\,\sin^2\pi b\,}{\,\sin\pi a\,}e_b
  \!-\!\frac{\,\sin\pi b\cos\pi b\,}{\,\sin\pi a\,}e_b\tan\pi s
  \!+\!\frac{\,\sin\pi A\,}{\,\sin\pi a\,}M_b\,.
  \label{sm:mixedidentity}
\end{equation}
For a common weight $w(s)$, the prescription on $L:\Re s=-3/4$ is summarized by
\begin{equation}
 \begin{array}{c|l}
 \text{integrand}&\text{terms added to its integral on }L\\\hline
 we_b\ \text{or}\ we_b\tan\pi s&+\res_{s=-b}(\text{integrand})\\
 wM_b&+\res_{s=-b}(wM_b)-\res_{s=a-1}(wM_b)\\
 wK_b&-\res_{s=a-1}(wK_b)
 \end{array}
 \label{sm:continuationtable}
\end{equation}
The $-b$ residues cancel in Eq.\,\eqref{sm:mixedidentity}, consistently with regularity of $K_b$ there. The tangent pole at $-1/2$ remains to the right of $L$.\

The remainder has the four-Gamma form
\begin{equation}
  M_b(s)=-2^{3-A}\Gamma(A\!+\!2s)\Gamma(-1\!-\!2s)
  \Gamma(a\!-\!1\!-\!s)\Gamma(b\!+\!s)\sin\pi(s\!-\!b)\,.
  \label{sm:fourGamma}
\end{equation}
For positive Mellin argument $z$, the binomial transform is
\begin{equation}
  \int\Gamma(A\!+\!2s)\Gamma(-1\!-\!2s)z^s\frac{\,\di s\,}{\,\ii\hp 2\pi\,}
  =\frac{\,\Gamma(A\!-\!1)\,}{\,2\sqrt{z}\,}(1\!+\!\sqrt{z}\,)^{1-A}\,.
  \label{sm:binomialMellin}
\end{equation}
There is no common real convergence strip for both Gamma pairs at the final balance.\  To justify their interchange, first replace $\Gamma(a\!-\!1\!-\!s)$ by $\Gamma(a\!-\!1\!+\!\epsilon_B\!-\!s)$. Its product with $\Gamma(b\!+\!s)$ has an Euler integral in an open convergence domain, for example $A\!=\!6/5,b\!=\!4/5,\epsilon_B\!=\!1,\Re s\!=\!-11/20$. Fubini's theorem applies in this domain.\  Split the sine into two exponential phases, use Eq.\,\eqref{sm:binomialMellin}, and continue the entire equality meromorphically to $\epsilon_B\!=\!0$ with the same crossed-pole residue contributions.\  The regulated identity thus defines the continued integral at the physical parameters.\

Multiplication by $s$ acts as a logarithmic derivative in the binomial transform.\  For the cubic $P_A$ this gives a rational function
\begin{equation}
  \Phi_{P_A}(y)=\frac{\,(A\!-\!1)(A\!+\!1)\,}{\,16(1\!-\!y)^3\,}
  \left[(A^2\!+\!A\!-\!2)y^2\!+\!(4\!-\!A^2)y\!+\!A\!-\!2\right]\,.
  \label{sm:rationalPhi}
\end{equation}
The difference of the two continued rays surrounds only $y\!=\!1$ in the right half plane.\  Hence
\begin{equation}
  \AC\int P_A(s)M_b(s)\frac{\,\di s\,}{\,\ii\hp 2\pi\,} =2^{3-A}\pi\Gamma(A\!-\!1)\Gamma(1\!-\!A) \res_{y=1}\left[y^{2b-2}(1\!+\!y)^{A-1}\Phi_{P_A}(y)\right]=-\frac{\,\pi^2(A\!+\!1)\mathcal N\,}{\,16\sin\pi A\,}\,. \label{sm:polyresidue}
\end{equation}
Since Eq.\,\eqref{sm:rationalPhi} has a third-order pole, a second derivative gives the residue and hence the elementary polynomial $\mathcal N$ in Eq.\,\eqref{eq:softfinite}. The derivation remains valid at zeros of $\mathcal N$, since it never divides by this polynomial.\

\subsection{From the Common Kernel to Ordinary Meijer G}
The reduction uses the normalization of $\mathcal S$ in Eq.\,\eqref{sm:Snormal}. Define $S_{\rm{red}}$ by
\begin{equation}
  \mathcal S=-\frac{2^A\pi^2S_{\rm{red}}}{A(A\!-\!1)(A+2)\sinh(\pi\Delta)\sinh(\pi\Sigma)}
  =-\frac{2^{A-1}\pi\m^2|\mathrm B(a,b)|^2S_{\rm{red}}}{\tnu A(A\!-\!1)(A+2)\sinh(2\pi\tnu)}.
  \label{sm:Srednormal}
\end{equation}
The equality follows from Gamma reflection and the gaps in Eq.\,\eqref{sm:gapsdef}. Combining equal and mixed channels with the thermal weights in Eq.\,\eqref{sm:Stwokernels} gives
\begin{equation}
  S_{\rm{red}}=\mathcal N\!+\!\frac{\,16\,}{\,\pi^2(A\!+\!1)\,}
  \sum_{c=a,b}\!\sin\pi c\,
  \AC\int P_A(s)e_c(s)[\ii\omega\!+\!\cos\pi c\tan\pi s]\frac{\,\di s\,}{\,\ii\hp 2\pi\,}\,.
  \label{sm:commonresult}
\end{equation}
The weights combine through the elementary identities
\begin{equation}
  s_B\sin\pi b\!-\!\sin\pi a=\ii\omega,
  \qquad s_A\sin\pi a\!-\!\sin\pi b=\ii\omega\,.
  \label{sm:thermalsum}
\end{equation}
They follow from the hyperbolic addition formulas and Eqs.\,\eqref{eq:finitepars} and \eqref{sm:Stwokernels}.

Resolve the periodic multiplier into contributions from the two sides of the branch cut:
\begin{equation}
  \ii\omega\!+\!\cos\pi c\tan\pi s
  =\frac{\,\ii\,}{\,2\cos\pi s\,}
  [(\omega\!-\!\cos\pi c)e^{\ii\pi s}\!+\!(\omega\!+\!\cos\pi c)e^{-\ii\pi s}]\,.
  \label{sm:sides}
\end{equation}
Reflection and recurrence give
\begin{subequations}
  \begin{align}
    \frac{\,\pi P_A(s)e_c(s)\,}{\,\cos\pi s\,}
    ={}                      & \frac{\,(s\!+\!h)\Gamma(-s)\Gamma(1\!-\!s)\Gamma(-1/2\!-\!s)
    \Gamma(1\!+\!A/2\!+\!s)\Gamma((A\!+\!1)/2\!+\!s)\Gamma(1\!+\!s)\Gamma(c\!+\!s)\,}{\,\Gamma(A\!+\!c\!+\!s)\,}\,,\label{sm:finalkernel} \\
    (s\!+\!h)\Gamma(1\!+\!s) & =\Gamma(2\!+\!s)\!+\!\frac{\,A(1\!-\!A)\,}{\,4\,}\Gamma(1\!+\!s)\,.
    \label{sm:integerpair}
  \end{align}
\end{subequations}
The two contiguous $G$ values retain the moving-pole contribution: their weights at $s=-c$ sum to $h-c$.\

The lists $\mathbf a_1,\mathbf a_2$ contain the upper Meijer-$G$ parameters and $\mathbf b_1,\mathbf b_2$ the lower parameters. We use the following Barnes definition for the grouped Meijer $G$ notation. Its argument $z$, later evaluated at $-1$, is distinct from the radial variable $z=k_3/(p+q)$ used in the soft integral:
\begin{equation}
  G_z\!\left[\begin{matrix}\mathbf a_1;\mathbf a_2\\\mathbf b_1;\mathbf b_2\end{matrix}\right]
  =\frac{\,1\,}{\,\ii\hp 2\pi\,}\int
  \frac{\,\prod_{b\in\mathbf b_1}\Gamma(b\!-\!s)\prod_{a\in\mathbf a_1}\Gamma(1\!-\!a\!+\!s)\,}{\,\prod_{b\in\mathbf b_2}\Gamma(1\!-\!b\!+\!s)\prod_{a\in\mathbf a_2}\Gamma(a\!-\!s)\,}z^s\di s\,.
  \label{sm:Gdefinition}
\end{equation}
The upward contour initially separates $s=b+n$ from $s=a-1-n$ for the indicated first lower and upper groups, $n\ge0$; elsewhere retain all crossed residues under continuation. Empty products are one, and $z^s=\exp(s\operatorname{Log}z)$ uses $-\pi<\arg z<\pi$. In Eq.\,\eqref{sm:finalkernel}, $\Gamma(-s),\Gamma(1-s),\Gamma(-1/2-s)$ give the first lower group $(0,1,-1/2)$; the other numerator factors give the four first-upper entries; $\Gamma(A+c+s)$ gives the final lower entry $1-A-c$. This is the parameter list of Eq.\,\eqref{eq:Hfinite}.\

For clarity, the two boundary values used in Eq.\,\eqref{eq:Tfinite} are
\begin{equation}
 \mathcal H_+(A,c)=\mathcal H(A,c),\qquad
 \mathcal H_-(A,c)=\overline{\mathcal H(\overline A,\overline c)},\qquad
 \operatorname{Log}(-1\pm\ii0)=\pm\ii\pi.
 \label{sm:Glips}
\end{equation}
The subscripts label the two boundary values. Complex conjugation exchanges the lips and conjugates the parameters; evaluating first at $A^*,c^*$ and then conjugating the value therefore gives the opposite lip at the original $A,c$. Equations\,\eqref{sm:commonresult}--\eqref{sm:Gdefinition} identify $S_{\rm{red}}$ with the bracket in Eq.\,\eqref{eq:softfinite}. Substituting Eq.\,\eqref{sm:Srednormal} into Eq.\,\eqref{sm:Snormal}, with $|\mathrm B|^2/\mathrm B=\overline{\mathrm B}$, supplies its prefactor and completes Eqs.\,\eqref{eq:softfinite}--\eqref{eq:Hfinite}, including the finite polynomial and branch prescription.

\section{Inflaton Correlator in the Large-Chemical-Potential Limit}
\label{sm:asym}
\label{App:F}
We derive the large-$\tlambda$ expansion of $\Cc$ and $\fNLCC$ from the complete hard and soft factors. The analysis starts at fixed gap $\Delta=\tnu-\tlambda$ and controls the remainder as the gap closes, permitting the limit at fixed nonzero mass. The resulting expression, Eq.\,\eqref{eq:asymnew}, makes explicit how the mass scaling and exponential suppression differ from the nonlocal approximation in Eq.\,\eqref{eq:asymold}.

\subsection{Fixed Gap and Fixed Mass}
To distinguish the two large-parameter limits of Eqs.\,\eqref{eq:hardfinite} and \eqref{eq:softfinite}--\eqref{eq:Hfinite}, define the positive gap
\begin{equation}
  \Delta=\tnu\!-\!\tlambda=\frac{\,\m^2\,}{\,\tnu\!+\!\tlambda\,},\qquad
  2\ii\tnu=A\!-\!2,\qquad a=-\ii\Delta,\qquad b=-2\ii\tnu\!-\!a,\qquad
  \m^2=a(2\ii\tnu\!+\!a)\,.
  \label{sm:asymgap}
\end{equation}
At fixed $\Delta>0$, taking $\tnu\to\infty$ makes the mass grow as $\m\sim\sqrt{2\tnu\Delta}$. At fixed $\m>0$, taking $\tlambda\to\infty$ instead closes the gap, $\Delta\to0$. A fixed-gap error estimate alone need not remain bounded on this second path. We therefore derive the fixed-gap expansion, establish a remainder bound uniform as the gap closes, and then specialize to fixed mass. In both cases, the leading squeezed coefficient has already been extracted at fixed physical parameters.

The leading contribution and its modulus are
\begin{subequations}
  \begin{align}
    \mathbb C_B^{(0)}={} & -\frac{\,\m^2\,}{\,16\pi\Lambda^3\,}e^{-11\pi\Delta/2}
    (2\tnu)^{\ii\Delta}
    \frac{\,\Gamma(1\!+\!\ii\Delta)\,}{\,(1\!+\!\ii\Delta)^2\Gamma(2\!+\!\ii\hp 2\Delta)\,}\,,
    \label{sm:CBleading}                                                                                         \\
    G(\Delta)={}         & \frac{\,\sqrt{\cosh(\pi\Delta)}\,}{\,(1\!+\!\Delta^2)\sqrt{1\!+\!4\Delta^2}\,},\qquad
    f_B=\frac{\,5\pi\,}{\,18\,}\Pz\m^2\tlambda^3e^{-11\pi\Delta/2}G(\Delta)\,.
    \label{sm:gapamplitude}
  \end{align}
\end{subequations}
The subscript $B$ identifies the channel in Eq.\,\eqref{sm:channelTable}, with the common normalization of $\Cc$. To verify the modulus in Eq.\,\eqref{sm:gapamplitude}, use $|\Gamma(1\!+\!\ii x)|^2\!=\!\pi x/\sinh\pi x$ and $\Gamma(2\!+\!\ii\hp 2 x)\!=\!(1\!+\!\ii\hp 2 x)\Gamma(1\!+\!\ii\hp 2 x)$. In particular $G(0)\!=\!1$.

For $\Delta$ in a compact positive interval, the first additional term is
\begin{subequations}
  \begin{align}
    \frac{\,\Cc\,}{\,\mathbb C_B^{(0)}\,} & =1\!+\!\Xi(\Delta,\tnu)\!+\!O(\tnu^{-1})\,,\label{sm:gapexpansion} \\
    \Xi(\Delta,\tnu)                       & =
    \frac{\,\sqrt{\pi}\, e^{\pi\Delta}\tanh(\pi\Delta)(1\!+\!\ii\hp 2\Delta)\,}{\,9\sqrt{2\tnu}\,\Gamma(1\!-\!\ii\hp 2\Delta)\,}
    \exp\!\left\{\ii\left[\frac{\,3\pi\,}{\,4\,}\!-\!4\tnu\log2\!-\!2\Delta\log\tnu\right]\right\}\,.
    \label{sm:Xi}
  \end{align}
\end{subequations}
The correction $\Xi$ describes interference at the level of the complex coefficient. Since $\Xi=O(\tnu^{-1/2})$ at fixed gap, $|1+\Xi|=1+\Re\Xi+O(\tnu^{-1})$: its real part gives the first relative amplitude correction, while its imaginary part gives the first phase shift. The modulus is taken only after combining these complex contributions.

\subsection{Endpoint and Stationary-Phase Contributions}
The two orders in Eq.\,\eqref{sm:gapexpansion} arise from different integration regions.\  Using the square moments $\mathcal J$ of Eq.\,\eqref{sm:squaremoment}, set
\begin{equation}
  \mathcal U_c=\mathcal J_{c,c}[w_U],\qquad
  \mathcal V=\mathcal J_{a,b}[w_V],\qquad
  \mathcal S=-2s_B\mathcal U_a\!-\!2s_A\mathcal U_b\!+\!\mathcal V\,.
  \label{sm:asymmoments}
\end{equation}
These are the three exact terms of the soft result.\  Here $K(a)$ is the leading endpoint coefficient. With $Y\!=\!2\tnu$, their asymptotic estimates are
\begin{subequations}
  \begin{align}
    -2s_B\mathcal U_a & =2e^{-2\pi\Delta}K(a)A^{-1-2a}[1\!+\!O(Y^{-1})],
                      & K(a)                                                                            & =-\frac{\,4a\Gamma(a)^2\Gamma(1\!-\!a)^2\,}{\,\Gamma(2\!-\!2a)\,}\,,\label{sm:equalasym} \\
    -2s_A\mathcal U_b & =\frac{\,8\pi^{\frac{3}{2}}\,}{\,9\,}\,4^{-a}e^{-\ii\hp 2\pi a}Y^{-\frac{3}{2}}
    e^{-\ii\pi/4-\ii\hp 2 Y\log2}\!+\!O(Y^{-2}),
                      & \mathcal V                                                                      & =O(Y^{-3})\,.\label{sm:saddleasym}
  \end{align}
\end{subequations}
The leading term comes from an endpoint of the Euler kernel; the next term comes from a saddle of the continued Barnes kernel.\  Dividing the second by the first and multiplying the exact common normalization gives Eq.\,\eqref{sm:Xi}.

The saddle can be found analytically.\  Use a contiguous relation to remove the equal-$b$ kernel's pole at $s\!=\!-A/2$, then shift its Barnes contour to $\Re s\!=\!-7/6$. The crossed poles at $s\!=\!-1$ and $s\!=\!-b\!-\!1$ contribute $O(Y^{-3})$ and an exponentially small term, respectively, including their original weights.\  With $s\!=\!-7/6\!+\!\ii Yr$, the unsuppressed interval is $-1/2<r<0$. The phase and stationary point are
\begin{subequations}
  \begin{gather}
    \varphi(r)=2g(-r)\!+\!2g(r\!+\!1/2)\!+\!g(r\!-\!1)\!-\!g(r),\qquad
    g(u)=u(\log|u|\!-\!1)\,,\label{sm:saddlephase}\\
    \varphi'(r)=\log\frac{\,(r\!+\!1/2)^2(1\!-\!r)\,}{\,(-r)^3\,},\qquad
    (r\!+\!1/2)^2(1\!-\!r)\!-\!(-r)^3=\frac{\,3r\!+\!1\,}{\,4\,}\,,\label{sm:saddlelocation}\\
    r_*=-\frac{\,1\,}{\,3\,},\qquad \varphi(r_*)=-3\log2,\qquad
    \varphi''(r_*)=\frac{\,81\,}{\,4\,}\,.
    \label{sm:saddledata}
  \end{gather}
\end{subequations}
The weighted amplitude at the saddle is $\sqrt{2}\,4^{-a}$. The usual one-dimensional stationary-phase formula gives the coefficient in Eq.\,\eqref{sm:saddleasym}, after restoring the Barnes prefactor.\  Away from $r_*$ its endpoint powers are $(r\!+\!1/2)^{1/6}$ and $(-r)^{\frac{1}{2}+a}$. They vanish at the endpoints and have integrable first derivatives.\  Integration by parts therefore bounds the nonsaddle piece by the stated $O(Y^{-2})$. In layers of width $1/Y$ near either endpoint, use the Gamma functions at their bounded arguments.\  Their contributions are $O(Y^{-19/6})$ and $O(Y^{-\frac{7}{2}})$. Outside this interval the Gamma exponential suppresses the contour.\  This specifies the endpoint control behind the saddle approximation.\

The mixed term requires its continued large-variable tail to be subtracted and restored, just as in Eq.\,\eqref{sm:finitepart}. On the damped imaginary ray its small-variable integral is $O(Y^{-1})$ before the $O(Y^{-2})$ prefactor.\  The other ray is multiplied by an exponentially small connection coefficient.\  The restored power tail is exponentially suppressed.\  Including the restored tail gives the mixed contribution $O(Y^{-3})$.\

\subsection{Uniform Approach to Zero Gap}
To reach the fixed-mass path, we must control the remainder as $a=-\ii\Delta\to0$. The endpoint coefficient $K(a)$ in Eq.\,\eqref{sm:equalasym} contains an explicit pole there. We isolate this known pole before estimating the remainder, so that the error bound can stay uniform. In the endpoint coordinates $u\!=\!1-R$, $v\!=\!1-S$, the subtraction uses the following identity for denominator powers $j\!=\!1,2,3$:
\begin{equation}
  c\int_0^1v^{c-1}(1\!-\!uv)^{-j}\di v
  =1\!+\!c\int_0^1v^{c-1}\big[(1\!-\!uv)^{-j}\!-\!1\big]\di v\,.
  \label{sm:polenormalize}
\end{equation}
The bracket vanishes linearly at $v\!=\!0$, so the right-hand side is regular near $c\!=\!0$. Fix an upper gap $D>0$. In $O_D$ the error constant may depend on $D$ but is uniform for $0<\Delta\leqslant D$. Applying the identity at $c\!=\!-\ii\Delta$ gives
\begin{equation}
  c\mathcal U_c=cK(c)A^{-1-2c}
  \left[1\!-\!\frac{\,(c\!+\!1)(2c\!+\!1)\,}{\,A\,}\right]
  \!-\!\frac{\,8c\Gamma(c)\,}{\,c\!-\!1\,}A^{-2-c}\!+\!O_D(Y^{-3}),
  \qquad cK(c)\longrightarrow-4\,.
  \label{sm:uniformendpoint}
\end{equation}
One way to bound the error is to subtract the two displayed endpoint powers from the explicit Euler kernel.\  The remaining amplitude and its first three derivatives have the corresponding cubic and quadratic endpoint bounds.\  Split the integration at $u\!=\!1/Y$; use those bounds below it and integrate by parts three times above it.\  Retaining all boundary terms gives $O_D(Y^{-3})$. The normalization by $c$ keeps the constants bounded as $c\to0$.

The apparent $1/Y$ error cancels at zero gap.\  Before expanding at small $a$, write the channel-$B$ ratio as
\begin{subequations}
  \begin{align}
    \frac{\,\mathbb C_B\,}{\,\mathbb C_B^{(0)}\,}
           & =1\!+\!\frac{\,g_1(a)\,}{\,2\ii\tnu\,}\!+\!h_1(a)(2\ii\tnu)^{a-1}\!+\!O_D(Y^{-2})\,,\label{sm:cancellationexpansion} \\
    g_1(a) & =1\!-\!\frac{\,19a\,}{\,2\,}\!-\!\frac{\,3a^2\,}{\,2\,}\,,\label{sm:g1Definition}                    \\
    h_1(a) & =-\frac{\,2\Gamma(2\!-\!2a)\,}{\,(1\!-\!a)\Gamma(1\!+\!a)\Gamma(1\!-\!a)^2\,}
    \!+\!e^{\ii\pi a}\Gamma(2\!-\!a),\qquad g_1(0)=1,\quad h_1(0)=-1\,.
    \label{sm:cancellationcoefficients}
  \end{align}
\end{subequations}
The bound $|(2\ii\tnu)^a\!-\!1|\leqslant C_D\Delta(1\!+\!\log Y)$ combines the two terms into $O_D(\Delta(1\!+\!\log Y)/Y)$, with a constant controlled at $\Delta\!=\!0$. The pole-normalized mixed term and the nonsaddle remainder satisfy the same uniform bounds.\  Hence
\begin{equation}
  \frac{\,\Cc\,}{\,\mathbb C_B^{(0)}\,}=
  1\!+\!\Xi(\Delta,\tnu)
  \!+\!O_D\!\left(\frac{\,\Delta(1\!+\!\log\tnu)\,}{\,\tnu\,}\!+\!\frac{\,1\,}{\,\tnu^2\,}\right)\,.
  \label{sm:uniformasym}
\end{equation}
All powers of $2\ii\tnu$ use $\log(2\ii\tnu)\!=\!\log(2\tnu)\!+\!\ii\pi/2$, as fixed by the physical continuation.\

\subsection{Fixed Mass and the Earlier Nonlocal Approximation}
For fixed $0<\m\leqslant M$, with $M$ an upper bound on the dimensionless mass, insert the following expansions. The constants in $O_M$ may depend on $M$:
\begin{equation}
  \Delta=\frac{\,\m^2\,}{\,2\tlambda\,}\!+\!O_M(\tlambda^{-3}),\qquad
  G(\Delta)=1\!+\!O_M(\tlambda^{-2})
  \label{sm:massgap}
\end{equation}
into the uniform result.\  The amplitude is
\begin{equation}
  \fNLCC=\frac{\,5\pi\,}{\,18\,}\Pz\m^2\tlambda^3e^{-11\pi\m^2/(4\tlambda)}
  \left[1\!+\!\frac{\,\pi^{\frac{3}{2}}\m^2\,}{\,9(2\tlambda)^{\frac{3}{2}}\,}
    \cos\!\left(4\tlambda\log2\!-\!\frac{\,3\pi\,}{\,4\,}\right)
    \!+\!O_M\!\left(\frac{\,\log\tlambda\,}{\,\tlambda^2\,}\right)\right]\,.
  \label{sm:massasym}
\end{equation}
The bracket in Eq.\,\eqref{sm:massasym} tends to one at fixed nonzero mass. Retaining its leading term gives
\begin{equation}
 \fNLCC\simeq\frac{5\pi}{18}\Pz\m^2\tlambda^3 e^{-11\pi\m^2/(4\tlambda)}\,.
 \label{sm:asymMain}
\end{equation}
This is the approximation used in Eq.\,\eqref{eq:asymnew}.

In particular, $\Cc\sim-\m^2/(16\pi\Lambda^3)$ with the phase convention of Eq.\,\eqref{eq:target}. The mass factor vanishes in the massless limit.\

For comparison, Ref.\cite{You:2026CC} gives the late-time nonlocal coefficient for one oscillatory power. Denoting it by $\mathcal A_{\ominus}^{\rm{tot}}$ as in that work, its modulus and the amplitude in our convention obey
\begin{equation}
 |\mathcal A_{\ominus}^{\rm{tot}}|\simeq
 \frac{H^3\pi^{7/2}\Pz^{3/2}\m^4\tlambda^{3/2}}{32\sqrt{2}}e^{-9\pi\Delta/2},
 \qquad f_{\rm{NL}}^{\rm{NLoc}}=\frac{5}{9\pi H^3\sqrt{\Pz}}|\mathcal A_{\ominus}^{\rm{tot}}| .
 \label{sm:previousInput}
\end{equation}
Here $|\mathcal A_{\ominus}^{\rm{tot}}|$ includes the overall $H^6$ in Eq.\,\eqref{eq:target}. Multiplication cancels $H^3$; using the gap expansion in Eq.\,\eqref{sm:massgap} then gives
\begin{equation}
  f_{\rm{NL}}^{\rm{NLoc}}\simeq
  \frac{\,5\pi^{\frac{5}{2}}\,}{\,288\sqrt{2}\,}\Pz\m^4\tlambda^{\frac{3}{2}}e^{-9\pi\Delta/2}
  \simeq\frac{\,5\pi^{\frac{5}{2}}\,}{\,288\sqrt{2}\,}\Pz\m^4\tlambda^{\frac{3}{2}}
  e^{-9\pi\m^2/(4\tlambda)}\,.
  \label{sm:previousasym}
\end{equation}
The last expression is the fixed-mass result used in Eq.\,\eqref{eq:asymold}.\  The ratio of the displayed leading approximants is
\begin{equation}
  \frac{\,f_B\,}{\,f_{\rm{NL}}^{\rm{NLoc}}\,}\simeq
  \frac{\,16\sqrt{2}\,}{\,\pi^{\frac{3}{2}}\,}\frac{\,\tlambda^{\frac{3}{2}}\,}{\,\m^2\,}e^{-\pi\Delta}G(\Delta)\,.
  \label{sm:asymratio}
\end{equation}
This ratio compares the leading asymptotic amplitudes. The finite formula determines the mass-dependent interference.\  Fig.\ref{fig:fnl} compares the two finite-parameter results with their large-$\tlambda$ approximations.\

\subsection{Origin of the Mass Powers and Exponential Weights}
\label{sm:factorOrigins}
We compare magnitudes at fixed $\m\!>\!0$ as $\tlambda\to\infty$.
Since $\Delta\!=\!\m^2/\Sigma$ and $\Sigma\!=\!\tnu\!+\!\tlambda$
[Eq.\,\eqref{sm:gapsdef}], a term linear in $\Delta$ in the logarithm of a
magnitude is of order $1/\tlambda$. We retain this term in the exponential,
as in Eqs.\,\eqref{eq:asymnew} and \eqref{eq:asymold}.
To read off the powers quoted in Sec.~5, we hold $H$ and $\dot\phi_0\!>\!0$
fixed and vary the cutoff. Since $\tlambda\!=\!\lambda/H$ in Eq.\,\eqref{eq:physical},
we use $\Lambda\!=\!\frac{\,\dot\phi_0\,}{\,\lambda\,}\!=\!\frac{\,\dot\phi_0\,}{\,H\tlambda\,}$
in the last step for each factor.

\emph{Hard vertices.} The two suppression factors in Eq.\,\eqref{eq:hardfinite}
can be read off separately. Applying $|\Gamma(\ii x)|^2\!=\!\pi/[x\sinh(\pi x)]$
to the three Gamma functions in $\mathrm B(a,b)$ gives
\begin{equation}
 \frac{1}{|\mathrm B(a,b)|^2}
 =\frac{\Delta\Sigma}{2\pi\tnu}\,
  \frac{\sinh(\pi\Delta)\sinh(\pi\Sigma)}{\sinh(2\pi\tnu)}
 =\Delta^2 e^{-\pi\Delta}\big[1+O_{\m}(\tlambda^{-2})\big].
 \label{sm:inverseBetaGap}
\end{equation}
Here $\Sigma\!=\!2\tnu\!-\!\Delta$; the other exponential comes from
$\sinh(\pi\tlambda)/\sinh(\pi\tnu)\simeq e^{-\pi\Delta}$ in the first
term of the square bracket in Eq.\,\eqref{eq:hardfinite}.
Its leading polynomial is $A^2\sim-4\tnu^2$.
To check the second term, the Gamma ratio in $\mathrm B(a,b)$ gives
$\mathrm B(a,b)\!=\!(\ii/\Delta)[1+O_{\m}(\log\tnu/\tnu)]$.
The second term divided by the first is then
$-\ii/(2\tnu)+O_{\m}(\log\tnu/\tnu^2)$: at order $1/\tnu$ it changes
only the phase. The rational factors outside the bracket also have no
linear correction to their moduli. Thus
\begin{equation}
 |\Ch|\simeq\frac{\pi\tnu(2\tnu)^2}{8\Lambda^2}\,\Delta^2 e^{-2\pi\Delta}
 \simeq\frac{\pi\m^4\tlambda}{8\Lambda^2}\,e^{-2\pi\Delta}
 =\frac{\,\pi H^2\m^4\tlambda^3\,}{\,8\dot\phi_0^2\,}\,e^{-2\pi\Delta}.
 \label{sm:hardGapWeight}
\end{equation}

\emph{Complete soft vertex.} Equation\,\eqref{sm:Snormal} divides the soft
moment by $8\pi^2\Lambda\m^2\mathrm B(a,b)$. Its leading endpoint in
Eq.\,\eqref{sm:equalasym} is $2e^{-2\pi\Delta}K(a)A^{-1-2a}$:
the first exponential is the soft SK weight $-s_B$ of Eq.\,\eqref{sm:Stwokernels}.
The other factors follow from $a\!=\!-\ii\Delta$ and $\arg A\to\pi/2$:
\begin{equation}
 |K(a)|=\frac{4}{\Delta}\,[1+O(\Delta^2)],\qquad
 |A^{-1-2a}|=\frac{e^{-2\Delta\arg A}}{|A|}\simeq\frac{e^{-\pi\Delta}}{2\tnu},
 \qquad |\mathrm B(a,b)|^{-1}\simeq\Delta e^{-\pi\Delta/2}.
 \label{sm:softGapPieces}
\end{equation}
The $1/\Delta$ in $K(a)$ cancels the $\Delta$ in $|\mathrm B|^{-1}$,
leaving the explicit $1/\m^2$ in the normalization. Multiplying the
three exponential weights gives
\begin{equation}
 |\Cs|\simeq
 \frac{2e^{-2\pi\Delta}}{8\pi^2\Lambda\m^2}\,
 |K(a)|\,|A^{-1-2a}|\,|\mathrm B(a,b)|^{-1}
 \simeq\frac{e^{-7\pi\Delta/2}}{2\pi^2\Lambda\m^2\tlambda}
 =\frac{\,H\,e^{-7\pi\Delta/2}\,}{\,2\pi^2\dot\phi_0\m^2\,}.
 \label{sm:completeSoftGapWeight}
\end{equation}
The two hard vertices supply two powers of $\tlambda$ through $\Lambda^{-2}$.
At the soft vertex the single factor $\Lambda^{-1}$ cancels the explicit
$\tlambda^{-1}$, leaving no chemical-potential power in the algebraic prefactor.
Equations\,\eqref{sm:uniformasym} and \eqref{sm:massgap} control the
omitted terms at fixed mass: $G(\Delta)$ has no linear term, and the
relative interference correction is $O_{\m}(\tlambda^{-3/2})$.
The same inverse mass power can be seen before summing the poles.
In Eq.\,\eqref{sm:LLBexact}, $|\Gamma(a)|\sim1/\Delta$ while the two
denominators $|t|$ and $|b|$ each grow as $2\tnu$. The first local pair
therefore gives $|\mathbb C_+^{\rm{LL}(0,0);++-}|/|\Ch|
\sim1/(8\pi^2\Lambda\tnu^2\Delta)\sim H/(4\pi^2\dot\phi_0\m^2)$, one half of the leading soft magnitude
in Eq.\,\eqref{sm:completeSoftGapWeight}. Resumming all higher local poles on the $p$ line at fixed $q$-line order zero
supplies the other half
[Eq.\,\eqref{sm:resummedlimits}].

\emph{Nonlocal late-time result.} Appendix~E of Ref.\cite{You:2026CC} groups its diagrams
differently. In its notation $C_\ell$ multiplies the hard subdiagram,
$C_{\mathcal{BR}}$ the loop momentum and soft vertex, and $\chi$ is the
relative factor in the four diagram pairs. With their Wick signs, the
four weights sum to $\chi+\chi+(\chi\!-\!1)+(1\!+\!\chi)\!=\!4\chi$.
The magnitudes of the three factors in that calculation are
\begin{equation}
 |C_\ell|\simeq 2\pi^{5/2}\Pz H^2\tlambda^{9/2},\qquad
 |C_{\mathcal{BR}}|\simeq
 \frac{\pi H\sqrt{\Pz}\,\m^4}{256\sqrt{2}\,\tlambda^3}e^{-3\pi\Delta},
 \qquad |\chi|\simeq e^{-3\pi\Delta/2}.
 \label{sm:previousNativeWeights}
\end{equation}
Thus $|\mathcal A_{\ominus}^{\rm{tot}}|\simeq4|C_\ell C_{\mathcal{BR}}\chi|$:
the mass and chemical-potential powers are $\m^4\tlambda^{9/2-3}$,
and the exponent is $-(3+3/2)\pi\Delta$, as in Eq.\,\eqref{sm:previousInput}.
To compare in the normalization of the present paper, remove $H^6$ as
in Eq.\,\eqref{eq:target} and divide by Eq.\,\eqref{sm:hardGapWeight}.
Using $H/\Lambda\!=\!2\pi\sqrt{\Pz}\,\tlambda$ gives
\begin{equation}
 \frac{|\mathcal A_{\ominus}^{\rm{tot}}|}{H^6|\Ch|}
 \simeq
 \frac{\sqrt{\pi}\,\m^4 e^{-9\pi\Delta/2}}{256\sqrt{2}\,\Lambda^3\tlambda^{3/2}}\,
 \frac{8\Lambda^2 e^{2\pi\Delta}}{\pi\m^4\tlambda}
 =\frac{e^{-5\pi\Delta/2}}{32\sqrt{2\pi}\,\Lambda\tlambda^{5/2}}
 =\frac{\,H\,e^{-5\pi\Delta/2}\,}{\,32\sqrt{2\pi}\,\dot\phi_0\tlambda^{3/2}\,}.
 \label{sm:previousSoftGapWeight}
\end{equation}
This quotient is the effective soft magnitude quoted in Sec.~5;
the original bubble--right factor itself carries $e^{-3\pi\Delta}$.
The product of the hard and soft factors now includes all three vertex couplings.
To obtain either $f_{\rm{NL}}$, Eq.\,\eqref{eq:fnl} multiplies this product
by $5H^3/(9\pi\sqrt{\Pz})\!=\!10H\dot\phi_0/9$, using the tree power in
Eq.\,\eqref{sm:scanbackground}. In terms of the inflaton velocity this gives
\begin{equation}
 \fNLCC\simeq
 \frac{\,5H^4\,}{\,72\pi\dot\phi_0^2\,}\,\m^2\tlambda^3 e^{-11\pi\Delta/2},
 \qquad
 f_{\rm{NL}}^{\rm{NLoc}}\simeq
 \frac{\,5\sqrt{\pi}\,H^4\,}{\,1152\sqrt{2}\,\dot\phi_0^2\,}\,
 \m^4\tlambda^{3/2} e^{-9\pi\Delta/2}.
 \label{sm:factorOriginNormalization}
\end{equation}
Writing $H^4/\dot\phi_0^2\!=\!4\pi^2\Pz$ and
$\Delta\simeq\m^2/(2\tlambda)$ recovers Eqs.\,\eqref{eq:asymnew}
and \eqref{eq:asymold}, respectively.

\section{Analyzing Inflaton Correlator through the Mellin Poles}
\label{sm:poles}
\label{App:G}
We explain the large-chemical-potential result through the Mellin poles of the two soft propagators. At fixed nonzero mass, the first local-pole pair on the $++-$ SK path asymptotically supplies half of the full complex coefficient $\Cc$. Resumming the higher local poles on the $p$ line at finite parameters, with the $q$ line kept at its first local pole, supplies the other half in the same limit.

\subsection{The Poles of One Soft Propagator}
Pole ratios compare complex coefficients, with the complete $\Cc$ in the denominator unless a single-channel denominator is stated. The limits $1/2$ and $1$ are taken at fixed $\m>0$ as $\tlambda\to\infty$, with a $++-$ numerator. Finite-parameter amplitudes and phases appear in Appendix\,\ref{sm:checks}.\

To see where the two pole rows come from, first take $X>0$ and $\Re(\tfrac12-\kappa)>0$. The Whittaker relation to Tricomi's function gives $e^{X/2}W_{\kappa,\ii\tnu}(X)/\sqrt X=X^{\ii\tnu}U(\tfrac12-\kappa+\ii\tnu,1+2\ii\tnu,X)$. Insert the Euler integral for $U$ and take the Mellin transform in $X$. For $0<\Re s<\Re(\tfrac12-\kappa)$, the $X$ integral gives $\Gamma(s+\ii\tnu)$ times a power of the Euler variable; its remaining integral is a Beta function:
\begin{equation}
 \int_0^\infty X^{s-1}\frac{e^{X/2}W_{\kappa,\ii\tnu}(X)}{\sqrt X}\,\di X
 =\frac{\Gamma(s+\ii\tnu)}{\Gamma(\tfrac12-\kappa+\ii\tnu)}
 \mathrm B(\tfrac12-\kappa-s,\,s-\ii\tnu).
 \label{sm:Wtransform}
\end{equation}
Writing the Beta function as a Gamma ratio and inverting the transform gives
\begin{equation}
 W_{\kappa,\ii\tnu}(X)=\frac{e^{-X/2}}
 {\Gamma(\tfrac12-\kappa+\ii\tnu)\Gamma(\tfrac12-\kappa-\ii\tnu)}
 \int_{\mathcal C_{\rm {MB}}}\frac{\di s}{\ii\hp 2\pi}\,
 \Gamma(s+\ii\tnu)\Gamma(s-\ii\tnu)\Gamma(\tfrac12-\kappa-s)\,X^{1/2-s}.
 \label{sm:WmellinMain}
\end{equation}
This is the representation in Eq.\,\eqref{eq:Wmellin}. The contour $\mathcal C_{\rm {MB}}$ initially runs upwards along a vertical line in the stated strip; continuing to the mode indices of Eq.\,\eqref{sm:modes} preserves the separation of pole families by contour deformation or, equivalently, includes the crossed residues.

For the positive clock, set $s_h=-\ii\tnu$ at the hard endpoint. The two factors $1/\sqrt{-2p\tau}$ in the spinor modes remove the half powers. With coefficients and exponentials factored out, the remaining product is $(2p)^{-s_h-s_s}(-\tau_h)^{-s_h}(-\tau_3)^{-s_s}$. Inserting the first pole from each of the two soft rows gives
\begin{equation}
 \begin{aligned}
 s_s=+\ii\tnu:\quad&(\tau_h/\tau_3)^{\ii\tnu}\quad(\mathrm{local}),\\
 s_s=-\ii\tnu:\quad&(4p^2\tau_h\tau_3)^{\ii\tnu}\quad(\mathrm{nonlocal}).
 \end{aligned}
 \label{sm:pairingsMain}
\end{equation}
These are the two powers in Eq.\,\eqref{eq:endpointpairings}: opposite Mellin signs cancel the loop momentum, whereas equal signs retain $p^{2\ii\tnu}$.

Shifting the soft pole by a nonnegative integer $n$ in Eq.\,\eqref{sm:WmellinMain} gives the local row $s_s\!=\!+\ii\tnu\!-\!n$ and the nonlocal row $s_s\!=\!-\ii\tnu\!-\!n$. Fig.\ref{fig:poles} marks the nonlocal first pole retained in Ref.\cite{You:2026CC} and the omitted terms. Both soft rows remain available for either fixed external clock.

Removing the common Frobenius power separates the two rows. In channel $j$, the parameter $c$ is $\alpha_j$ on the $p$ line and $\beta_j$ on the $q$ line [Eq.\,\eqref{sm:channelTable}].
\begin{subequations}
  \begin{align}
    \Phi_c(X) & =e^{-X/2}U(c,1\!-\!2\ii\tnu,X)\nonumber                            \\
              & =e^{-X/2}\left[
      b_c(0)\,{}_1F_1(c;1\!-\!2\ii\tnu;X)
    \!+\!b_c(2\ii\tnu)X^{2\ii\tnu}\,{}_1F_1(c\!+\!2\ii\tnu;1\!+\!2\ii\tnu;X)\right]\,,\label{sm:fullwave} \\
    b_c(0)    & =\frac{\,\Gamma(2\ii\tnu)\,}{\,\Gamma(c\!+\!2\ii\tnu)\,},\qquad
    b_c(2\ii\tnu)=\frac{\,\Gamma(-2\ii\tnu)\,}{\,\Gamma(c)\,}\,.
    \label{sm:firstresidues}
  \end{align}
\end{subequations}
The terms are respectively the local and nonlocal rows for the chosen hard pole. Here $X=2pz_3/k_3$ or $2qz_3/k_3$. A first-pole truncation sets the corresponding ${}_1F_1$ to one but retains $e^{-X/2}$; the late-time truncation also sets that exponential to one. They agree at small $X$ but have different soft-time integrals at finite $X$. Ref.\cite{You:2026CC} retains the nonlocal first pole with this additional late-time step.\

The first parameters of the two functions in each channel are listed in Eq.\,\eqref{sm:channelTable}. In particular, channel $B$ has $1+a$ on the $p$ line and $a$ on the $q$ line; this difference determines their unequal descendant weights.\

\subsection{A Finite Formula for the Local First Poles on One SK Path}\label{sm:SKchoice}
To measure the contribution of the first local-pole pair, retain its exponential on each soft line and select the $++-$ assignment. The $++$ hard propagator still includes both time orderings. Let $h_{ab}$ denote the hard weight for branches $a,b$, $h_{\rm{sum}}$ their sum, and $Z_h$ the common Gamma-function factor:
\begin{subequations}\label{sm:SKseeds}
  \begin{gather}
    \alpha=\ii(\tnu\!-\!\tlambda)=-a,\qquad \beta=\ii(\tnu\!+\!\tlambda)=-b,\qquad
    Z_h=\frac{\,e^{-\pi(\tnu-\tlambda)}\Gamma(2\!+\!2\ii\tnu)\,}{\,\Gamma(2\!+\!\alpha)\Gamma(2\!+\!\beta)\,}\,,\label{sm:SKparameters}\\
    h_{++}=\ii Z_h\big[(1\!+\!\beta)^2\!-\!e^{-2\pi\tlambda}(1\!+\!\alpha)^2\big]\!-\!2\tlambda\,,\label{sm:SKpp}\\
    h_{--}=\ii Z_h\big[(1\!+\!\alpha)^2\!-\!e^{-2\pi\tlambda}(1\!+\!\beta)^2\big]\!+\!2\tlambda e^{-2\pi\tnu}\,,\label{sm:SKmm}\\
    h_{+-}=h_{-+}=\ii Z_h(1\!-\!e^{-2\pi\tlambda})(1\!+\!\alpha)(1\!+\!\beta),\qquad
    h_{\rm{sum}}=h_{++}\!+\!h_{--}\!+\!h_{+-}\!+\!h_{-+}\,.
    \label{sm:SKmixed}
  \end{gather}\end{subequations}
The ratio $h_{++}/h_{\rm{sum}}$ isolates the $++$ hard assignment. Its dominance follows directly from the powers of $\beta$ in Eq.\,\eqref{sm:SKseeds}: at fixed nonzero mass, $\alpha\to0$, $\beta\sim2\ii\tlambda$ and $Z_h\to1$. Thus $h_{++}\sim-4\ii\tlambda^2$, $h_{+-}\!=\!h_{-+}\sim-2\tlambda$ and $h_{--}\sim\ii$, giving
\begin{equation}
  \frac{\,h_{+-}\,}{\,h_{++}\,}=\frac{\,h_{-+}\,}{\,h_{++}\,}\sim-\frac{\,\ii\,}{\,2\tlambda\,},\qquad
  \frac{\,h_{--}\,}{\,h_{++}\,}\sim-\frac{\,1\,}{\,4\tlambda^2\,}\,.
  \label{sm:hardSKhierarchy}
\end{equation}
Hence $h_{++}/h_{\rm{sum}}\to1$. The suppression of the other hard assignments is algebraic; the finite expression below retains their exact relative weights.\

Let $\mathcal R_A,\mathcal R_B$ be the channel ratios defined in Eq.\,\eqref{sm:hardratios}. With both soft paths already summed, these weights are
\begin{equation}
  W_A=\mathcal R_A\Ch,\quad W_B=\mathcal R_B\Ch,\quad W_C=\Ch,\quad W_D=\Ch/\m^2\,.
  \label{sm:Wweights}
\end{equation}
The weight on the negative soft branch is $W_j^{(-)}$; $d_j$ sets the relative exponential weight of the two branches:
\begin{equation}
  W_j^{(-)}=\frac{\,W_j\,}{\,1\!-\!e^{2\pi d_j}\,},\qquad
  (d_A,d_B,d_C,d_D)=(\tnu\!+\!2\tlambda,\tnu\!-\!2\tlambda,\tnu,\tnu)\,.
  \label{sm:negativeweights}
\end{equation}
At a removable zero, evaluate the combined expression by continuity before separating its numerator and denominator. The positive-soft-path weights are $W_j^{(+)}\!=\!W_j\!-\!W_j^{(-)}$. Since the same local-first-pole moment multiplies either weight, Eq.\,\eqref{sm:negativeweights} gives
\begin{equation}
  \frac{\,W_j^{(+)}\,}{\,W_j^{(-)}\,}=-e^{2\pi d_j},\qquad
  \frac{\,W_B^{(+)}\,}{\,W_B^{(-)}\,}=-e^{-2\pi(2\tlambda\!-\!\tnu)}\longrightarrow0\,.
  \label{sm:softSKhierarchy}
\end{equation}
Since $\tnu-2\tlambda=-\tlambda+O_{\m}(\tlambda^{-1})$, the $B$ channel favors the negative soft branch. The channel ratios below check that the other components do not restore a constant term on the positive branch.\

The required local-first-pole coefficient is the finite Gamma expression
\begin{equation}
  \mathbb C_+^{\rm{LL}(0,0);++-}=
  \frac{\,h_{++}\,}{\,h_{\rm{sum}}\,}\frac{\,\Gamma(1\!-\!2\ii\tnu)\,}{\,2\pi^2\Lambda\,}
  \left\{\frac{\,1\!-\!2\ii\tnu\,}{\,2\ii\tnu(1\!+\!2\ii\tnu)\,}
  \big[W_A^{(-)}b_{1\!+\!b}(0)b_b(0)\!+\!W_B^{(-)}b_{1\!+\!a}(0)b_a(0)\big]
  \!-\!\frac{\,W_D^{(-)}\,}{\,3\,}b_a(0)b_b(0)\right\}\,.
  \label{sm:LLfinite}
\end{equation}
To perform the time and angular integrals, set $y\!=\!(p\!+\!q)/k_3\ge1$, $w\!=\!(q\!-\!p)/k_3\in[-1,1]$ and $v\!=\!z_3$. The external line and two retained exponentials give $e^{-(1+y)v}$. In channel $B$, the spin polynomial is $2(y^2\!-\!1)[w\!+\!(1\!+\!w)v]$. Its part odd in $w$ integrates to zero, leaving
\begin{equation}
  2\Gamma(2\!-\!2\ii\tnu)\FP\int_1^\infty\frac{\,y^2\!-\!1\,}{\,(1\!+\!y)^{2-2\ii\tnu}\,}\di y
  =\frac{\,2^{2+2\ii\tnu}\Gamma(2\!-\!2\ii\tnu)\,}{\,2\ii\tnu(1\!+\!2\ii\tnu)\,}\,.
  \label{sm:LLmoment}
\end{equation}
Put $s\!=\!1\!+\!y$ and integrate $s^{2\ii\tnu}\!-\!2s^{2\ii\tnu-1}$ first for $\Re(2\ii\tnu)<-1$, then continue.\  Channel $A$ has the same angular moment.\  For $C,D$, the identity
\begin{equation}
  b_{1\!+\!a}(0)b_{1\!+\!b}(0)=-\frac{\,b_a(0)b_b(0)\,}{\,\m^2\,}
  \label{sm:CDcancel}
\end{equation}
leaves the difference of their spin polynomials, $-4(1\!-\!w^2)v$. Its moment gives the last term of Eq.\,\eqref{sm:LLfinite}. This explains the cancellation before any large-parameter approximation.\

\subsection{Half of the Coefficient from the First Local Pair}
Gamma reflection puts the $B$ contribution to Eq.\,\eqref{sm:LLfinite} in the form
\begin{equation}
  \left.\mathbb C_+^{\rm{LL}(0,0);++-}\right|_B=
  \frac{\,h_{++}\,}{\,h_{\rm{sum}}\,}
  \frac{\,\Ch e^{-\pi(\tnu-2\tlambda)}\,}{\,4\pi^2\Lambda\sinh\pi\tnu\,}
  \frac{\,1\!-\!2\ii\tnu\,}{\,2\ii\tnu(1\!+\!2\ii\tnu)\,}
  \frac{\,\Gamma(a)\Gamma(2\ii\tnu)\,}{\,\Gamma(-b)\,}
  \frac{\,\sin\pi b\,}{\,b\sin(\pi 2\ii\tnu)\,}\,.
  \label{sm:LLBexact}
\end{equation}
Writing $r_T\!=\!e^{-2\pi\tlambda}\sinh(\pi\Delta)/\sinh[\pi(\tnu\!+\!\tlambda)]$, let $A/B$ and $(C+D)/B$ denote ratios of channel contributions to the same $\mathrm{LL}(0,0);++-$ coefficient in Eq.\,\eqref{sm:LLfinite}. They are
\begin{equation}
  \frac{\,A\,}{\,B\,}=r_T^2,\qquad
  \frac{\,C\!+\!D\,}{\,B\,}=r_T\frac{\,2\ii\tnu(1\!+\!2\ii\tnu)\,}{\,3(1\!-\!2\ii\tnu)\,}\,.
  \label{sm:LLotherchannels}
\end{equation}
Both vanish exponentially at fixed mass. To check the positive soft branch including all spinor channels, multiply these terms by the corresponding factors $-e^{2\pi d_j}$ in Eq.\,\eqref{sm:softSKhierarchy}. The common hard weight cancels, so
\begin{equation}\begin{aligned}[b]
  \frac{\,\mathbb C_+^{\rm{LL}(0,0);+++}\,}{\,\mathbb C_+^{\rm{LL}(0,0);++-}\,}
  &=-\frac{\,e^{2\pi(\tnu\!-\!2\tlambda)}\!+\!e^{2\pi(\tnu\!+\!2\tlambda)}r_T^2
  \!+\!e^{2\pi\tnu}r_T\frac{\,2\ii\tnu(1\!+\!2\ii\tnu)\,}{\,3(1\!-\!2\ii\tnu)\,}\,}
  {\,1\!+\!r_T^2\!+\!r_T\frac{\,2\ii\tnu(1\!+\!2\ii\tnu)\,}{\,3(1\!-\!2\ii\tnu)\,}\,}\\
  &=O_{\m}(e^{-2\pi\tlambda})\,.
  \label{sm:allsoftSKratio}
\end{aligned}\end{equation}
Here $r_T\!=\!O_{\m}(\tlambda^{-1}e^{-4\pi\tlambda})$ and $2\ii\tnu\!=\!O(\tlambda)$. Equations\,\eqref{sm:hardSKhierarchy} and \eqref{sm:allsoftSKratio} therefore establish that $++-$ is the only SK assignment of this double-local first-pole term with a nonzero constant limit for $\Cc$ at fixed $\Lambda$. The conjugate clock has the reversed assignment $--+$.\  Now $\Gamma(a)\sim\ii/\Delta$, $\Gamma(2\ii\tnu)/\Gamma(-b)\to1$, $\sin\pi b/\sin(\pi 2\ii\tnu)\to-1$, and the hard factor satisfies
\begin{equation}
  \Ch\sim\frac{\,\ii\pi\tnu(2\tnu)^2\Delta^2\,}{\,8\Lambda^2\,},\qquad
  \frac{\,h_{++}\,}{\,h_{\rm{sum}}\,}\to1\,.
  \label{sm:LLhardlimit}
\end{equation}
Substitution gives
\begin{equation}
  \mathbb C_+^{\rm{LL}(0,0);++-}\sim-\frac{\,\tnu\Delta\,}{\,16\pi\Lambda^3\,}
  \longrightarrow-\frac{\,\m^2\,}{\,32\pi\Lambda^3\,},\qquad
  \frac{\,\mathbb C_+^{\rm{LL}(0,0);++-}\,}{\,\Cc\,}\longrightarrow\frac{\,1\,}{\,2\,}\,.
  \label{sm:LLhalf}
\end{equation}
At fixed $\m>0$ and $\Lambda$, $\tnu\Delta=\m^2\tnu/(\tnu+\tlambda)\to\m^2/2$. For the full coefficient, $G(0)=1$ fixes the modulus, while the phase factors in Eq.\,\eqref{sm:CBleading} tend to one, including $(2\tnu)^{\ii\Delta}$ since $\Delta\ln(2\tnu)\to0$. With the error in Eq.\,\eqref{sm:uniformasym} tending to zero, the two results are
\begin{equation}
 \lim_{\tlambda\to\infty}\mathbb C_+^{\rm{LL}(0,0);++-}=-\frac{\m^2}{32\pi\Lambda^3},\qquad
 \lim_{\tlambda\to\infty}\Cc=-\frac{\m^2}{16\pi\Lambda^3}.
 \label{sm:halfMain}
\end{equation}
These are the two complex limits in Eq.\,\eqref{eq:half}; the ratio is one half, including the phase. The finite-parameter contributions are evaluated below.\

\subsection{Higher-Order Terms and Finite Resummation before the Limit}
In channel $B$, $a=-\ii\Delta\to0$ at fixed mass. Local-pole orders $n_p,n_q\ge0$ contribute

\begin{equation}
  \frac{\,(1\!+\!a)_{n_p}(a)_{n_q}\,}{\,n_p!\,n_q!\,(1\!-\!2\ii\tnu)_{n_p}(1\!-\!2\ii\tnu)_{n_q}\,}
  (y\!-\!w)^{n_p}(y\!+\!w)^{n_q} v^{n_p+n_q}\,.
  \label{sm:higherordercoefficient}
\end{equation}
For fixed $n_p$, $(1\!+\!a)_{n_p}/n_p!\to1$. For each fixed $n_q\ge1$, $(a)_{n_q}/n_q!\sim a/n_q$. The time integral supplies $\Gamma(1+n_p+n_q\!-\!2\ii\tnu)$ or $\Gamma(2+n_p+n_q\!-\!2\ii\tnu)$, which cancels the leading powers from $(1\!-\!2\ii\tnu)_{n_p}(1\!-\!2\ii\tnu)_{n_q}$. At the first nonzero radial endpoint term, $y\!=\!1$ and $x\!=\!p/k_3\!=\!(1\!-\!w)/2$. Its normalized angular moment is
\begin{equation}
  \frac12\int_{-1}^{1}\left(\frac{1\!-\!w}{2}\right)^{n_p}\left(\frac{1+w}{2}\right)^{n_q+1}\di w
  =\int_0^1x^{n_p}(1\!-\!x)^{n_q+1}\di x=\mathrm B(n_p+1,n_q+2).
  \label{sm:higherorderAngular}
\end{equation}
For $n_q=0$ write $n=n_p$ as in the main text. Combining the angular moment with the two Pochhammer limits gives
\begin{subequations}
  \begin{align}
    \lim_{\tlambda\to\infty}\frac{\mathbb C_+^{\rm{LL}(n,0);++-}}{\Cc} & =\int_0^1(1\!-\!x)x^n\di x=\frac{1}{(n\!+\!1)(n\!+\!2)}\,,\label{sm:higherorderweights} \\
    \frac{\,\mathbb C_+^{\rm{LL}(n_p,n_q);++-}\,}{\,\Cc\,} & =
    \frac{\,a\,}{\,n_q\,}\mathrm B(n_p\!+\!1,n_q\!+\!2)\!+\!o_{\m,n_p,n_q}(|a|),\qquad n_q\ge1\,.
    \label{sm:qhigherorders}
  \end{align}
\end{subequations}
Equation\,\eqref{sm:higherorderweights} is the result used in Eq.\,\eqref{eq:ladder}; integration gives $1/(n+1)-1/(n+2)$. Both limits above are taken at fixed pole orders. The first nonzero radial derivative of $y^2-1$ at $y=1$ supplies the angular weight; its value removes the lower-order term. To establish the infinite-row result, restore one complete local confluent hypergeometric function ${}_1F_1$ at finite parameters before taking the limit. Its time integral is

\begin{equation}
  \int_0^\infty v^{\ell-2\ii\tnu}e^{-(1+y)v}{}_1F_1(c;1\!-\!2\ii\tnu;\xi v)\di v
  =\frac{\,\Gamma(1\!+\!\ell\!-\!2\ii\tnu)\,}{\,(1\!+\!y)^{1+\ell-2\ii\tnu}\,}
  {}_2F_1\!\left(c,1\!+\!\ell\!-\!2\ii\tnu;1\!-\!2\ii\tnu;\frac{\,\xi\,}{\,1\!+\!y\,}\right),\quad \ell=0,1\,.
  \label{sm:singlelinetime}
\end{equation}
For the $p$ line, $c\!=\!1\!+\!a$ and $\xi\!=\!y\!-\!w$; for the $q$ line, $c\!=\!a$ and $\xi\!=\!y\!+\!w$. Put $Y\!=\!1\!+\!y$ and $d\!=\!Y\!-\!\xi$. Then $Y\ge2$ and $0\leqslant d\le2$. The Gauss function equals $(d/Y)^{-c}$ for $\ell\!=\!0$; for $\ell\!=\!1$ it has the additional factor $1\!+\!c(Y\!-\!d)/[(1\!-\!2\ii\tnu)d]$. With $P_0\!+\!vP_1\!=\!2(y^2\!-\!1)[w\!+\!(1\!+\!w)v]$, the remaining integral, apart from its common Gamma and normalization factors, is
\begin{equation}
  \int_2^\infty\di Y\,Y^{2\ii\tnu+c-1}\int_0^2\di d\,d^{-c}
  \left[P_0\!+\!\frac{\,cP_1\,}{\,d\,}\!+\!\frac{\,(1\!-\!2\ii\tnu\!-\!c)P_1\,}{\,Y\,}\right]\,.
  \label{sm:singlelinepowers}
\end{equation}
Every term is now a product of elementary powers.\  For a monomial $Y^h d^j$, including the common factor $2^{-1-2\ii\tnu}$, its continued integral is $-2^{h+j}/[(2\ii\tnu\!+\!c\!+\!h)(1\!-\!c\!+\!j)]$. It can be derived with a common radial regulator in a convergence domain and then continued with the same finite part as the full soft integral.\

Divide each result by the double-local-first-pole contribution in the \emph{same $B$ channel}. Restoring the $p$ or $q$ row gives, respectively,
\begin{subequations}
  \begin{align}
    F_p(c,2\ii\tnu) & =\frac{\,4\ii\tnu(2\ii\tnu\!+\!1)(2\ii\tnu c\!+\!c\!+\!(2\ii\tnu)^2\!+\!2\ii\tnu\!-\!2)\,}{\,(2\!-\!c)(2\ii\tnu\!+\!c)(2\ii\tnu\!-\!1)(2\ii\tnu\!+\!c\!+\!1)(2\ii\tnu\!+\!c\!+\!2)\,}\,,\label{sm:Fp}              \\
    F_q(c,2\ii\tnu) & =\frac{\,4\ii\tnu(2\ii\tnu\!+\!1)(-c^2\!+\!3c\!+\!(2\ii\tnu)^2\!+\!6\ii\tnu\!-\!2)\,}{\,(2\!-\!c)(1\!-\!c)(2\ii\tnu\!+\!c)(2\ii\tnu\!-\!1)(2\ii\tnu\!+\!c\!+\!1)(2\ii\tnu\!+\!c\!+\!2)\,}\,.\label{sm:Fq}
  \end{align}
\end{subequations}
These are finite-parameter rational identities.\  Applying $a(2\ii\tnu\!+\!a)\!=\!\m^2$ only afterward gives
\begin{equation}
  F_p(1\!+\!a,2\ii\tnu)=2\!+\!\frac{\,2(\m^2\!-\!2)\,}{\,2\ii\tnu\,}\!+\!O_{\m}((2\ii\tnu)^{-2}),\qquad
  F_q(a,2\ii\tnu)=1\!+\!\frac{\,3\m^2\!+\!4\,}{\,4\ii\tnu\,}\!+\!O_{\m}((2\ii\tnu)^{-2})\,.
  \label{sm:resummedlimits}
\end{equation}
The other channels are exponentially suppressed relative to channel $B$ in this limit.\  Restoring the $p$ row supplies the remaining half, whereas restoring only the $q$ row leaves the first-pole limit unchanged.\  The finite resummation establishes these results before taking the large-parameter limit.\

If the entire \emph{wave}, including its nonlocal row, is restored on that same line, the additional contribution divided by its double-local first term is
\begin{equation}
  -\frac{\,\Gamma(2\ii\tnu\!+\!c)\,}{\,2\ii\tnu\Gamma(2\ii\tnu)\Gamma(c)\,}
  \frac{\,K_{{\rm{N}},p/q}(c,2\ii\tnu)\,}{\,4(1\!-\!2\ii\tnu)/[2\ii\tnu(2\ii\tnu\!+\!1)]\,}\,,
  \label{sm:nonlocalratio}
\end{equation}
where $K_{{\rm{N}},p}$ and $K_{{\rm{N}},q}$ are the integrated nonlocal kernels for restoring the $p$ and $q$ line, respectively:
\begin{subequations}
  \begin{align}
    K_{{\rm{N}},p}(c,2\ii\tnu) & =\frac{\,16\ii\tnu\,}{\,(c\!-\!2)(2\ii\tnu\!+\!1)(2\ii\tnu\!+\!2)(2\ii\tnu\!+\!4)\,}\,,\label{sm:nonlocalp}                                            \\
    K_{{\rm{N}},q}(c,2\ii\tnu) & =\frac{\,16\ii\tnu(-c^2\!+\!2\ii\tnu c\!+\!5c\!-\!4\ii\tnu\!-\!8)\,}{\,c(c\!-\!2)(c\!-\!1)(2\ii\tnu\!+\!1)(2\ii\tnu\!+\!2)(2\ii\tnu\!+\!4)\,}\,.\label{sm:nonlocalq}
  \end{align}
\end{subequations}
Both normalized expressions are $-2/(2\ii\tnu)\!+\!O_{\m}((1\!+\!\log|2\ii\tnu|)/|2\ii\tnu|^2)$ with their respective $c$. The apparent $1/c$ in the second cancels the zero in $1/\Gamma(c)$. Neither changes Eq.\,\eqref{sm:resummedlimits}.

The mixed first-pole pairs are subleading.\  After summing all SK paths, their ratios to the full coefficient are
\begin{equation}
  \frac{\,\mathbb C_+^{\rm{LN}(0,0)}\,}{\,\Cc\,}=
  \frac{\,2\!+\!o_{\m}(1)\,}{\,(2\tnu)^2\,},\qquad
  \frac{\,\mathbb C_+^{\rm{NL}(0,0)}\,}{\,\Cc\,}=
  \frac{\,1\!+\!o_{\m}(1)\,}{\,(2\tnu)^2\,}\,.
  \label{sm:mixedfirst}
\end{equation}
The nonlocal--nonlocal first-pole coefficient is $\Lambda^{-3}O_{\m}(\tnu^{-1})$ and tends to zero at fixed $\Lambda$.\  Its radial phase has no stationary point; the spin polynomial vanishes at the lower boundary, so two integrations by parts suffice, with the original power-tail continuation retained.\  The other seven SK paths of the double-local first term also have no constant asymptotic contribution.\  The two halves therefore come from the local first-pole pair and the surviving local poles with $n\ge1$.\

\section{Benchmark Points and Numerical Analysis}
\label{sm:checks}
\label{App:H}
We relate $\Cc$ to the curvature-bispectrum amplitude and phase, specify the inputs for Fig.\ref{fig:fnl}, and present four benchmark evaluations. The analytic small-mass limit and alternative evaluations of the soft integral provide checks on the complex coefficient. We also use loop counting to estimate the size of the two-loop correction at the benchmark cutoffs $\Lambda/H=30$ and $60$.

\subsection{Normalization and the Mass Scan}\label{sm:normalization}
We obtain $\fNLCC$ by matching the coefficient of one oscillatory power in the loop calculation to the same coefficient in the bispectrum template.\ In Eq.\,\eqref{eq:target}, each external inflaton field gives one factor of $-\frac{\,H\,}{\,\dot\phi_0\,}$ when converted to $\zeta$.\ For $\dot\phi_0>0$ and $\rho\!=\!\frac{\,k_3\,}{\,k_1\,}$, the result is
\begin{equation}
  \langle\zeta_{\vct k_1}\zeta_{\vct k_2}\zeta_{\vct k_3}\rangle'_{\rm {osc,LO}}
  =-\frac{\,H^9\,}{\,\dot\phi_0^3 k_1^6\,}
  \left[\Cc\rho^{\ii\hp 2\tnu}\!+\!\Cc^*\rho^{-\ii\hp 2\tnu}\right]\,.
  \label{sm:curvatureclock}
\end{equation}

We use the template normalization of Eqs.\,(4.1)--(4.4) of Ref.\,\cite{You:2026CC}.\ It follows the Planck convention\,\cite{Planck9} for the primordial Bardeen potential $\Phi$, with $\Phi\!=\!\frac{\,3\,}{\,5\,}\zeta$ on superhorizon scales during matter domination.\ In the scale-invariant approximation, let $A_\Phi$ denote the power-spectrum amplitude of $\Phi$ (called $A$ in that reference), and let $\mathcal C(k_1,k_2,k_3)$ denote the dimensionless bispectrum template.\ Then
\begin{subequations}\label{sm:templateNormalization}
\begin{align}
 A_\Phi &\equiv k^3\langle\Phi_{\vct k}\Phi_{-\vct k}\rangle'
 =\left(\!\frac{\,3\,}{\,5\,}\!\right)^{\!\!2} k^3\langle\zeta_{\vct k}\zeta_{-\vct k}\rangle'
 =\left(\!\frac{\,3\,}{\,5\,}\!\right)^{\!\!2} 2\pi^2\Pz\,,
 \label{sm:bardeenPower}\\
 \langle\Phi_{\vct k_1}\Phi_{\vct k_2}\Phi_{\vct k_3}\rangle'
 &=\frac{\,6A_\Phi^2\,}{\,(k_1 k_2 k_3)^2\,}\,\fNL\,\mathcal C(k_1,k_2,k_3)\,.
 \label{sm:bardeenTemplate}
\end{align}
The factor $6$ sets the conventional amplitude unit.\ For example, in the local quadratic ansatz the quadratic field can occur on any of the three legs, with two Wick contractions per leg; at equal momenta all six terms have the same weight.\ We use this unit with the oscillatory shape below.\ Converting the three potentials in Eq.\,\eqref{sm:bardeenTemplate} to $\zeta$ gives
\begin{equation}
 \langle\zeta_{\vct k_1}\zeta_{\vct k_2}\zeta_{\vct k_3}\rangle'
 =\left(\!\frac{\,5\,}{\,3\,}\!\right)^{\!\!3}
  \frac{\,6A_\Phi^2\,}{\,(k_1 k_2 k_3)^2\,}\,\fNL\,\mathcal C(k_1,k_2,k_3)
 =\frac{\,72\pi^4\Pz^2\,}{\,5(k_1 k_2 k_3)^2\,}\,\fNL\,\mathcal C(k_1,k_2,k_3)\,.
 \label{sm:curvaturetemplate}
\end{equation}
Thus $\frac{\,72\pi^4\,}{\,5\,}$ comes from $6\left(\!\frac{\,5\,}{\,3\,}\!\right)^{\!\!3}\left(\!\frac{\,3\,}{\,5\,}\!\right)^{\!\!4}(2\pi^2)^2$: the four powers of $\frac{\,3\,}{\,5\,}$ enter through the two powers of $A_\Phi$, and the three inverse factors enter through the bispectrum.\ For the leading clock signal we set $\fNL\!=\!\fNLCC$ and use the squeezed template of Ref.\,\cite{You:2026CC},
\begin{equation}
 \mathcal C(k_1,k_2,k_3)\simeq
 e^{\ii\varphi_0}\left(\!\frac{\,k_1\,}{\,k_3\,}\!\right)^{\!\!\ii\hp 2\tnu-2}\!+\!\mathrm{c.c.}
 =2\rho^2\cos\!\left(2\tnu\ln\frac{\,k_1\,}{\,k_3\,}\!+\!\varphi_0\right)\,,
 \label{sm:squeezedClockShape}
\end{equation}
\end{subequations}
where $\mathrm{c.c.}$ denotes the complex conjugate of the preceding term.\ The two complex powers give the factor $2$ in the cosine amplitude.\ The remaining $\rho^2$ gives $\frac{\,\rho^2\,}{\,(k_1 k_2 k_3)^2\,}\!=\!\frac{\,1\,}{\,k_1^4 k_2^2\,}$, which equals $k_1^{-6}[1\!+\!O(\rho)]$ for $k_2\!=\!k_1\!+\!k_3$.\ Keeping the leading term gives the normalization used in Eq.\,\eqref{eq:fnltemplate}:
\begin{equation}
 \langle\zeta_{\vct k_1}\zeta_{\vct k_2}\zeta_{\vct k_3}\rangle'_{\rm {osc,LO}}
 =\frac{\,144\pi^4\Pz^2\,}{\,5k_1^6\,}\fNLCC
 \cos\!\left(2\tnu\ln\frac{\,k_1\,}{\,k_3\,}\!+\!\varphi_0\right)\,.
 \label{sm:cosMain}
\end{equation}

The template phase $e^{\ii\varphi_0}$ multiplies $\rho^{-\ii\hp 2\tnu}$.\ Matching this power in Eqs.\,\eqref{sm:curvatureclock} and \eqref{sm:curvaturetemplate}--\eqref{sm:squeezedClockShape} cancels $k_1^{-6}$ and gives
{\small
\begin{subequations}\label{sm:fnlmatch}
\begin{align}
 \frac{\,72\pi^4\Pz^2\,}{\,5\,}\,\fNLCC e^{\ii\varphi_0}
 &=-\frac{\,H^9\,}{\,\dot\phi_0^3\,}\Cc^*\,,
 \label{sm:oneClockPower}\\
 \fNLCC
 &=\frac{\,5H^9\,}{\,72\pi^4\Pz^2\dot\phi_0^3\,}|\Cc|
 =\frac{\,5H^3\,}{\,9\pi\sqrt{\Pz}\,}|\Cc|\,,
 \qquad \varphi_0=\arg(-\Cc^*)=\pi\!-\!\arg\Cc\pmod{2\pi}\,.
 \label{sm:clockModulusPhase}
\end{align}
\end{subequations}
}
{\par\parindent=2.5mm
\hspace*{-2.5mm}%
The last equality for the amplitude uses the tree-level power $\Pz\!=\!\frac{\,H^4\,}{\,4\pi^2\dot\phi_0^2\,}$, or $\dot\phi_0\!=\!\frac{\,H^2\,}{\,2\pi\sqrt{\Pz}\,}$.\ With $\Cc\!=\!\Ch\Cs$ from Eq.\,\eqref{sm:factorizedresult}, this reproduces Eq.\,\eqref{eq:fnl}:
\par}
\begin{equation}
 \boxed{\fNLCC=\frac{\,5H^3\,}{\,9\pi\sqrt{\Pz}\,}|\Ch\,\Cs|\,.}
 \label{sm:fNLMain}
\end{equation}
In Eq.\,(4.4) of Ref.\,\cite{You:2026CC}, the factor $\frac{\,5\,}{\,9\pi H^3\sqrt{\Pz}\,}$ multiplies $|\mathcal A_{\ominus}^{\rm {tot}}|$, the magnitude of the nonlocal inflaton coefficient.\ For the complete signal the coefficient of the same momentum power, $\frac{\,\rho^{-\ii\hp 2\tnu}\,}{\,k_1^6\,}$, is $H^6\Cc^*$ in Eq.\,\eqref{eq:target}.\ This accounts for the $H^3$ in the numerator of Eq.\,\eqref{sm:fNLMain}.\ 

The curves in Fig.\ref{fig:fnl} use the tree power as their common reference.\  At each mass the background is fixed by
\begin{equation}
  \Pz=e^{3.047}\times10^{-10},\qquad
  \frac{\,\dot\phi_0\,}{\,H^2\,}=\frac{\,1\,}{\,2\pi\sqrt{\Pz}\,},\qquad
  \tlambda=\frac{\,1\,}{\,2\pi\sqrt{\Pz}\,(\Lambda/H)\,},\qquad \Lambda/H=30,60\,.
  \label{sm:scanbackground}
\end{equation}
At each cutoff, both finite curves use these inputs.\  The solid curves include the phases of all contributions to the complete coefficient, and the dashed curves use the finite-parameter nonlocal expression of Ref.\cite{You:2026CC}. Red and blue denote $\Lambda/H\!=\!30$ and $60$, respectively. The black dotted curves are the corresponding large-$\tlambda$ approximations in Eqs.\,\eqref{eq:asymnew} and \eqref{eq:asymold}. The mass grid is $\m\!=\!0.1,0.2,\ldots,10$, with interpolation used only to draw the curves between grid points.\

For the main scan at $\Lambda/H\!=\!30$, the complete and nonlocal amplitudes attain their grid maxima at $\m\!=\!3.7$ and $5.8$, respectively, as discussed in Sec.\,5. Changing the numerical resolution changes the complex complete coefficient by at most $1.4\times10^{-7}$ over this scan. An independent adaptive integration at the lowest-mass point agrees within $1.6\times10^{-6}$. These resolution comparisons characterize numerical stability of the leading coefficient; the finite-$\rho$ remainder of the squeezed expansion is a separate issue.\

\subsection{Complex Coefficients at Four Benchmark Points}
The four benchmarks in Table~\ref{tab:benchmarks} compare the two predictions at the same mass and background. For $\Lambda/H\!=\!30$ we use the unrounded background of Fig.\ref{fig:fnl}. For the three nominal $\Lambda/H\!=\!60$ benchmarks we retain $\tlambda\!=\!57.81$ and use Eq.\,\eqref{sm:scanbackground} to set the vertex factors.

\par\medskip\noindent\begin{minipage}{\linewidth}
\makeatletter\def\@captype{table}\makeatother
\centering
{\renewcommand{\arraystretch}{1.48}\setlength{\tabcolsep}{8pt}
\begin{tabular}{c|c||c|c|c|c}
\hline\hline
$\Lambda/H$ & $\m$ & $H^3|\Cc|$ & $\arg\Cc$ (rad) & {\color{red!82!black}$\fNLCC$} & {\color{blue!80!black}$f_{\rm{NL}}^{\rm{NLoc}}$} \\
\hline\hline
$60$ & $4$ & $1.35\times10^{-7}$ & $-2.97$ & {\color{red!82!black}$5.20\times10^{-4}$} & {\color{blue!80!black}$7.33\times10^{-6}$} \\
\hline
$30$ & $4$ & $3.58\times10^{-6}$ & $-3.00$ & {\color{red!82!black}$1.38\times10^{-2}$} & {\color{blue!80!black}$5.44\times10^{-5}$} \\
\hline
$60$ & $8$ & $4.96\times10^{-10}$ & $-2.30$ & {\color{red!82!black}$1.91\times10^{-6}$} & {\color{blue!80!black}$4.74\times10^{-7}$} \\
\hline
$60$ & $10$ & $4.90\times10^{-12}$ & $2.04$ & {\color{red!82!black}$1.89\times10^{-8}$} & {\color{blue!80!black}$3.40\times10^{-8}$} \\
\hline\hline
\end{tabular}\par}
\vspace{3pt}
\caption[Complex coefficients and oscillatory amplitudes at four benchmarks]{Leading oscillatory signal at four choices of cutoff $\Lambda/H$ and mass $\m\!=\!M_R/H$. The middle columns give $H^3|\Cc|$ and the phase in radians of the coefficient multiplying $\rho^{2\ii\tnu}$ in Eq.\,\eqref{eq:target}. The final columns give $f_{\rm{NL}}$ with the same tree-level power normalization: the red column ($\fNLCC$) uses the full modes on both soft lines; the blue column ($f_{\rm{NL}}^{\rm{NLoc}}$) keeps only the leading nonlocal late-time term on each line, as in Ref.\cite{You:2026CC}.}
\label{tab:benchmarks}
\end{minipage}\par\medskip

At the last benchmark point, restoring the omitted modes reduces the magnitude through destructive interference.\  This finite-mass cancellation explains the reduction relative to the earlier result, while Eq.\,\eqref{sm:asymratio} describes the asymptotic enhancement.\

To see how much each soft line contributes, Table~\ref{tab:singlelinecheck} restores one line at a time at the same four benchmarks.

\par\medskip\noindent\begin{minipage}{\linewidth}
\makeatletter\def\@captype{table}\makeatother
\centering
{\renewcommand{\arraystretch}{1.22}\setlength{\tabcolsep}{5pt}
\begin{tabular}{cc *{4}{r@{\hspace{.12em}}c@{\hspace{.12em}}l}}
\toprule
$\Lambda/H$ & $\m$ & \multicolumn{3}{c}{$p$: local poles} & \multicolumn{3}{c}{$q$: local poles} & \multicolumn{3}{c}{$p$: full mode} & \multicolumn{3}{c}{$q$: full mode} \\
\midrule
60 & 4 & $0.973$ & $+$ & $\ii\hp 0.153$ & $0.491$ & $+$ & $\ii\hp 0.026$ & $0.978$ & $+$ & $\ii\hp 0.162$ & $0.497$ & $+$ & $\ii\hp 0.035$ \\
30 & 4 & $0.991$ & $+$ & $\ii\hp 0.069$ & $0.497$ & $+$ & $\ii\hp 0.009$ & $0.992$ & $+$ & $\ii\hp 0.074$ & $0.498$ & $+$ & $\ii\hp 0.013$ \\
60 & 8 & $0.564$ & $+$ & $\ii\hp 0.404$ & $0.319$ & $+$ & $\ii\hp 0.105$ & $0.578$ & $+$ & $\ii\hp 0.393$ & $0.333$ & $+$ & $\ii\hp 0.094$ \\
60 & 10 & $-0.538$ & $-$ & $\ii\hp 0.305$ & $-0.286$ & $-$ & $\ii\hp 0.022$ & $-0.517$ & $-$ & $\ii\hp 0.280$ & $-0.263$ & $+$ & $\ii\hp 0.003$ \\
\bottomrule
\end{tabular}\par}
\vspace{3pt}
\caption[Restoring the soft modes one line at a time]{Complex ratios obtained by restoring one soft fermion line at a time. Let $\vct q$ be the loop momentum and $\vct k_3$ the external momentum at vertex-3. The line joining vertices-2 and-3 has momentum magnitude $p\!=\!|\vct q\!-\!\vct k_3|$; the line joining vertices~3 and~1 has magnitude $q\!=\!|\vct q|$. The columns labeled \emph{local poles} sum all local poles on the indicated line at the soft vertex; the columns labeled \emph{full mode} also include its nonlocal poles. The other line retains its first local pole ($n\!=\!0$) and its exponential. Each entry divides the sum of the four spinor channels on the $++-$ contour by the complete $\Cc$ summed over all contours at the same parameters.}
\label{tab:singlelinecheck}
\end{minipage}\par\medskip

At $\m=4$, the local-pole ratios are close to $1$ and $1/2$, consistent with the limits discussed after Eq.\,\eqref{sm:resummedlimits}. At $\m=10$, all four ratios have negative real parts, reflecting relative phases. Observable amplitudes depend on the modulus of the combined coefficient.

\begin{samepage}
\subsection{An Analytic Check at Small Mass}\label{sm:smallmasscheck}
At each fixed $\tlambda>0$, the coefficient of the leading mass term in the complete result is
\begin{equation}
 \Cc=-\frac{\m^2}{16\pi\Lambda^3}\tanh(\pi\tlambda)
 \frac{(2\ii\tlambda+4)(2\ii\tlambda-1)}{(2\ii\tlambda)(2\ii\tlambda+3)}
 +\frac{O_{\tlambda}(\m^4)}{\Lambda^3}.
 \label{sm:smallmasscoefficient}
\end{equation}
The remainder is defined at fixed $\tlambda$. In the channel assembly, $A,C$ start at $\m^4$, whereas $B,D$ start at $\m^2$. Relative to the common factor $-\m^2\tanh(\pi\tlambda)/(16\pi\Lambda^3)$, their leading rational factors are $1-2/[(2\ii\tlambda)(2\ii\tlambda+3)]$ and $-2/[(2\ii\tlambda)(2\ii\tlambda+3)]$, respectively. Their sum gives Eq.\,\eqref{sm:smallmasscoefficient} and agrees with the exact massless zero in Eq.\,\eqref{sm:masslesszero}. This checks the $\m^2$ law and finite-$\tlambda$ phase. The $D$ term is relatively $O(\tlambda^{-2})$ only at large chemical potential. In that limit, the mass coefficient tends to $-1/(16\pi\Lambda^3)$; a fixed-mass conclusion still requires the uniform bound of Appendix\,\ref{sm:asym}.\

\end{samepage}
\subsection{Independent Checks and Numerical Evaluation}
Gamma recurrence and reflection identities verify the single-Beta form of the hard factor. Multiplying the prefactors of Eqs.\,\eqref{eq:hardfinite} and \eqref{eq:softfinite} gives
\begin{equation}
  \frac{\,\ii2^{A-8}\,}{\,\Lambda^3 A(A\!+\!2)(A\!-\!1\!-\!\m^2)\cosh^2(\pi\tnu)\mathrm B(a,b)\,}\,.
  \label{sm:prefactorproduct}
\end{equation}
This identity checks the normalization and phase of the combined hard and soft factors. The endpoint-subtracted soft integral and its equivalent first-order differential system yield the coefficient without Meijer $G$ reduction, retaining the finite terms restored in Eq.\,\eqref{sm:finitepart}. Comparing their complex values tests the integrated soft response and normalization; the prefactor identity tests only the assembly.

The two moments in Eq.\,\eqref{eq:Wseed} were checked by direct quadrature in the auxiliary domain $k_3>p+q$. This tests the connection coefficients, powers and Gamma factors; equality in the physical triangle follows from analytic continuation (Appendix\,\ref{sm:Wcontinuation}).

Evaluate Eqs.\,\eqref{eq:hardfinite} and \eqref{eq:softfinite}--\eqref{eq:Hfinite} with exact or high-precision inputs; sum complex terms before taking the modulus. Under strong cancellations, test stability by varying precision and integration resolution independently. Let $\Cc^{(1)}$ and $\Cc^{(2)}$ denote two numerical evaluations of the same one-loop coefficient at the same physical point with different numerical settings. Compare $|\Cc^{(1)}-\Cc^{(2)}|$ and $|\Cc^{(1)}-\Cc^{(2)}|/|\Cc^{(2)}|$; near a zero, use the absolute difference and leave the phase undefined. These tests use the leading squeezed coefficient at fixed parameters, normalized by Eq.\,\eqref{sm:fnlmatch}.

\subsection{On the Validity of Loop Expansion}
\label{sm:cutoffcounting}

\begin{figure}[!t]
\centering
\begin{tikzpicture}[
  neutrino/.style={line width=.9pt,postaction={decorate},
    decoration={markings,mark=at position .12 with {\arrow{>}}}},
  inflaton/.style={line width=.8pt,dash pattern=on 3pt off 2pt},
  every node/.style={font=\small}]
  \foreach \shiftx/\panel in {0/a,5.3/b}{
    \begin{scope}[shift={(\shiftx,0)}]
      \draw[neutrino] (0,0) circle (.80);
      \foreach \theta in {90,210,330}{
        \draw[inflaton] (\theta:.80)--(\theta:1.43);
        \fill (\theta:.80) circle (1.6pt);
      }
      \def\firstpanel{a}
      \ifx\panel\firstpanel
        \node at (0,0) {$N_{\!R}$};
        \node at (0,-1.09) {(a)};
      \else
        \draw[inflaton] (150:.80) to[bend right=22] (30:.80);
        \foreach \theta in {150,30}{\fill (\theta:.80) circle (1.6pt);}
        \node at (0,-.14) {$\delta\phi$};
        \node at (0,-1.09) {(b)};
      \fi
    \end{scope}
  }
\end{tikzpicture}\par
\caption[One-loop triangle and two-loop inflaton exchange]{%
Three-point inflaton correlation function from the one-loop and two-loop contributions.\  
Solid lines denote the right-handed Majorana neutrino $N_R^{}$ and dashed lines the inflaton 
fluctuation $\delta\phi$.\ The internal dashed line in plot\,(b) connects the two additional 
derivative vertices.}
\label{fig:loopcounting}
\end{figure}
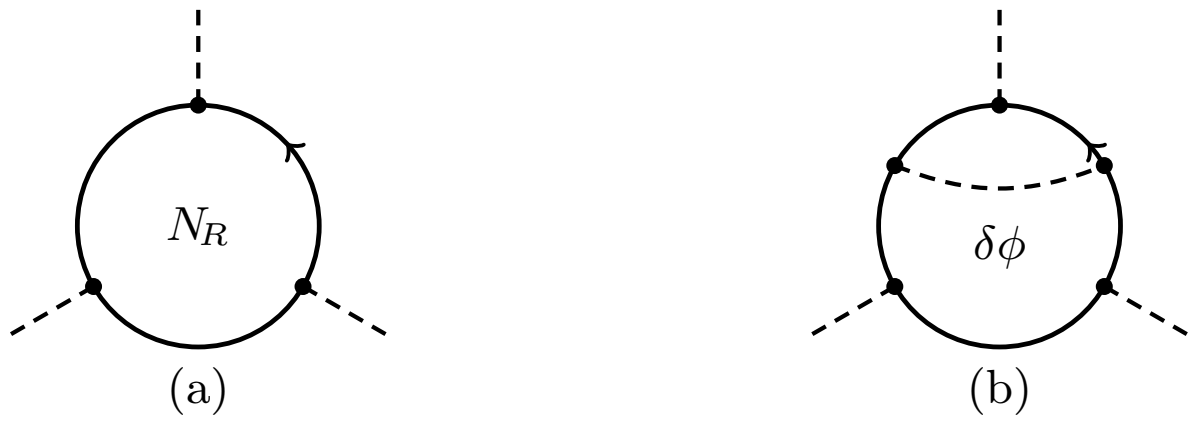

In Fig.\,\ref{fig:loopcounting}, we show the one-loop triangle contribution
to the inflaton three-point function and a representative two-loop correction.\
In panel (b), the exchanged inflaton connects two segments of the same neutrino loop.\
Each additional vertex splits a neutrino propagator into two,
so the exchange adds two neutrino propagators, one inflaton propagator,
and one independent loop-momentum integration.

\vs 

In the following, we estimate the sizes of the one-loop versus two-loop contributions 
by proper power counting.\ 
Let $p_*$ be the characteristic physical momentum of the exchange and $R_{21}^{}$ 
the ratio of the two-loop over one-loop contributions.\ 
The following counting gives a parametric estimate of this ratio:
\begin{equation}
  R_{21}\sim
  \underbrace{\frac{p_*^4}{16\pi^2}}_{\text{one loop integral}}\,
  \times\underbrace{\frac{1}{p_*^2}}_{\text{two neutrino lines}}\,
  \times\underbrace{\frac{1}{p_*^2}}_{\text{one inflaton line}}\,
  \times\underbrace{\frac{p_*^2}{\Lambda^2}}_{\text{two derivative vertices}}
  =\frac{p_*^2}{16\pi^2\Lambda^2}\,.
  \label{sm:extraloop}
\end{equation}

For the numerical analysis in Fig.\,\ref{fig:fnl}, 
we have $p_*\!\sim\!|\lambda|\!=\!|\dot\phi_0|/\Lambda\hs$, 
where $p_*$ represents the typical momentum scale of neutrino production.\ 
Thus, we can further obtain the following estimate: 
\begin{equation}
  R_{21}^{} \sim \frac{1}{16\pi^2}
  \frac{|\lambda|^2}{\,\Lambda^2\,}
  =\frac{\dot\phi_0^{\,2}}{16\pi^2\Lambda^4}
  \equiv \epsilon_{\rm{loop}} \,, 
\label{sm:loopcounting}
\end{equation}
where rolling speed of the inflaton background field is $|\dot\phi_0|\hsm\simeq\hsm (60H)^2$.\ 
The validity the perturbative loop expansion requies $R_{21}^{} \!<\! 1$,
which imposes a lower bound on the cutoff scale $\Lambda$.\ 
The condition $R_{21}^{} \!<\! 1$ 
is presented in the first row of Table\,\ref{tab:cutoffcounting}.\  
For comparison, we also illustrate the stronger conditions
such as $R_{21}^{} \!<\! 0.1$ (the second row) and $R_{21}^{} \!<\! 0.01$ (the third row).
This table shows that the condition $R_{21}^{} \!<\! 1$ only places a mild lower bound
on the cutoff scale $\Lambda\hs$.\

\vspace*{5mm}
\par\smallskip\noindent\begin{minipage}{\linewidth}
\makeatletter\def\@captype{table}\makeatother
\centering
{\renewcommand{\arraystretch}{1.15}
\setlength{\tabcolsep}{10pt}
\begin{tabular}{c|c|c}
\hline\hline
Perturbation Condition
& Lower Bound on $\Lambda/H$
&  Benchmarks of $\Lambda$ in Fig.\,\ref{fig:fnl} \\
\hline
$\epsilon_{\rm{loop}}\lesssim 1\phantom{.0}$
& $\gtrsim 17$ & $30,\;60$ \\
$\epsilon_{\rm{loop}}\lesssim 0.1$
& $\gtrsim 30$ & $30,\;60$ \\
\hline\hline
\end{tabular}\par}
\vspace{3pt}
\caption[Loop-counting criteria for the two benchmark cutoffs]{%
Estimated lower bounds on the cutoff scale $\Lambda\hs$ as imposed by requiring
the validity of the perturbative loop-expansion.}
\label{tab:cutoffcounting}
\end{minipage}\par
\end{document}